\documentclass[]{aa}

\usepackage{natbib}
\usepackage{graphics}
\usepackage{graphicx}
\usepackage{array}
\usepackage{epstopdf}
\usepackage{url}
\usepackage[T1]{fontenc}
\usepackage[utf8]{inputenc}  % accents 8 bits dans le source

\def\msunyr{M$_{\odot}$~yr$^{-1}$}
\def\ha{\ifmmode {\rm H{\alpha}} \else $\rm H{\alpha}$\fi}
\def\hb{\ifmmode {\rm H{\beta}} \else $\rm H{\beta}$\fi}
\def\oii{[O\,{\sc ii}] $\lambda$3727}

\def\oiii{[O\,{\sc iii}] $\lambda\lambda$4959,5007}

\def\oiiib{[O\,{\sc iii}] $\lambda$5007}

\def\H2{H{\sc ii}}

\newcommand{\msun}{\,{\rm M_\odot}}
\newcommand{\kms}{ km~s$^{-1}$}
\newcommand{\vf}{velocity field}

\def\degr{\hbox{$^\circ$}}

\usepackage[pdftex]{hyperref}
\def\jnl@style{\it}
\def\aaref@jnl#1{{\jnl@style#1}}
\def\aaref@jnl#1{{\jnl@style#1}}

\def\aj{\aaref@jnl{AJ}}                   % Astronomical Journal
\def\araa{\aaref@jnl{ARA\&A}}             % Annual Review of Astron and Astrophys
\def\apj{\aaref@jnl{ApJ}}                 % Astrophysical Journal
\def\apjl{\aaref@jnl{ApJ}}                % Astrophysical Journal, Letters
\def\apjs{\aaref@jnl{ApJS}}               % Astrophysical Journal, Supplement
\def\ao{\aaref@jnl{Appl.~Opt.}}           % Applied Optics
\def\apss{\aaref@jnl{Ap\&SS}}             % Astrophysics and Space Science
\def\aap{\aaref@jnl{A\&A}}                % Astronomy and Astrophysics
\def\aapr{\aaref@jnl{A\&A~Rev.}}          % Astronomy and Astrophysics Reviews
\def\aaps{\aaref@jnl{A\&AS}}              % Astronomy and Astrophysics, Supplement
\def\azh{\aaref@jnl{AZh}}                 % Astronomicheskii Zhurnal
\def\baas{\aaref@jnl{BAAS}}               % Bulletin of the AAS
\def\jrasc{\aaref@jnl{JRASC}}             % Journal of the RAS of Canada
\def\memras{\aaref@jnl{MmRAS}}            % Memoirs of the RAS
\def\mnras{\aaref@jnl{MNRAS}}             % Monthly Notices of the RAS
\def\pra{\aaref@jnl{Phys.~Rev.~A}}        % Physical Review A: General Physics
\def\prb{\aaref@jnl{Phys.~Rev.~B}}        % Physical Review B: Solid State
\def\prc{\aaref@jnl{Phys.~Rev.~C}}        % Physical Review C
\def\prd{\aaref@jnl{Phys.~Rev.~D}}        % Physical Review D
\def\pre{\aaref@jnl{Phys.~Rev.~E}}        % Physical Review E
\def\prl{\aaref@jnl{Phys.~Rev.~Lett.}}    % Physical Review Letters
\def\pasp{\aaref@jnl{PASP}}               % Publications of the ASP
\def\pasj{\aaref@jnl{PASJ}}               % Publications of the ASJ
\def\qjras{\aaref@jnl{QJRAS}}             % Quarterly Journal of the RAS
\def\skytel{\aaref@jnl{S\&T}}             % Sky and Telescope
\def\solphys{\aaref@jnl{Sol.~Phys.}}      % Solar Physics
\def\sovast{\aaref@jnl{Soviet~Ast.}}      % Soviet Astronomy
\def\ssr{\aaref@jnl{Space~Sci.~Rev.}}     % Space Science Reviews
\def\zap{\aaref@jnl{ZAp}}                 % Zeitschrift fuer Astrophysik
\def\nat{\aaref@jnl{Nature}}              % Nature
\def\iaucirc{\aaref@jnl{IAU~Circ.}}       % IAU Cirulars
\def\aplett{\aaref@jnl{Astrophys.~Lett.}} % Astrophysics Letters
\def\apspr{\aaref@jnl{Astrophys.~Space~Phys.~Res.}}
\def\bain{\aaref@jnl{Bull.~Astron.~Inst.~Netherlands}} 
\def\fcp{\aaref@jnl{Fund.~Cosmic~Phys.}}  % Fundamental Cosmic Physics
\def\gca{\aaref@jnl{Geochim.~Cosmochim.~Acta}}   % Geochimica Cosmochimica Acta
\def\grl{\aaref@jnl{Geophys.~Res.~Lett.}} % Geophysics Research Letters
\def\jcp{\aaref@jnl{J.~Chem.~Phys.}}      % Journal of Chemical Physics
\def\jgr{\aaref@jnl{J.~Geophys.~Res.}}    % Journal of Geophysics Research
\def\jqsrt{\aaref@jnl{J.~Quant.~Spec.~Radiat.~Transf.}}
\def\memsai{\aaref@jnl{Mem.~Soc.~Astron.~Italiana}}
\def\nphysa{\aaref@jnl{Nucl.~Phys.~A}}   % Nuclear Physics A
\def\physrep{\aaref@jnl{Phys.~Rep.}}   % Physics Reports
\def\physscr{\aaref@jnl{Phys.~Scr}}   % Physica Scripta
\def\planss{\aaref@jnl{Planet.~Space~Sci.}}   % Planetary Space Science
\def\procspie{\aaref@jnl{Proc.~SPIE}}   % Proceedings of the SPIE

\begin{document}

%    \title{Mass \textsc{a}ssembly \textsc{s}urvey with \textsc{Sinfoni} \textsc{}n \textsc{Vvds} \thanks{Based on observations collected at the European Southern Observatory (ESO) Very Large Telescope, Paranal, Chile, as part of the Large Program 179.A-0823}}
   \title{MASSIV: Mass Assembly Survey with SINFONI in VVDS
\thanks{Based on observations collected at the European Southern Observatory (ESO) Very Large Telescope, Paranal, Chile, as part of the Programs 179.A-0823, 177.A-0837, 78.A-0177, 75.A-0318, and 70.A-9007.}
\thanks{Based on observations obtained with MegaPrime/MegaCam, a joint project of CFHT and CEA/DAPNIA, at the Canada-France-Hawaii Telescope (CFHT) which is operated by the National Research Council (NRC) of Canada, the Institut National des Sciences de l'Univers of the Centre National de la Recherche Scientifique (CNRS) of France, and the University of Hawaii. This work is based in part on data products produced at TERAPIX and the Canadian Astronomy Data Centre as part of the Canada-France-Hawaii Telescope Legacy Survey, a collaborative project of NRC and CNRS.}
% \thanks{VVDS}
% \thanks{VIMOS}
}
   \subtitle{II. Kinematics and close environment classification}
\author{
% 	 B. Epinat et al. \inst{1,2}\thanks{E-mail: benoit.epinat@ast.obs-mip.fr}
% % 
	 B. Epinat \inst{1,2,3}\thanks{E-mail: benoit.epinat@oamp.fr}
% 	\and T. Contini\inst{1,2}
	\and L. Tasca\inst{3}
	\and P. Amram\inst{3}
	\and T. Contini\inst{1,2}
	\and O. Le F\`evre\inst{3}
	\and J. Queyrel\inst{1,2}
% 	\and L. Tasca\inst{3}
	\and D. Vergani\inst{4}
	\and B. Garilli\inst{5}
	\and M. Kissler-Patig\inst{6}
	\and J. Moultaka\inst{1,2}
	\and L. Paioro\inst{5}
	\and L. Tresse\inst{3}
	\and F. Bournaud\inst{7}
	\and C. L\'{o}pez-Sanjuan\inst{3}
	\and V. Perret\inst{3}
	}
\institute{Institut de Recherche en Astrophysique et Planétologie (IRAP), CNRS, 14, avenue \'Edouard Belin, F-31400 Toulouse, France
	\and IRAP, Universit\'e de Toulouse, UPS-OMP, Toulouse, France
	\and Laboratoire d'Astrophysique de Marseille, Universit\'e d'Aix-Marseille \& CNRS, UMR7326, 38 rue F. Joliot-Curie, F-13388 Marseille Cedex 13, France 
	\and INAF-Osservatorio Astronomico di Bologna, via Ranzani 1, I-40127, Bologna, Italy
	\and IASF-INAF, Via Bassini 15, I-20133, Milano, Italy
	\and ESO, Karl-Schwarzschild-Str.2, D-85748 Garching b. M\"unchen, Germany
 	\and Laboratoire AIM Paris-Saclay, CEA/IRFU/SAp, Université Paris Diderot, CNRS, F-91191 Gif-sur-Yvette Cedex, France.
	}

\abstract
{Processes driving mass assembly are expected to evolve on different timescales along cosmic time.
% From high to low redshifts, galaxy mass assembly processes are thought to change.
%  In particular cold gas accretion might be less efficient at low redshift.
A transition might happen around $z\sim1$ as the cosmic star formation rate starts its decrease.}
{We aim to identify the dynamical nature of galaxies in a
%  statistical and
representative sample to be able to infer and compare the mass assembly mechanisms across cosmic time.}
{We present an analysis of the kinematics properties of 50 galaxies with redshifts $0.9<z<1.6$ from the MASSIV sample observed
%  using integral field techniques
with SINFONI/VLT with a mass range from $4.5 \times 10^{9} \msun$ to $1.7 \times 10^{11} \msun$ and a star formation rate from 6 \msunyr\ to 300 \msunyr. This is the largest sample with 2D kinematics in this redshift range. We provide a classification based on kinematics as well as on close galaxy environment.}
{We find that a significant fraction of galaxies in our sample (29\%) experience merging or have close companions that may be gravitationally linked. This places a lower limit on the fraction of interacting galaxies because ongoing mergers are probably also present but harder to identify.
% observed but hardly distinguished.
We find that at least 44\% of the galaxies in our sample display ordered rotation, whereas at least 35\% are non-rotating objects.
% We also find 44\% of non-rotating galaxies, the others clearly displaying ordered rotation.
All rotators except one are compatible with rotation-dominated ($V_{max}/\sigma>1$) systems. Non-rotating objects are mainly small objects ($R_e < 4$ kpc). They  show an anti-correlation of their velocity dispersion and their effective radius.
% These objects may be on-going mergers in a transient state, galaxies with only one unresolved star-forming region, galaxies with unstable gaseous disks or less probably spheroids, since they are not massive ($\log{M_{star}}<10.5$).
These low-mass objects ($\log{M_{star}}<10.5$) may be ongoing mergers in a transient state, galaxies with only one unresolved star-forming region, galaxies with an unstable gaseous phase or, less probably, spheroids.
% , since they are not massive ($\log{M_{star}}<10.5$).
% Compared to
Combining our sample with other 3D-spectroscopy samples, we find that the local velocity dispersion of the ionized gas component decreases continuously from $z\sim 3$ to $z=0$. The proportion of disks also seems to be increasing in star-forming galaxies when the redshift decreases.
The number of interacting galaxies seems to be at a maximum at $z\sim 1.2$.
%  as well as the number of interacting galaxies.
}
{These results draw a picture in which cold gas accretion may still be efficient at $z\sim1.2$ but in which mergers may play a much more significant role at $z\sim1.2$ than at higher redshift. From a dynamical point of view, the redshift range $1<z<2$ therefore appears as a transition period in the galaxy mass assembly process. All the data published in this paper are publicly available at the time of the publication following this link: \url{http://cosmosdb.lambrate.inaf.it/VVDS-SINFONI}.}

\keywords{galaxies: evolution -- galaxies: formation -- galaxies: kinematics and dynamics -- galaxies: high-redshift}

\maketitle

\clearpage
\section{Introduction}
\label{intro}

During the last decade the first observations of distant galaxies with Integral Field Unit spectrographs (IFU) have led to the construction of several galaxy samples from $z\sim0.5$ to $z\sim3$. These IFU observations have given new insights into the resolved physical properties of galaxy populations at various redshifts but a clear view of the transition epoch between young, unstable and clumpy galaxies and evolved and stable galaxies that form the Hubble Sequence is still missing.

In the redshift range $0.5<z<3$, large quantities of gas are converted into stars, producing a peak in the cosmic star formation at these epochs (see e.g. \citealp{Hopkins:2006,Tresse:2007}). Large gas reservoirs are being accreted onto galaxies via various mechanisms: isolated events such as galaxy major and minor mergers (e.g. \citealp{deRavel:2009,Conselice:2008,Lin:2008,Lopez-Sanjuan:2011}) and cold gas accretion along cosmic filaments, a more continuous process
% and more continuously thanks to cold gas accretion along cosmic filaments as suggested by recent cosmological simulations
(e.g. \citealp{Keres:2005,Dekel:2006,Genel:2008,Dekel:2009}).
Cosmological numerical simulations have helped to outline the importance that these processes may play in the build-up of the present-day Hubble sequence. 
% These processes that may lead to the build-up of the modern Hubble sequence are evidenced from cosmological numerical simulations.
% suggested by cosmological simulations.
Indeed, on the one hand, merging galaxies are a natural mass assembly mechanism expected in the $\Lambda$CDM framework where dark matter halos grow from hierarchical assembly.
% since it predicts a hierarchical growth of dark matter halos in which galaxies reside.
Some authors have shown that a peak of major merger activity might occur around $1<z<2$ \citep{Ryan:2008,Conselice:2008,Lopez-Sanjuan:2009}.
% 
% On the other hand, fresh cold gas accretion along cosmic web filaments may play a significant role at $z>2$ but might be less usual at $z<1$ (e.g. \citealp{Dekel:2009,Keres:2009}). This process seems to be efficient for halo masses larger than $\sim 10^{10} \msun$ \citep{Bouche:2010}.
On the other hand, cold gas accretion may play a significant role at $z>2$ but might be less usual at $z<1$ (e.g. \citealp{Dekel:2009,Keres:2009}). This process seems to be efficient for halo masses larger than $\sim 10^{11} \msun$ \citep{Bouche:2010}.
%  higher redshifts (e.g. \citealp{Dekel:2009}). \citet{Keres:2009} have highlighted that the latter might be efficient at $z>2$ but not at $z<1$.

% concordance, hierarchical, ,

% Epoque de formation de clusters. Convertion du gas en étoiles, réservoirs de gas importants (cf CO a plus grand z), stabilization des disques.

% Avant, galaxies plus bleues, plus de gaz, disques épais. Quels sont les précurseurs des elliptiques? Quel est le rôle de l'accrétion? Des fusions?

% but due to instrumental limits, $1<z<2$ is known as the redshift desert. The VVDS (\textsc{Vimos-Vlt} Deep Survey) has been a unique survey to find efficiently galaxies and has thus enabled to build MASSIV a representative sample in that redshift range.

Recent IFU studies aim at understanding the role of these different processes
% and to determine these
in the precursors of local ellipticals and spirals.
From the IMAGES sample at $0.4<z<0.75$ \citep{Yang:2008,Neichel:2008,Puech:2008,Rodrigues:2008}, it has been shown that regular rotating disks are quite similar to local rotators and that merging is playing a significant role in galaxy mass assembly.
% The IMAGES sample \citep{Yang:2008,Neichel:2008,Puech:2008,Rodrigues:2008} is covering the redshift range $0.4<z<0.75$. It has shown that since $z\sim 0.6$, galaxies have formed a significant fraction of their stellar content. This survey has also shown that regular rotating disks are rather similar to local rotators whereas merging is still playing a significant role in galaxy mass assembly.
% 
At higher redshift, the SINS sample at $z\sim 2.2$ \citep{Genzel:2008,Shapiro:2008,Forster-Schreiber:2009,Shapiro:2009,Cresci:2009}, the LSD/AMAZE sample at $z\sim3.3$ \citep{Gnerucci:2011} and the sample built by \citet{Law:2009} at $2<z<3$ are all containing many galaxies with high gaseous turbulence.
% Their interpretation is} that continuous cold gas accretion along cosmic web filaments is playing a significant role in mass assembly. They also show that merging is at play.
% and cold gas accretion are acting efficiently.
Based on numerical simulation results of unstable gas-rich disks (e.g. \citealp{Bournaud:2007,Dekel:2009}), these authors claim that continuous cold gas accretion along cosmic web filaments is playing a significant role in mass assembly at these redshifts because they find that half the disks are dispersion-dominated and that even rotation-dominated disks show a high gaseous local velocity dispersion (above 60\kms), uncommon for local galaxies \citep{Epinat:2010}.
\citet{Lehnert:2009} and \citet{Letiran:2011} suggested that this high gaseous local velocity dispersion is not directly powered by shocks or Jeans instabilities due to cold gas accretion but might be related to a vigorous star formation at both high- and low-redshift. This was also supported by \citet{Green:2010} who observed a few local analogs to these high-redshift galaxies that all have a star formation rate above 15$\msun/yr$. \citet{Goncalves:2010} also observed galaxies with high gaseous local velocity dispersion among a sample of $z\sim 0.2$ Lyman-break analogs that are also forming stars very efficiently. It has been suggested that this intense star formation could also be triggered by merging \citep{Basu-Zych:2007,Basu-Zych:2009a,Basu-Zych:2009b,Overzier:2008}.

% \citet{Green:2010} suggested that this high gaseous local velocity dispersion might be related to a vigorous star formation since they have observed a few local analogues to these high redshift galaxies that all have a star formation rate above 15$\msun/yr$. \citet{Goncalves:2010} also observed galaxies with high gaseous local velocity dispersion among a sample of $z\sim 0.2$ Lyman-break analogues that are also forming stars very efficiently.

Among these various samples (IMAGES, SINS and LSD/AMAZE), the redshift range $0.9<z<1.8$, at the peak of the cosmic star formation history, is still poorly explored.
% there is still an unexplored redshift range ($0.75<z<1.3$) using statistical samples.
The MASSIV survey has been built to study this redshift range, which seems to correspond to the period where the modern Hubble Sequence is being built (e.g. \citealp{Bell:2004,Williams:2009}).
%  and maybe a transition period when cold gas accretion is becoming less efficient and the contribution of major mergers to mass assembly becomes more important.
% but also where major merger peaks and cold gas accretion is being less efficient.

% disk stabilization could happen.

The first goal of this paper is to study the dynamical nature of 50 galaxies with $0.9<z<1.6$ from MASSIV and the evolution with redshift of the fraction of galaxies in a given dynamical state.
The second objective is to identify the main physical mechanisms responsible for the transition between $z\sim 2$, where many disks show high intrinsic gaseous velocity dispersions, and $z\sim 0.5$ where the disks seem to be more ``stable''.
The general presentation of the MASSIV sample selection and data acquisition strategy is the subject of a companion paper \citep{Contini:2011}.
An analysis focused on the dynamical properties of rotators is presented in a second companion paper (\citeauthor{Vergani:2012} submitted) and the analysis of the spatially-resolved metallicity of this first set of 50 MASSIV galaxies is discussed in an third paper \citep{Queyrel:2011}.

The paper is organized as follows. In section \ref{sample_obs_dr}, the MASSIV ``first epoch'' sample, the observations, and the data reduction are presented. In section \ref{models}, galaxy morphology and kinematics models are described.
% The impact of spectral resolution on local velocity dispersion is investigated in section \ref{veldispersion}.
The galaxy classification scheme is detailed in section \ref{classification} and the discussion of this classification is given in section \ref{discussion}.
Appendix \ref{comments} contains detailed informations and comments on each galaxy.

In this paper, we use the cosmological parameters $\Omega_m=0.3$, $\Omega_{\Lambda}=0.7$ and $H_0=70$ km s$^{-1}$ Mpc$^{-1}$.

% sigma en fonction de la couleur?

\section{Observations and data reduction}
\label{sample_obs_dr}

\subsection{Sample selection}

A complete description of the MASSIV sample selection and general properties is presented in \citet{Contini:2011}. Here we only present a brief summary.

The MASSIV sample contains 84 star-forming galaxies in the redshift range $0.9<z<1.8$.
It was constructed from the VVDS (VIMOS VLT Deep Survey) spectroscopic survey \citep{Lefevre:2005}.
This survey is I-band magnitude limited and complete up to magnitude $I_{AB}<24$ for VVDS-Deep (RA$=02$h, \citealp{Lefevre:2005}), $I_{AB}<24.75$ for VVDS-Ultra-Deep (RA$=02$h, \citealp{Cassata:2011}; Le F\`{e}vre et al. in prep)
and $I_{AB}<22.5$ for VVDS-Wide (RA$=14$h and RA$=22$h, \citealp{Garilli:2008}). VVDS provides low-resolution ($R\sim 230$) spectra that lead to accurate spectroscopic redshifts
% and to estimates of the star formation activity from the colours and from the \oii\ emission line
for 4446 galaxies in the redshift range $0.9<z<2$.

The selection of the MASSIV sample was based on star formation activity traced by \oii\ emission line equivalent width for galaxies with $z<1.46$ and from their observed photometric $UBVRIK$ spectral energy distribution for galaxies with $z>1.46$. These criteria ensure that the selected targets are star-forming galaxies for which strong emission lines can be studied using SINFONI to trace the kinematics. The \oii\ selection criteria has been tested on a pilot sample \citep{Epinat:2009} and has proven to be very efficient. Resulting from this selection function, the MASSIV sample provides a good representation of star-forming galaxies with $SFR \geq 5$ \msunyr\ at $z\sim 1.5$ in the stellar mass regime $10^9 - 10^{11}\msun$ (see \citealp{Contini:2011} for a detailed study).

The MASSIV ``first epoch'' sample discussed in this paper is a subsample of the MASSIV sample.
% It contains the 49 galaxies that have been observed before January 2010.
It contains the 50 galaxies with $0.9<z<1.6$ (the median redshift is 1.24) that ware observed before January 2010.
Except for two galaxies observed in the VVDS-Ultra-Deep, the galaxies studied in this paper are from the VVDS-Deep and VVDS-Wide samples.

\subsection{Observations}

SINFONI \citep{Eisenhauer:2003,Bonnet:2004} was used to obtain 2D spatially resolved spectroscopy in the NIR of the MASSIV galaxies around the \ha\ line or around the \oiii\ line for four galaxies.
% Out of an initial sample of 49 galaxies no line was detected for only 3 galaxies.
Out of an initial sample of 50 galaxies no line was detected for only four galaxies. The success rate is much better when the \ha\ line is targeted (44/46).
VVDS220148046 was observed at $z=2.244$, whereas it was expected at $z\sim1.371$ from the VIMOS spectrum. The redshift determination for this galaxy is based on the observation of \oiii\ and \hb\ lines in our SINFONI data.
The observations were obtained in service mode from period P79 to period P82 as part of the Large Programme 179.A-0823 (P.I. T. Contini) and during a pilot program presented in \citet{Epinat:2009} (ESO runs 75.A-0318 and 78.A-0177). Galaxies were observed in two or three observing blocks.

For the galaxies with $z<1.1$, \ha\ was observed with the J grism over the spectral range 1.08--1.41$\mu$m, whereas for galaxies with $z>1.2$, \ha\ was observed with the H grism over the spectral range 1.43--1.86$\mu$m. \oiii\ was observed in the J-band except for VVDS220148046 for which it was observed in the H-band. SINFONI spectral resolution in J- and H-bands reaches $\sim 2000$ and $\sim 2500$, respectively.

For seeing-limited observations, we offset the target alternatingly from one corner of the field-of-view of the instrument to the opposite one (object nodding) to maximize the observing efficiency during the nights. This observing strategy allowed us to avoid sky frame acquisition.
For AO observations, we used one sky frame for four object frames to maximize the time spent on sources (details in \citealp{Contini:2011}).
% For AO observations, we performed a OSOOOSO pattern in order to maximize the time spent on sources.
%
In addition, we also applied a sub-dithering to avoid the return of the target to the same position on the chip.
To allow for an accurate on-source pointing of our galaxies, we acquired them through a blind offset from a bright nearby star (PSF star used to measure the spatial PSF) to our target. We also observed standard (STD) stars for flux calibration during the same night. Individual exposures were 300s, 600s or 900s with a total on-source integration time that ranges between 1h and 2h.

Most of the data were observed in seeing-limited mode using the $0.125''\times 0.25''$ pixel scale leading to a 8''$\times$8'' field of view with a mean seeing of $0.68 \pm 0.12$'', considering only detected galaxies.
However, a subset of seven galaxies were observed with the laser guide star (LGS) adaptive optics system using the $0.05''\times 0.10''$ pixel scale with a 3.2''$\times$3.2'' field of view. Two of them were not detected (VVDS020126402 and VVDS220071601) and the PSF star was missed for VVDS220386469. The mean spatial resolution for the four other galaxies is 0.23''. The observing details along with the periods at which the galaxies were observed are listed in Table \ref{log}.

\subsection{Data reduction}

The data reduction was performed using the ESO-SINFONI pipeline (version 2.0.0, \citealp{Modigliani:2007}) complemented with additional IDL and PYTHON routines to perform the data processing homogeneously among the reducers and to improve some reduction steps.

First, the PSF stars and STD stars observations were reduced using standard data reduction.

Then, for each object science frame, the following steps where applied:
\begin{enumerate}
 \item bad line removal in the raw data;
 \item dark-current, sky-background and night-sky line subtraction from the raw data using the contiguous frame (with the target in the opposite corner due to the observing strategy or with only sky for AO data);
 \item flat-field correction using an internal lamp;
 \item wavelength calibration using arc-lamps;
 \item flux calibration using the STD telluric star spectrum with standard techniques to convert counts into flux units.
This enables one to correct for the atmospheric transmission and instrumental response;
 \item cube reconstruction (with and without sky lines) in counts;
 \item adjustment of the astrometry to match our I-band reference images (CFHT12k survey, \citealp{McCracken:2003}) using the acquisition PSF star and the offsets of object observations with respect to this star. This method was not possible on AO data due to missing meta-informations. For some galaxies of the pilot program, no PSF star was observed and in other cases, PSF stars were saturated in the CFHT images, which provided a deteriorated astrometry.
\end{enumerate}
In the pipeline, the spaxels are resampled so that they are square (0.125\arcsec\ or 0.05\arcsec).

All cubes obtained for each science frame for a given object were then combined using the SINFO\_UTL\_CUBE\_COMBINE recipe to obtain the final cubes using an additional median filtering to remove sky line residuals as accurately as possible from the final cube.

Mono-dimensional sky spectra were extracted from the sky cubes.
They were used to
\begin{itemize}
 \item determine the effective spectral resolution: using a Gaussian fit to approximate sky lines, the spectral resolution element is found to be fairly constant over the whole wavelength range in the J- and H-bands and the dispersion of the Gaussian was estimated to $\sigma \sim2.8 \pm0.2$ \AA;
 \item quantify the noise that was considered as a Poissonian noise.
\end{itemize}

The data from the pilot study \citep{Epinat:2009} were reduced again using this new procedure.

\subsection{Map extraction}
\label{maps_extraction}

The ionized gas kinematics of MASSIV galaxies is studied through the brightest emission line available in the NIR spectra, the \ha\ line or the \oiiib\ line in a few cases.
IDL routines, based on the \texttt{mpfit} routine \citep{Markwardt:2009}, were used to extract the kinematic maps from the SINFONI data.
First, to enhance the signal-to-noise ratio (S/N) without degrading the spatial resolution of the data, a subresolution 2D spatial Gaussian smoothing (FWHM of two pixels) was applied on the data cubes. In addition, the residual cosmic rays were rejected using a 15$\sigma$ clipping on 3$\times$3 pixel boxes.
Then, for each spatial pixel, the spectrum around \ha\ was fitted by a single Gaussian profile and a constant continuum (four parameters in total). To minimize the effects of noise induced by sky lines but also of sky line residuals on the line parameter determination, the 1D sky spectrum was used as an estimate of the noise to weight the contribution of each spectral element.
From these fitting techniques it was possible to recover the line flux map, the velocity field and the velocity dispersion map for each source. The instrumental spectral PSF was taken into account to compute the velocity dispersion: $\sigma^2=\sigma_{obs}^2-\sigma_{PSF}^2$.
During the line fitting procedure, the velocity dispersion was allowed to vary in the range 40\kms (to avoid fitting noise) to 250\kms (to avoid fitting a continuum).
2D error maps were also derived for each quantity from the fitting procedure. These are statistical errors that take into account the error spectrum and that indicate the accuracy of the fit for each parameter. An S/N map was computed. The computation of this map was refined since the study of the pilot program \citep{Epinat:2009}. The presence of sky lines is now taken into account to modulate the confidence on the line detection: the inverse of the sky spectrum is used for weighting ($w$).
The signal ($S$) is computed as the weighted flux of the line divided by the dispersion of the line ($\sigma_{\lambda}$) multiplied by $\sqrt{2\pi}$. In the following equations, the weight was normalized ($\sum{w}=1$), $n_z$ is the number of spectral elements, $\Delta\lambda$ refers to the spectral sampling and is expressed in the same unit as $\sigma_{\lambda}$, $m$ is the modeled spectrum, $c$ is the continuum of the modeled spectrum and $l$ is the observed line spectrum:
$$S=\frac{n_z \Delta\lambda \sum{w  (m-c)} }{\sigma_{\lambda}\sqrt{2 \pi}}\;.$$
The noise ($N$) is computed as the square root of the weighted variance of the residual spectrum:
$$N=\sqrt{\sum{w (l-m)^2}}\;.$$
This yields the S/N:
\begin{equation}
\frac{S}{N}=\frac{n_z \Delta\lambda \sum{w (m-c)}}{\sigma_{\lambda}\sqrt{2 \pi \sum{w (l-m)^2} } }\;.
\label{snr_def}
\end{equation}
% Where the weight $w$ has been normalized ($\sum{w}=1$), $\Delta\lambda$ and $\sigma_{\lambda}$ are respectively the spectral sampling and the dispersion of the line expressed in the same unit and $n_z$ is the number of spectral elements.
% Where $w=1/sky/\sum(sky)$.
If the weight is constant, then, according to equation \ref{snr_def}, the S/N is simply the ratio of the intensity of a Gaussian line over the RMS of the spectrum.
This criterion is robust and an S/N threshold of three was used to clean the kinematic maps (see Appendix \ref{kin_maps}). Extra-cleaning was performed manually to remove isolated pixels (less than about five adjacent pixels) or groups of pixels associated with data artifacts that could have passed the criteria. The existence of any I-band counterpart was checked when there were groups of more than four pixels.

\begin{table*}
\caption{MASSIV galaxies ``first epoch'' sample and SINFONI observation setups}
\begin{center}
\begin{tabular}{cccccccccccc}
\hline
VVDS ID & R.A. & Dec. & z & Scale & Pixel scale & Band & Line & t$_{exp}$ & Seeing & R & Period \\
 & [J2000] & [J2000] &   & [kpc/\arcsec] & [\arcsec] & & & [minutes] & [\arcsec] & & \\
 (1) & (2) & (3) & (4) & (5) & (6) & (7) & (8) & (9) & (10) & (11) & (12) \\
\hline
020106882 & 02:25:21.819 & $-$04:46:18.35 &   1.3991 &     8.43 &  0.125$\times$0.25 & H & H$\alpha$ &  80 &   0.49 &  2477 & P82G \\
020116027 & 02:25:51.085 & $-$04:45:06.08 &   1.5302 &     8.47 &  0.125$\times$0.25 & H & H$\alpha$ &  75 &   0.60 &  2609 & P75A \\
020126402$^1$ & 02:25:11.658 & $-$04:43:40.12 & 1.2332 &  8.33 &  0.050$\times$0.10 & J & [O\textsc{iii}] &  60 &   $-$ & $-$ & P79B \\
020147106 & 02:26:45.362 & $-$04:40:47.50 &   1.5195 &     8.47 &  0.125$\times$0.25 & H & H$\alpha$ & 120 &   0.65 &  2598 & P75A \\
020149061 & 02:27:05.226 & $-$04:40:29.21 &   1.2905 &     8.37 &  0.125$\times$0.25 & H & H$\alpha$ &  80 &   0.85 &  2362 & P82G \\
020164388 & 02:26:50.942 & $-$04:38:20.72 &   1.3547 &     8.41 &  0.125$\times$0.25 & H & H$\alpha$ &  80 &   0.83 &  2430 & P82G \\
020167131 & 02:26:47.307 & $-$04:37:55.36 &   1.2246 &     8.32 &  0.125$\times$0.25 & J & [O\textsc{iii}] & 120 &   0.68 &  2295 & P79B \\
020182331 & 02:26:44.242 & $-$04:35:52.01 &   1.2290 &     8.32 &  0.125$\times$0.25 & H & H$\alpha$ & 180 &   0.74 &  2302 & P78A \\
020193070 & 02:25:18.713 & $-$04:34:19.77 &   1.0279 &     8.06 &  0.125$\times$0.25 & J & H$\alpha$ & 120 &   0.58 &  2094 & P82G \\
020208482 & 02:25:16.739 & $-$04:32:11.92 &   1.0375 &     8.08 &  0.125$\times$0.25 & J & H$\alpha$ & 120 &   0.58 &  2104 & P82G \\
020214655 & 02:26:23.441 & $-$04:31:22.78 &   1.0395 &     8.08 &  0.125$\times$0.25 & J & H$\alpha$ &  80 &   0.87 &  2101 & P82G \\
020217890$^1$ & 02:26:27.162 & $-$04:30:51.83 & 1.5129 &  8.46 &  0.125$\times$0.25 & H & H$\alpha$ & 120 &   $-$ & $-$ & P78A \\
020239133 & 02:26:43.006 & $-$04:28:31.20 &   1.0194 &     8.04 &  0.125$\times$0.25 & J & H$\alpha$ &  80 &   0.79 &  2084 & P82G \\
020240675 & 02:26:54.140 & $-$04:28:17.64 &   1.3270 &     8.40 &  0.125$\times$0.25 & H & H$\alpha$ &  80 &   0.85 &  2402 & P82G \\
020255799 & 02:26:45.859 & $-$04:26:15.80 &   1.0351 &     8.07 &  0.125$\times$0.25 & J & H$\alpha$ &  80 &   0.76 &  2101 & P82G \\
020261328 & 02:27:11.023 & $-$04:25:31.57 &   1.5290 &     8.47 &  0.125$\times$0.25 & H & H$\alpha$ &  60 &   0.62 &  2609 & P75A \\
020278667 & 02:25:58.203 & $-$04:23:11.67 &   1.0516 &     8.10 &  0.125$\times$0.25 & J & H$\alpha$ & 120 &   0.65 &  2115 & P82G \\
020283083 & 02:26:30.832 & $-$04:22:35.82 &   1.2818 &     8.36 &  0.125$\times$0.25 & H & H$\alpha$ &  80 &   0.78 &  2353 & P82G \\
020283830 & 02:26:28.926 & $-$04:22:31.14 &   1.3949 &     8.43 &  0.125$\times$0.25 & H & H$\alpha$ & 120 &   0.77 &  2472 & P82G \\
020294045 & 02:25:47.152 & $-$04:21:07.41 &   1.0028 &     8.01 &  0.125$\times$0.25 & J & H$\alpha$ & 120 &   0.59 &  2067 & P79B \\
020306817$^1$ & 02:25:50.316 & $-$04:19:22.93 & 1.2225 &  8.32 &  0.125$\times$0.25 & J & [O\textsc{iii}] & 120 & $-$ & $-$ & P79B \\
020363717 & 02:26:23.709 & $-$04:11:57.87 &   1.3339 &     8.40 &  0.125$\times$0.25 & H & H$\alpha$ &  80 &   0.64 &  2407 & P82G \\
020370467 & 02:26:14.690 & $-$04:11:05.44 &   1.3338 &     8.40 &  0.125$\times$0.25 & H & H$\alpha$ &  80 &   0.71 &  2407 & P82G \\
020386743 & 02:27:13.989 & $-$04:08:59.73 &   1.0487 &     8.09 &  0.125$\times$0.25 & J & H$\alpha$ & 120 &   0.73 &  2111 & P79B \\
020461235 & 02:26:47.102 & $-$04:23:55.70 &   1.0349 &     8.07 &  0.125$\times$0.25 & J & H$\alpha$ & 120 &   0.63 &  2101 & P79B \\
020461893 & 02:27:12.252 & $-$04:23:11.28 &   1.0486 &     8.09 &  0.125$\times$0.25 & J & H$\alpha$ &  80 &   0.60 &  2115 & P82G \\
020465775 & 02:26:59.366 & $-$04:19:00.08 &   1.3583 &     8.41 &  0.125$\times$0.25 & H & H$\alpha$ &  80 &   0.88 &  2434 & P82G \\
140083410 & 13:57:50.595 & $+$04:17:38.71 &   0.9435 &     7.89 &  0.125$\times$0.25 & J & H$\alpha$ &  80 &   0.69 &  2005 & P81D \\
140096645 & 13:58:26.336 & $+$04:19:47.75 &   0.9655 &     7.94 &  0.125$\times$0.25 & J & H$\alpha$ & 120 &   0.56 &  2024 & P81D \\
140123568 & 13:55:57.628 & $+$04:24:20.11 &   1.0012 &     8.01 &  0.125$\times$0.25 & J & H$\alpha$ & 120 &   0.76 &  2067 & P79A \\
140137235 & 13:56:12.729 & $+$04:26:31.74 &   1.0445 &     8.09 &  0.125$\times$0.25 & J & H$\alpha$ &  80 &   0.76 &  2111 & P79A \\
140217425 & 13:57:56.405 & $+$04:38:37.00 &   0.9792 &     7.97 &  0.125$\times$0.25 & J & H$\alpha$ & 100 &   0.95 &  2040 & P81D \\
140258511 & 14:00:19.658 & $+$04:44:45.86 &   1.2423 &     8.33 &  0.125$\times$0.25 & H & H$\alpha$ &  80 &   0.49 &  2315 & P79A \\
140262766 & 13:59:55.518 & $+$04:45:30.04 &   1.2836 &     8.37 &  0.125$\times$0.25 & H & H$\alpha$ & 120 &   0.51 &  2353 & P79A \\
140545062 & 13:59:35.598 & $+$05:30:31.11 &   1.0408 &     8.08 &  0.125$\times$0.25 & J & H$\alpha$ & 120 &   0.70 &  2104 & P81D \\
220014252 & 22:17:45.677 & $+$00:28:39.52 &   1.3105 &     8.38 &  0.125$\times$0.25 & H & H$\alpha$ & 120 &   0.70 &  2384 & P75A \\
220015726 & 22:15:42.435 & $+$00:29:03.58 &   1.2933 &     8.37 &  0.125$\times$0.25 & H & H$\alpha$ & 120 &   0.46 &  2384 & P79B \\
220071601$^1$ & 22:18:01.569 & $+$00:45:34.69 & 1.3538 &  8.41 &  0.050$\times$0.10 & H & H$\alpha$ &  80 &   $-$ & $-$ & P79B \\
220148046 & 22:14:37.904 & $+$01:08:20.65 &   2.2442 &     8.24 &  0.050$\times$0.10 & H & [O\textsc{iii}] &  80 &   0.27 &  2450 & P81E \\
220376206 & 22:20:05.772 & $-$00:08:21.74 &   1.2445 &     8.34 &  0.125$\times$0.25 & H & H$\alpha$ & 120 &   0.50 &  2315 & P79B \\
220386469 & 22:19:56.603 & $-$00:03:03.78 &   1.0226 &     8.05 &  0.050$\times$0.10 & J & H$\alpha$ &  40 & 0.23$^2$ &  2090 & P79B \\
220397579 & 22:20:36.512 & $+$00:01:46.85 &   1.0379 &     8.08 &  0.125$\times$0.25 & J & H$\alpha$ & 120 &   0.64 &  2101 & P79B \\
220544103 & 22:15:25.689 & $+$00:06:40.31 &   1.3973 &     8.43 &  0.125$\times$0.25 & H & H$\alpha$ & 120 &   0.76 &  2472 & P75A \\
220544394 & 22:14:24.153 & $+$00:06:46.67 &   1.0101 &     8.03 &  0.125$\times$0.25 & J & H$\alpha$ & 120 &   0.58 &  2073 & P79B \\
220576226 & 22:16:11.417 & $+$00:16:30.46 &   1.0217 &     8.05 &  0.125$\times$0.25 & J & H$\alpha$ & 120 &   0.58 &  2087 & P79B \\
220578040 & 22:17:04.113 & $+$00:16:56.80 &   1.0462 &     8.09 &  0.125$\times$0.25 & J & H$\alpha$ & 120 &   0.62 &  2111 & P79B \\
220584167 & 22:15:22.917 & $+$00:18:48.82 &   1.4655 &     8.45 &  0.125$\times$0.25 & H & H$\alpha$ & 120 &   0.75 &  2541 & P75A \\
220596913 & 22:14:29.179 & $+$00:22:18.93 &   1.2658 &     8.35 &  0.050$\times$0.10 & H & H$\alpha$ & 120 &   0.18 &  2340 & P79B \\
910193711 & 02:25:46.285 & $-$04:32:33.43 &   1.5564 &     8.47 &  0.050$\times$0.10 & H & H$\alpha$ &  80 &   0.27 &  2636 & P82F \\
910279515 & 02:25:36.233 & $-$04:21:16.13 &   1.4013 &     8.43 &  0.050$\times$0.10 & H & H$\alpha$ &  80 &   0.21 &  2477 & P82F \\
\hline
\end{tabular}
\end{center}
\footnotesize{
The coordinates of the galaxies are given in column (2) and (3). The redshift (4) is determined from SINFONI data. The physical scale (5) is computed from the redshift. The SINFONI pixel scale is given in column (6).  In column (7), the wavelength band used to to observe the emission line (8) is given. Column (9) is the on-source exposure time. The seeing of SINFONI observations (10) is derived from observed PSF stars. Column (11) is the spectral resolution. The ESO observing period is given in column (12).\\
$^1$ These galaxies are those for which no line was detected in SINFONI data. The redshift is the one derived from VIMOS spectra, the line is the targeted one and both spectral and spatial resolution have not been measured.\\
$^2$ No PSF star was observed: the resolution is the mean resolution of AO observations.}
\label{log}
\end{table*}

% \begin{table*}
% \caption{Galaxies and SINFONI observation setups}
% \begin{center}
% \begin{tabular}{ccccccccccc}
% \hline
% Object & R.A. & Dec. & z & Scale & Pixel scale & Band & Line & t$_{exp}$ & Seeing & Period \\ %& Field\\
%  & J2000 & J2000 &   & kpc/\arcsec & \arcsec & & & min & \arcsec & \\
% \hline
% VVDS020116027 & 02:25:51.133 & -04:45:04.48 & 1.526 & xxx & 0.125$\times$0.25 & H & \ha & 100 & 0.65 & 075.A-0318(A) \\ %& VVDS-02h
% \hline
% \end{tabular}
% \end{center}
% Add AO/NOAO, date?, change format of pixel scale?, exposure in secondes?
% \label{log}
% \end{table*}

\section{Modeling}
\label{models}

\subsection{Morphology modeling}
\label{model_morph}

\begin{table*}[t]
\caption{CFHT observations and morphological parameters from GALFIT modeling}
\begin{center}
% \begin{tabular}{ccccccccc}
\begin{tabular}{cccccccc}
\hline
% VVDS ID & Survey & Pixel scale [\arcsec] & Seeing [\arcsec]  & $PA$ [\degr]  & $b/a$ & $r_e$ [kpc]  & $n$ & $mag$ \\
% VVDS ID & Survey & Pixel scale [\arcsec] & Seeing [\arcsec]  & $PA$ [\degr]  & $b/a$ & $r_e$ [kpc]  & $n$ \\
VVDS ID & Survey & Pixel scale & Seeing  & $PA$  & $b/a$ & $R_e$  & $n$ \\
 &  & [\arcsec] & [\arcsec]  & [\degr]  &  & [kpc]  & \\
 (1)  & (2) & (3) & (4)  & (5) & (6)  & (7)  & (8) \\
\hline
020106882 & LS & 0.186 & 0.604 &   294 $\pm$     4 &   0.62 $\pm$   0.03 &  3.51 $\pm$  0.14 &  0.77 $\pm$  0.16 \\
020116027 & LS & 0.186 & 0.604 &   184 $\pm$     2 &   0.38 $\pm$   0.03 &  4.27 $\pm$  0.16 &  0.70 $\pm$  0.15 \\
% 020116027s & LS & 0.186 & 0.604 &  136 $\pm$    19 &   0.75 $\pm$   0.14 & 9.52 $\pm$  20.61 &  3.58 $\pm$  4.88 \\
020126402 & LS & 0.186 & 0.615 &    15 $\pm$     8 &   0.56 $\pm$   0.10 &  2.12 $\pm$  0.31 &  0.59 $\pm$  0.58 \\
020147106 & LS & 0.186 & 0.615 &   310 $\pm$     1 &   0.19 $\pm$   0.15 &  1.17 $\pm$  0.41 &  0.41 $\pm$  1.82 \\
020149061 & LS & 0.186 & 0.649 &   201 $\pm$   369 &   0.93 $\pm$   0.83 &  1.09 $\pm$  0.69 &  0.11 $\pm$  6.68 \\
020164388 & LS & 0.186 & 0.618 &   162 $\pm$     3 &   0.71 $\pm$   0.02 &  2.72 $\pm$  0.05 &  0.85 $\pm$  0.09 \\
020167131 & LS & 0.186 & 0.618 &   272 $\pm$     3 &   0.60 $\pm$   0.03 &  2.60 $\pm$  0.09 &  1.64 $\pm$  0.26 \\
020167131s & LS & 0.186 & 0.618 &   57 $\pm$     1 &   0.37 $\pm$   0.02 &  4.11 $\pm$  0.08 &  0.31 $\pm$  0.09 \\
020182331 & LS & 0.186 & 0.618 &   268 $\pm$     4 &   0.49 $\pm$   0.04 &  4.19 $\pm$  0.17 &  0.63 $\pm$  0.21 \\
020193070 & LS & 0.186 & 0.615 &   215 $\pm$     2 &   0.33 $\pm$   0.03 &  3.49 $\pm$  0.09 &  0.47 $\pm$  0.13 \\
020208482 & LS & 0.186 & 0.600 &   355 $\pm$     7 &   0.70 $\pm$   0.05 &  3.67 $\pm$  0.20 &  0.82 $\pm$  0.21 \\
020214655 & LS & 0.186 & 0.604 &    32 $\pm$     4 &   0.41 $\pm$   0.06 &  1.47 $\pm$  0.08 &  1.97 $\pm$  0.53 \\
020217890 & LS & 0.186 & 0.604 &   171 $\pm$     2 &   0.21 $\pm$   0.04 &  3.57 $\pm$  0.14 &  1.13 $\pm$  0.22 \\
020239133 & LS & 0.186 & 0.608 &   123 $\pm$     2 &   0.39 $\pm$   0.03 &  3.14 $\pm$  0.09 &  0.25 $\pm$  0.13 \\
020240675 & LS & 0.186 & 0.619 &   190 $\pm$    13 &   0.62 $\pm$   0.13 &  1.09 $\pm$  0.16 &  2.03 $\pm$  1.19 \\
020255799 & LS & 0.186 & 0.607 &   164 $\pm$    12 &   0.80 $\pm$   0.08 &  1.89 $\pm$  0.26 &  0.24 $\pm$  0.44 \\
020261328 & LS & 0.186 & 0.634 &   171 $\pm$     8 &   0.51 $\pm$   0.11 &  1.83 $\pm$  0.27 &  0.64 $\pm$  0.53 \\
020278667 & LS & 0.186 & 0.596 &   140 $\pm$    12 &   0.74 $\pm$   0.09 &  2.82 $\pm$  1.66 &  9.60 $\pm$  5.68 \\
020283083 & LS & 0.186 & 0.596 &   301 $\pm$     3 &   0.38 $\pm$   0.05 &  4.29 $\pm$  0.20 &  0.46 $\pm$  0.18 \\
020283830 & LS & 0.186 & 0.596 &   142 $\pm$     1 &   0.35 $\pm$   0.01 &  6.82 $\pm$  0.16 &  0.09 $\pm$  0.06 \\
% 020283830s & LS & 0.186 & 0.596 &  112 $\pm$    10 &   0.21 $\pm$   0.15 &  7.12 $\pm$ 41.46 & 19.90 $\pm$ 41.74 \\
020294045 & LS & 0.186 & 0.606 &     1 $\pm$     3 &   0.61 $\pm$   0.03 &  2.89 $\pm$  0.10 &  1.00 $\pm$  0.17 \\
020306817 & LS & 0.186 & 0.606 &    86 $\pm$     2 &   0.54 $\pm$   0.02 &  4.22 $\pm$  0.09 &  0.27 $\pm$  0.08 \\
020363717 & LS & 0.186 & 0.626 &   154 $\pm$     6 &   0.55 $\pm$   0.08 &  0.72 $\pm$  0.09 &  3.59 $\pm$  1.01 \\
020370467 & LS & 0.186 & 0.626 &    50 $\pm$    24 &   0.78 $\pm$   0.18 &  1.31 $\pm$  0.25 &  0.54 $\pm$  0.97 \\
020386743 & LS & 0.186 & 0.639 &   203 $\pm$     8 &   0.64 $\pm$   0.07 &  2.68 $\pm$  0.21 &  1.15 $\pm$  0.42 \\
020461235 & LS & 0.186 & 0.597 &   332 $\pm$     3 &   0.57 $\pm$   0.02 &  3.96 $\pm$  0.12 &  0.47 $\pm$  0.11 \\
% 020461235s & LS & 0.186 & 0.597 &   71 $\pm$     8 &   0.04 $\pm$   0.11 & 10.81 $\pm$ 36.31 &  7.85 $\pm$ 21.37 \\
020461893 & LS & 0.186 & 0.634 &   283 $\pm$     3 &   0.46 $\pm$   0.04 &  2.66 $\pm$  0.11 &  1.04 $\pm$  0.23 \\
020465775 & LS & 0.186 & 0.623 &   146 $\pm$     3 &   0.51 $\pm$   0.03 &  4.04 $\pm$  0.16 &  0.09 $\pm$  0.17 \\
140083410 & 12K & 0.204 & 0.958 &   311 $\pm$   10 &   0.67 $\pm$   0.09 &  1.93 $\pm$  0.19 &  1.90 $\pm$  0.82 \\
140096645 & 12K & 0.204 & 0.747 &   217 $\pm$  114 &   0.93 $\pm$   0.28 &  1.80 $\pm$  0.40 &  0.85 $\pm$  1.26 \\
140123568 & 12K & 0.204 & 0.968 &   227 $\pm$  229 &   0.80 $\pm$   1.14 &  1.39 $\pm$  0.98 &  0.08 $\pm$ 32.82 \\
140137235 & 12K & 0.204 & 0.920 &   116 $\pm$   24 &   0.47 $\pm$   0.31 &  5.45 $\pm$  2.03 &  0.88 $\pm$  1.73 \\
140217425 & 12K & 0.204 & 0.713 &   260 $\pm$    1 &   0.25 $\pm$   0.01 &  8.96 $\pm$  0.13 &  0.33 $\pm$  0.04 \\
140258511 & 12K & 0.204 & 0.768 &   265 $\pm$   42 &   0.54 $\pm$   0.51 &  2.62 $\pm$  2.11 &  0.40 $\pm$  3.22 \\
140262766 & 12K & 0.204 & 0.613 &   142 $\pm$   28 &   0.65 $\pm$   0.26 &  2.07 $\pm$  0.70 &  0.63 $\pm$  1.34 \\
140545062 & 12K & 0.204 & 0.724 &   216 $\pm$   41 &   0.46 $\pm$   0.35 &  2.89 $\pm$  6.86 &  0.06 $\pm$  4.03 \\
220014252 & LS & 0.186 & 0.687 &   136 $\pm$     2 &   0.22 $\pm$   0.06 &  3.32 $\pm$  0.12 &  0.94 $\pm$  0.23 \\
220015726 & LS & 0.186 & 0.686 &   193 $\pm$    12 &   0.78 $\pm$   0.07 &  2.76 $\pm$  0.25 &  0.26 $\pm$  0.35 \\
220071601 & LS & 0.186 & 0.665 &    74 $\pm$     1 &   0.31 $\pm$   0.01 &  8.13 $\pm$  0.13 &  0.21 $\pm$  0.04 \\
220148046 & LS & 0.186 & 0.800 &   236 $\pm$    10 &   0.10 $\pm$   0.87 &  1.98 $\pm$  0.38 &  2.10 $\pm$  2.36 \\
220376206 & LS & 0.186 & 0.749 &   235 $\pm$     1 &   0.46 $\pm$   0.01 &  5.29 $\pm$  0.08 &  0.21 $\pm$  0.05 \\
% 220376206s & LS & 0.186 & 0.749 &   32 $\pm$     9 &   0.03 $\pm$   0.17 &  1.81 $\pm$  1.29 &  5.63 $\pm$ 13.61 \\
220386469 & LS & 0.186 & 0.725 &   168 $\pm$     7 &   0.70 $\pm$   0.05 &  2.90 $\pm$  0.13 &  0.73 $\pm$  0.26 \\
220397579 & LS & 0.186 & 0.707 &   331 $\pm$     5 &   0.43 $\pm$   0.06 &  3.02 $\pm$  0.32 &  0.21 $\pm$  0.37 \\
220397579s & LS & 0.186 & 0.707 &   332 $\pm$    2 &   0.14 $\pm$   0.03 &  9.77 $\pm$  3.29 &  3.77 $\pm$  1.49 \\
220544103 & LS & 0.186 & 0.686 &   208 $\pm$    55 &   0.08 $\pm$   0.16 &  5.61 $\pm$ 19.89 &  0.02 $\pm$  0.91 \\
220544103s & LS & 0.186 & 0.686 &   161 $\pm$    5 &   0.26 $\pm$   0.06 &  5.47 $\pm$  0.44 &  1.67 $\pm$  0.77 \\
220544394 & LS & 0.186 & 0.624 &   234 $\pm$     7 &   0.70 $\pm$   0.04 &  3.40 $\pm$  0.15 &  0.48 $\pm$  0.13 \\
220544394s & LS & 0.186 & 0.624 &   233 $\pm$   10 &   0.30 $\pm$   0.17 &  3.69 $\pm$  0.70 &  1.18 $\pm$  1.02 \\
220576226 & LS & 0.186 & 0.675 &   237 $\pm$    15 &   0.85 $\pm$   0.06 &  2.16 $\pm$  0.12 &  1.58 $\pm$  0.36 \\
220578040 & LS & 0.186 & 0.658 &   112 $\pm$    14 &   0.89 $\pm$   0.04 &  3.81 $\pm$  0.12 &  0.42 $\pm$  0.10 \\
220584167 & LS & 0.186 & 0.684 &   193 $\pm$     2 &   0.52 $\pm$   0.01 &  7.17 $\pm$  0.19 &  0.14 $\pm$  0.06 \\
220596913 & LS & 0.186 & 0.623 &   255 $\pm$     1 &   0.18 $\pm$   0.02 &  9.49 $\pm$ 11.54 &  0.04 $\pm$  0.12 \\
910193711 & LS & 0.186 & 0.597 &     9 $\pm$     3 &   0.50 $\pm$   0.03 &  2.27 $\pm$  0.06 &  0.50 $\pm$  0.17 \\
910279515 & LS & 0.186 & 0.602 &   131 $\pm$     3 &   0.61 $\pm$   0.02 &  4.44 $\pm$  0.11 &  0.28 $\pm$  0.08 \\
\hline
\end{tabular}
\end{center}
\footnotesize{
In column (2), ``LS'' refers to the CFHT Legacy Survey (\url{http://www.cfht.hawaii.edu/Science/CFHLS/}) whereas ``12K'' refers to the CFH12K/CFHT survey \citep{McCracken:2003}. The pixel scale of the CFHT image and the seeing determined from close stars are respectively given in column (3) and (4). The parameters of the model are the position angle of the major axis (5), the axis raio (6), the effective radius (7), the Sersic index (7) and the magnitude (not given here due to a zero point magnitude mismatch between the two CFHT surveys).\\
The suffix ``s'' refers to secondary objects which have been detected both in H$\alpha$ and in the I-band image. The companion's parameters are only given  when they are resolved.}
% Je laisse tomber 045 et 830 car les compagnons sont trop proches pour être contraints correctement
\label{chft_table}
\end{table*}

The stellar continuum is barely detected in the SINFONI data. To study the morphology of the stellar component, we used the best CFHT I-band images available, i.e. from the CFH12K/CFHT survey \citep{McCracken:2003} for the galaxies in the 14h field and from the CFHT Legacy Survey\footnote{\url{http://www.cfht.hawaii.edu/Science/CFHLS/}} with the best seeing for the galaxies in the 02h and 22h fields.
We ran GALFIT \citep{Peng:2002} on those images using a Sersic profile \citep{Sersic:1968} to recover the morphological parameters: the center, the position angle of the major axis ($PA$), the axis ratio ($b/a$), the effective radius ($R_e$), the Sersic index ($n$) and the total magnitude.
GALFIT convolves the model to the spatial PSF to converge into a set of beam-smearing corrected parameters.

To obtain robust estimates of the parameters and of their associated error bars, it is mandatory to control the PSF as well as possible. Indeed, for objects with sizes comparable to the resolution, the use of an overestimated PSF leads to low axis-ratios with small error bars since no strong elongation may be reproduced: this would lead to systematic underestimating of structures in the galaxies.
We are indeed in most cases in this situation when the size on the objects is of the same order as the spatial resolution for seeing-limited ground-based imaging surveys.
Thus, in each CFHT field, we randomly selected stars to characterize and follow the variation of the PSF. The detection of stars is based on color and morphological criteria for the 02h and 22h fields and on morphological criteria only for the 14h field. From these stars, the width, the axis ratio and the orientation of the PSF are found to vary much across the 14h and 22h fields. The most affected field is the 14h one where the seeing smoothly varies from 0.6\arcsec\ to 1.0\arcsec.
Therefore, we selected specific PSFs in each field, located at 45\arcsec\ from each galaxy on average.
The study of the PSF distribution over the various fields enabled us to determine the uncertainty on the PSF FWHM to be $\sim0.1$ pixel ($\sim 20$ mas). It is estimated as the deviation from large-scale variations.

GALFIT produces residual maps that where used to check the convergence of the fits.
It was sometimes necessary to fit secondary objects to have a correct fit (cf. Appendix \ref{comments}).
The parameters of the fits are summarized in Table \ref{chft_table}.

We tried in a second step to fit the morphology while fixing the position angle of the major axis to the value determined from the kinematics modeling (cf. Section \ref{model_kin}) as was done in the pilot study \citep{Epinat:2009}. However, this leads to a bias toward high axis ratios because GALFIT cannot match the elongation when the position angle is fixed and finds that round morphologies match the data better. Therefore, we adopt the morphologies as derived in an unconstrained way from the CFHT images in the following.

\subsection{Kinematics modeling}
\label{model_kin}

\begin{table*}
\caption{Physical parameters from kinematic modeling}
\begin{center}
\begin{tabular}{ccccccccccc}
\hline
VVDS ID & $i$ & $PA_k$ & $r_t$ & $V_t$ & $V_t/r_t$  & Res$_V$ & $\chi^2$ & $V_{max}$ & $\sigma$ & $R_{last}$ \\
       & [\degr] & [\degr]  & [kpc] & [km s$^{-1}$] & [km s$^{-1}$ kpc$^{-1}$] & [km s$^{-1}$] &   & [km s$^{-1}$] & [km s$^{-1}$] & [kpc] \\
 (1)  & (2) & (3) & (4)  & (5) & (6)  & (7)  & (8) & (9) & (10) & (11)  \\
\hline
020106882 &    52 $\pm$     3 &   317 $\pm$     1 &      1.1 $\pm$      0.1 &   133 $\pm$     3 &   126 &    12 &   1.6 &   133 $\pm$    25 &    41 $\pm$    31 &   5.1 \\
020116027 &    68 $\pm$     4 &   207 $\pm$     6 &      4.3 $\pm$      3.0 &    27 $\pm$     6 &     6 &    10 &   3.7 &    27 $\pm$    10 &    47 $\pm$    15 &   6.5 \\
020147106 &    60 $\pm$    24 &   317 $\pm$     3 &      1.6 $\pm$      4.6 &    26 $\pm$     3 &    16 &     6 &   1.9 &    26 $\pm$    51 &    81 $\pm$    10 &   7.8 \\
020149061 &    60 $\pm$    24 &   235 $\pm$     3 &     42.0 &   976 &    23 &    15 &   1.9 &   112 $\pm$   216 &    73 $\pm$    20 &   4.8 \\
020164388 &    45 $\pm$     4 &    97 $\pm$     4 &      7.2 $\pm$      4.9 &    79 $\pm$    44 &    11 &     8 &   0.9 &    79 $\pm$    19 &    52 $\pm$    19 &   8.2 \\
020167131 &    53 $\pm$     4 &   185 $\pm$    10 &      1.0 $\pm$      0.1 &   127 $\pm$    43 &   123 &    13 &   0.9 &   127 $\pm$    29 &    26 $\pm$    37 &   1.8 \\
020182331 &    61 $\pm$     2 &   234 $\pm$     4 &     27.0 &   647 &    24 &    11 &   0.9 &   132 $\pm$    26 &    67 $\pm$    29 &   5.5 \\
020193070 &    71 $\pm$     4 &   184 $\pm$     2 &    108.7 &  3217 &    30 &    15 &   1.8 &   117 $\pm$    23 &    33 $\pm$    27 &   3.9 \\
020208482 &    46 $\pm$     3 &   336 $\pm$    10 &      1.0 $\pm$      0.1 &   158 $\pm$    38 &   157 &    11 &   0.6 &   158 $\pm$    31 &     7 $\pm$    11 &   1.4 \\
020214655 &    66 $\pm$    11 &   341 $\pm$     4 &      1.0 $\pm$      0.1 &    52 $\pm$     4 &    51 &     8 &   1.1 &    52 $\pm$    14 &    63 $\pm$    25 &   5.7 \\
020239133 &    67 $\pm$     4 &   109 $\pm$     6 &     24.6 &   766 &    31 &    13 &   0.9 &   149 $\pm$    33 &    75 $\pm$    35 &   4.8 \\
020240675 &    60 $\pm$    24 &   183 $\pm$    12 &      8.9 &   116 &    13 &    13 &   1.5 &    50 $\pm$    97 &    33 $\pm$    17 &   3.8 \\
020255799 &    37 $\pm$    15 &    89 $\pm$    76 &      1.0 $\pm$      0.1 &    14 $\pm$    16 &    14 &    10 &   0.6 &    14 $\pm$    26 &    76 $\pm$    23 &   4.0 \\
020261328 &    59 $\pm$    11 &   179 $\pm$     2 &      6.2 &   154 &    25 &    10 &   1.6 &   127 $\pm$    35 &    54 $\pm$    19 &   5.1 \\
020278667 &    42 $\pm$    20 &   175 $\pm$     6 &      3.7 &   224 &    60 &    16 &   1.5 &    77 $\pm$   189 &    52 $\pm$    37 &   1.3 \\
020283083 &    68 $\pm$     2 &   359 $\pm$     4 &      1.0 $\pm$      0.1 &    59 $\pm$     5 &    57 &    11 &   1.4 &    59 $\pm$    12 &    39 $\pm$    23 &   5.6 \\
020283830 &    70 $\pm$     1 &   156 $\pm$     2 &      1.9 $\pm$      3.4 &   186 $\pm$     5 &    96 &    21 &   4.7 &   186 $\pm$    30 &    17 $\pm$    24 &   7.9 \\
020294045 &    52 $\pm$     5 &     3 $\pm$     1 &     31.1 &  1332 &    43 &    33 &  12.4 &   234 $\pm$    51 &    60 $\pm$    45 &   5.5 \\
020363717 &    60 $\pm$    24 &   106 $\pm$     6 &     16.4 &   123 &     8 &     9 &   1.3 &    45 $\pm$    86 &    91 $\pm$    14 &   6.0 \\
020370467 &    39 $\pm$    14 &    19 $\pm$    11 &      1.0 $\pm$      0.1 &    51 $\pm$    10 &    49 &    18 &   1.2 &    51 $\pm$    64 &    86 $\pm$    35 &   5.5 \\
020386743 &    50 $\pm$     4 &   136 $\pm$     3 &     37.0 &   286 &     8 &     9 &   2.5 &    42 $\pm$    10 &    53 $\pm$    18 &   5.4 \\
020461235 &    55 $\pm$     1 &   351 $\pm$     2 &      1.2 $\pm$      7.7 &    82 $\pm$     4 &    67 &    11 &   1.8 &    82 $\pm$    16 &    24 $\pm$    22 &   5.4 \\
020461893 &    63 $\pm$     4 &   279 $\pm$     4 &      1.2 $\pm$      9.1 &    58 $\pm$     6 &    46 &     8 &   1.1 &    58 $\pm$    13 &    67 $\pm$    22 &   6.5 \\
020465775 &    59 $\pm$     2 &   178 $\pm$     5 &      1.1 $\pm$      0.1 &    68 $\pm$     7 &    65 &    13 &   1.5 &    68 $\pm$    15 &    84 $\pm$    30 &   4.9 \\
140083410 &    48 $\pm$    16 &    39 $\pm$     6 &      1.0 $\pm$      0.1 &    30 $\pm$     4 &    30 &    10 &   2.3 &    30 $\pm$    33 &    66 $\pm$    21 &   5.3 \\
140096645 &    22 $\pm$    15 &   197 $\pm$     1 &      1.1 $\pm$      1.1 &   295 $\pm$    12 &   258 &    12 &   6.1 &   295 $\pm$   709 &    77 $\pm$    27 &   4.5 \\
140123568 &    60 $\pm$    24 &   184 $\pm$    14 &      1.0 &   108 &   108 &     6 &   0.7 &    50 $\pm$    99 &    73 $\pm$    24 &   0.5 \\
140137235 &    62 $\pm$     1 &   123 $\pm$    16 &      2.2 &   200 &    91 &     4 &   0.6 &    62 $\pm$    11 &    17 $\pm$    27 &   0.7 \\
140217425 &    76 $\pm$     2 &   258 $\pm$     1 &     16.1 &   499 &    31 &    29 &  20.7 &   320 $\pm$    46 $^1$ &    45 $\pm$    33 &  14.5 \\
140258511 &    57 $\pm$     3 &   213 $\pm$     1 &      1.0 $\pm$      0.1 &   124 $\pm$     4 &   119 &    19 &   5.2 &   124 $\pm$    26 &    25 $\pm$    29 &   5.2 \\
140262766 &    60 $\pm$    24 &   175 $\pm$     2 &     18.1 &   524 &    29 &     9 &   1.2 &   119 $\pm$   231 &    39 $\pm$    16 &   4.1 \\
140545062 &    63 $\pm$     5 &   229 $\pm$     1 &     13.9 &   378 &    27 &    13 &   3.6 &   204 $\pm$    46 &    67 $\pm$    29 &   7.5 \\
220014252 &    77 $\pm$     2 &   141 $\pm$     1 &      1.0 $\pm$      0.1 &   129 $\pm$     1 &   123 &    15 &   4.5 &   129 $\pm$    27 &    90 $\pm$    28 &  10.3 \\
220015726 &    39 $\pm$    15 &   186 $\pm$     1 &      1.5 $\pm$      0.4 &   231 $\pm$     4 &   155 &    10 &   3.1 &   231 $\pm$   356 &    62 $\pm$    22 &   3.7 \\
220148046 &    60 $\pm$    24 &   261 $\pm$    15 &      1.3 &    61 &    49 &    10 &   2.9 &    42 $\pm$    83 &    46 $\pm$    21 &   0.9 \\
220376206 &    63 $\pm$     2 &   225 $\pm$     1 &      8.3 $\pm$      0.2 &   201 $\pm$     4 &    24 &    15 &   4.8 &   201 $\pm$    27 &    73 $\pm$    26 &  10.0 \\
220386469 &    46 $\pm$     7 &   151 $\pm$    10 &     63.0 &   973 &    15 &    16 &   2.0 &    40 $\pm$    11 &    43 $\pm$    25 &   2.6 \\
220397579 &    65 $\pm$     6 &     1 $\pm$     9 &     35.9 &    32 &     1 &     9 &   7.0 &     9 $\pm$    10 &    59 $\pm$    17 &  10.2 \\
220397579s &    80 $\pm$     5 &   344 $\pm$     2 &      3.8 $\pm$      1.2 &   222 $\pm$     9 &    59 &    17 &   2.6 &   222 $\pm$    15 &    27 $\pm$    37 &   6.2 \\
220544103 &    80 $\pm$     4 &   198 $\pm$     1 &      1.1 $\pm$      0.1 &   137 $\pm$     2 &   130 &    12 &   4.4 &   137 $\pm$    24 &    71 $\pm$    19 &   7.6 \\
220544394 &    46 $\pm$     2 &   180 $\pm$     2 &      2.0 $\pm$      1.5 &    55 $\pm$     4 &    27 &     5 &   0.8 &    55 $\pm$    11 &    49 $\pm$    17 &   5.0 \\
220544394s &    73 $\pm$     5 &   198 $\pm$     4 &      1.0 $\pm$      0.1 &    94 $\pm$     8 &    94 &    16 &   1.7 &    94 $\pm$    18 &    32 $\pm$    28 &   4.5 \\
220576226 &    32 $\pm$     6 &   283 $\pm$     3 &      1.0 $\pm$      0.1 &    30 $\pm$     1 &    30 &     4 &   1.1 &    30 $\pm$    12 &    51 $\pm$    14 &   6.1 \\
220578040 &    27 $\pm$     9 &   103 $\pm$     2 &     18.5 &   654 &    35 &    16 &   5.1 &   247 $\pm$   205 &    50 $\pm$    23 &   7.0 \\
220584167 &    59 $\pm$     1 &   178 $\pm$     1 &      9.7 $\pm$      0.2 &   234 $\pm$     2 &    24 &    15 &   7.9 &   234 $\pm$    35 &    49 $\pm$    21 &  13.1 \\
220596913 &    80 $\pm$     2 &   247 $\pm$     1 &      0.9 $\pm$      0.1 &   141 $\pm$     2 &   153 &    22 &   4.3 &   141 $\pm$    10 &    38 $\pm$    28 &   9.3 \\
910193711 &    60 $\pm$     8 &    39 $\pm$     4 &      3.5 $\pm$      0.7 &    63 $\pm$     8 &    18 &    17 &   2.6 &    63 $\pm$    12 &    80 $\pm$    37 &   4.1 \\
910279515 &    52 $\pm$     4 &   108 $\pm$     7 &      3.8 &   265 &    70 &    34 &   5.3 &   186 $\pm$    14 &    47 $\pm$    40 &   2.7 \\
\hline
\end{tabular}
\end{center}
% Rajouter une colonne Beam smearing?
\footnotesize{The parameters of the kinematics modeling are the inclination (2), the position angle of the major axis (3), the turnover radius (4) and velocity (5). The inner slope is given in column (6). The mean residual of the velocity field is given in column (7) and the fit chi square in column (8). The maximum rotational velocity (9) and the mean velocity dispersion corrected from beam smearing (10) are computed after from the results of the fit. The extent of the velocity field (11) is derived using a S/N threshold of 3.\\
The suffix ``s'' refers to secondary objects which have been detected in H$\alpha$ and large enough to perform kinematics modeling. No error is given for $r_t$ and $V_t$ when the plateau is not reached (see Section \ref{other_parameters}).\\
$^1$ $V_{max}$ is not coming from the model but from a detailed analysis (see Appendix \ref{comments}).}
\label{kinem_table}
\end{table*}

Among the various dynamical states of galaxies, the easiest to probe is that of the rotating disk. We therefore tested the likelihood of this hypothesis for the galaxies in our sample and recovered the fundamental dynamical parameters within this hypothesis.
The \vf\ is accordingly fitted with a model that assumes that the ionized gas is located in an infinitely thin rotating disk, as in \citet{Epinat:2009}. The rotation curve is described by a linear slope in the inner parts and a plateau in the outer parts. The velocity along the line of sight is computed taking into account geometrical position effects.
The model parameters are
\begin{itemize}
 \item $x_c$, $y_c$: the center coordinates;
 \item $z$: the redshift corresponding to the systemic velocity;
 \item $i$: the inclination of the gaseous disk;
 \item $PA_k$: the position angle of the major axis;
 \item $V_t$: the plateau rotation velocity;
 \item $r_t$: the turnover radius at which the plateau is reached.
\end{itemize}

The method used to adjust the models is described in detail in \citet{Epinat:2010}. It is based on a $\chi^2$ minimization and takes the velocity error map into account to minimize the contribution of the regions with low S/N.
The spatial PSF is taken into account in these models and is described with a 2D Gaussian. Its FWHM is computed on the PSF stars associated to each observation.
To compute the model \vf, a higher resolution \vf\ is constructed (with at least eight pixels in the PSF FWHM) from the analytical model and a high-resolution line flux map has also to be built.
Indeed, in a final low spatial resolution element, the contribution of the line of sight velocity at higher resolution is weighted by the true line flux distribution.
We used a linear interpolation of the observed flux map to avoid making any assumption on the real flux distribution. This is one of the major uncertainties of our models because the real line flux distribution could be more clumpy than observed.

Our model allows us to compute the circular velocity within the disk hypothesis.
This assumption is not realistic for all objects in the MASSIV sample in which we also expect mergers (ongoing or late stage), spheroids, or structures with chaotic motions.
However, this hypothesis allows us to compute a map that contains only the beam-smearing effect on the velocity dispersion, due to the blurring of large-scale motions. Thus, by subtracting quadratically the map deduced from the model to the observed velocity dispersion map, a velocity dispersion map corrected for the beam-smearing effect is obtained (noted ``$\sigma$ residuals'' in Appendix \ref{kin_maps}).
This correction is also valid at first order for non-rotating objects, even if it can overestimate the correction in the inner parts. Details of the method are given in Appendix A of \citet{Epinat:2010}.
The maps of the models are shown in Appendix \ref{kin_maps} and the resulting parameters are given in Table \ref{kinem_table}.
For non-isolated galaxies (see Section \ref{classification}), we fitted the various components separately when they were sufficiently extended (VVDS220397579). In the table, companions have the suffix ``s''.

The model rotation curve only reproduces the \vf\ and is not based on a gravitational potential model (as for instance in \citealp{Forster-Schreiber:2006,Cresci:2009,Gnerucci:2011}) since the shape of the gravitational potential is unknown and can probably not be described by a stellar component only. Note that a gaseous thin disk in rotation is not incompatible with a spheroidal stellar distribution such as was observed in local ellipticals (e.g. \citealp{Sarzi:2006}) or as suggested from numerical simulations (e.g. \citealp{Bournaud:2007,Xu:2010}).

As described in the following sections, some parameters are difficult to constrain from the kinematics. To reduce the number of free parameters, we constrained the center and the inclination from the morphology to model the kinematics, assuming that the stars and the ionized gas follow a common distribution. Thus only four parameters remain free and can be reasonably constrained from our observed velocity fields.
Using these constraints, \citet{Epinat:2010} have shown from 137 galaxies of the GHASP sample (the largest 2D kinematics sample of nearby late-type galaxies described in \citealp{Epinat:2008_1,Epinat:2008_2}) projected at $z \sim 1.7$, that this method enables one to recover the other model parameters statistically.

\subsubsection{Center}

Owing to the low spatial resolution of our data, the kinematic center of a given galaxy is barely constrained from the kinematics \citep{Epinat:2010}. Therefore the centers are taken from the I-band morphology since we expect that in the inner parts of a galaxy at these redshifts, the stars dominate the gravitational potential.

Thanks to the method we used, the astrometry in the SINFONI data cubes matches the I-band astrometry. However, a post correction was applied to match the outer isophotes of the galaxy in the CFHT images and in the \ha\ maps.
The median offset of this post correction is 0.18\arcsec, which agrees with the accuracy of the SINFONI pointing system, which is estimated to be 0.1-0.2\arcsec\ (see SINFONI manual, using a guiding star instead of offsetting from a bright star).
This is also the final accuracy of our astrometry. This offset was computed for 38 galaxies. The other galaxies were excluded (i) when no PSF star was observed or (ii) when they were observed with AO because the observing sequence did not allow us to compute an astrometry correction.
We find a good agreement between the centers derived from the I-band images and the peak in the \ha\ maps.

\subsubsection{Inclination}
\label{inclination}

Disk inclination is a critical parameter to estimate because it is directly linked to the rotational velocity ($V_{\theta}$) of the disks. Indeed, we measured the velocities projected along the line of sight: $V_{los}=V_{\theta} \times \sin{i}$. Consequently, the observed velocity has to be corrected for the inclination and this correction is larger for galaxies with low inclination.
Owing to this degeneracy between inclination and rotation velocity in rotating disk models, the inclination is not well constrained from the kinematics alone.
% It is thus necessary to constrain the inclination in the models using information decoupled from the kinematics. Indeed, the inclination is not well constrained from the kinematics alone since rotating disk models present a degeneracy between the inclination and the rotation velocity
% because we measure the velocities projected along the line of sight: $V_{los}=V_{\theta}\times \sin{i}$.
% due to these projection effects.
This degeneracy can theoretically be solved for high-resolution observations but not for observations with strong beam-smearing \citep{Epinat:2010}: this would lead to an erroneous rotational velocity.
Therefore the inclination was constrained from the morphology axis ratio. In addition, since the uncertainty on the inclination (its sine) directly impacts the uncertainty on the rotation velocity, we paid special attention in deriving realistic error bars on the inclination.
The thickness of the disk, which is considered as null in our models, could impact the determination of the inclination, mainly for edge-on galaxies. However, for these objects, the deprojection has a negligible impact on the determination of $V_{max}$ (less than 3\% considering a thickness leading to $b/a\sim 0.2$).

\begin{figure}
\includegraphics[width=9cm]{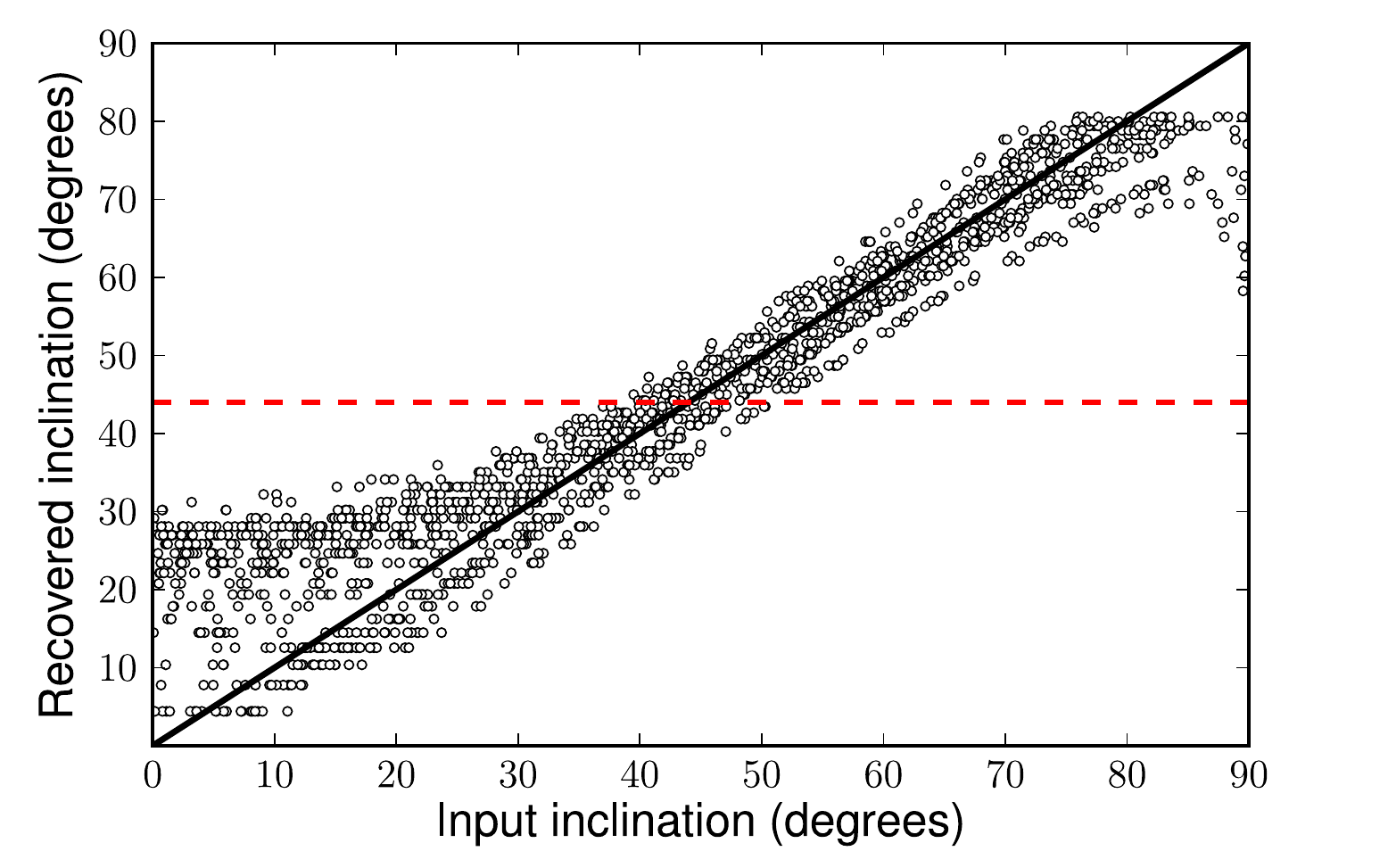}
\caption{Example of the method used to estimate the uncertainty on the inclination for galaxy VVDS020164388. The x-axis corresponds to the input inclination of the modeled galaxies, the y-axis corresponds to the inclination recovered by GALFIT for each modeled galaxy and the red dashed line corresponds to the inclination of the real galaxy. The uncertainty is measured along this line as the range that contains 68\% of the points (1$\sigma$).
}
\label{incl_carlos}
\end{figure}

To take into account both the uncertainty on the model and the uncertainty on the PSF used to recover galaxy parameters in GALFIT, we used a Monte Carlo method. For each galaxy we simulated 2000 synthetic sources with the same structural parameters (magnitude, Sersic index, effective radius, position angle), but a random value of $b/a$. The PSF to simulate these sources was also randomly chosen among different stars in the fields. Poisson noise was added and the simulated sources were placed in a nearby piece of sky from the original image (three pieces for each galaxy). The simulated sources were then modeled with GALFIT using the PSF used for the real galaxy. The final uncertainty was estimated from the range of input inclinations that led to the inclination that was measured with GALFIT on the real galaxy (see Figure \ref{incl_carlos}).
For the smallest galaxies, the inclination was not constrained enough. We therefore decided to use an inclination of 60\degr, the median value for randomly distributed disks, and an uncertainty of 24\degr, which yields a probability of 0.68 (1$\sigma$).
%  (see Appendix \ref{proba_computing}).

% The error bars on the inclination combine quadratically the statistical error on the axis ratio resulting from GALFIT ($\Delta b/a$) and the error linked to the uncertainty on the PSF ($\Delta s$). We compute lower and upper estimates for the axis ratio based on the error of the PSF:
% \begin{equation}
% (b/a)_l=\sqrt{\frac{b^2+s^2-(s+\Delta s)^2}{a^2+s^2-(s-\Delta s)^2}}
% \end{equation}
% \begin{equation}
% (b/a)_u=\sqrt{\frac{b^2+s^2-(s-\Delta s)^2}{a^2+s^2-(s+\Delta s)^2}}
% \end{equation}
% Where $s$ is the seeing. These errors thus increase for small galaxies with size comparable to the PSF.
% The final errors are (lower limit):
% $$\Delta(b/a)_l=\sqrt{(\Delta b/a)^2+((b/a)-(b/a)_l)^2}$$
% and (upper limit):
% $$\Delta(b/a)_u=\sqrt{(\Delta b/a)^2+((b/a)_u-(b/a))^2}$$

\begin{figure}
\includegraphics[width=8cm]{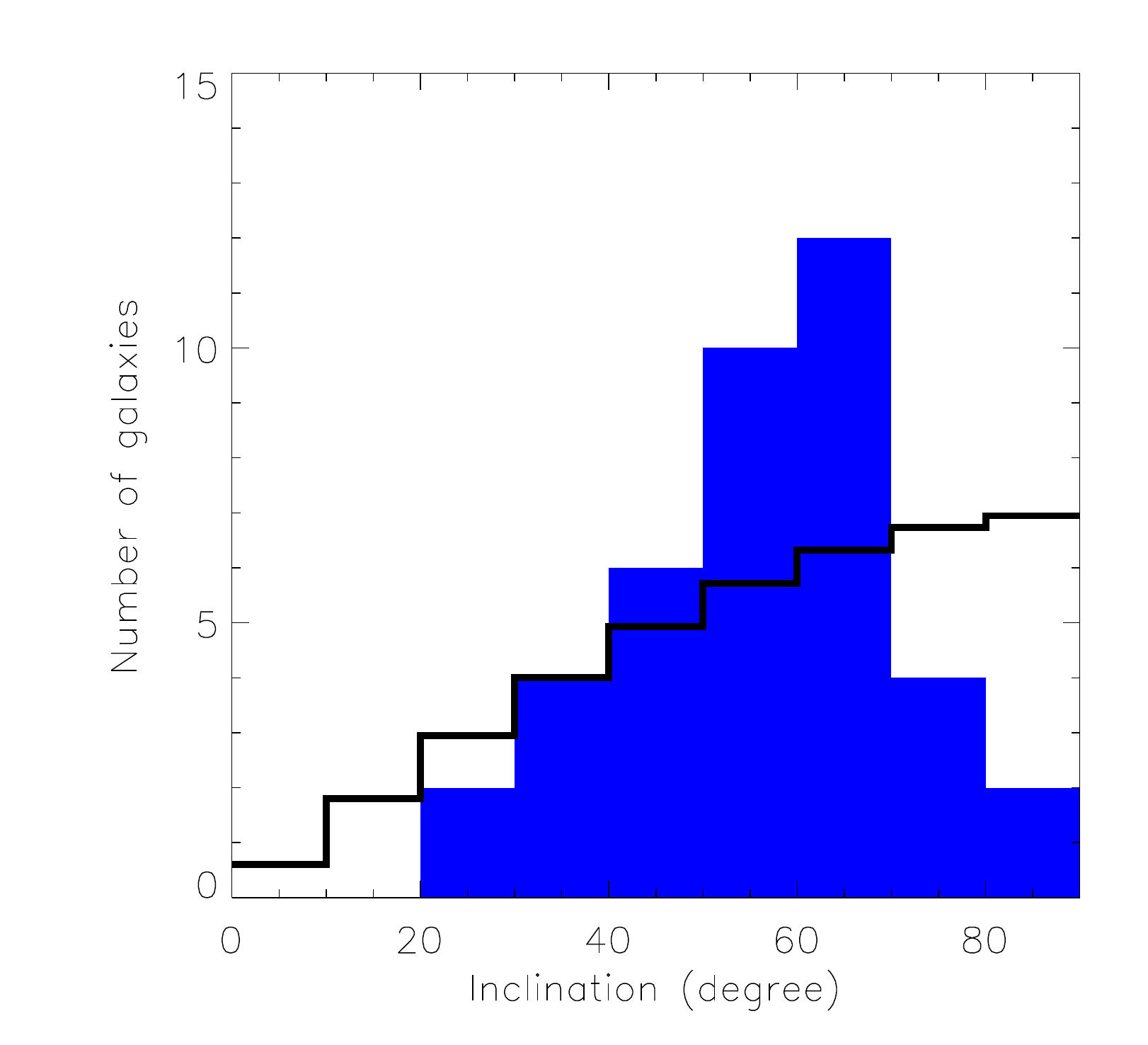}
\caption{Distribution of the inclinations for the 40 galaxies of the MASSIV sample (with a constrained inclination) using a Sersic profile with GALFIT and assuming they are thin disks (filled blue histogram) compared with the theoretical distribution for a randomly oriented thin disk (black histogram).
}
\label{histo_incl}
\end{figure}

We also checked that there was no bias by studying the inclination distribution for the MASSIV sample.
Theoretically, for uniformly randomly oriented thin disks (i.e. with a null thickness) the probability to observe a disk with an inclination between $\theta_1$ and $\theta_2$ is equal to $\mid\cos{\theta_1}-\cos{\theta_2}\mid$, leading to
%  we should observe the inclination
distribution as displayed in black in Figure \ref{histo_incl}
%  (see Appendix \ref{proba_computing}),
with a median value of 60\degr.
% since the probability to observe a disks with inclination $0 \le \theta_1 \le \theta \le \theta_2 \le \pi/2$ is :
% $$P(\theta_2)-P(\theta_1)=\cos{\theta_1}-\cos{\theta_2}$$
In Figure \ref{histo_incl}, we observe that the distribution of MASSIV galaxies with constrained inclinations (filled blue  histogram) misses face-on objects ($i=0$\degr) and edge-on objects ($i=90$\degr) and has an excess of objects with intermediate inclinations compared to the theoretical distribution. The most plausible explanation for the lack of edge-on morphologies is that disks are thick. It could also be that extinction is higher in these galaxies, inducing a non-detection of the \oii\ line in the VVDS spectra.
Figure \ref{incl_carlos} clearly shows that GALFIT cannot recover extreme inclinations.
However, the lack of face-on objects could also be attributed to the small numbers expected or to a computing bias. It could also be that galaxies do not have regular morphologies and that, due to surface brightness dimming, one can only observe the clumpy irregular emission. Indeed, I-band morphology could be in some cases contaminated by gaseous emission lines like \oii\ because the MegaCam/MegaPrime i' band filter covers the 0.70 to 0.84 $\mu m$ spectral range, which corresponds to \oii\ redshifted between $0.87$ and $1.26$. Unfortunately, the CFHT I-band imaging resolution does not allow us to conclude about the clumpy emission.
The peak around 60\degr\ could also be attributed to (i) the thickness of disks and (ii) the observation of galaxies that are not disks (mergers for example).
The median of the distribution for our sample (59\degr) is compatible with the median expected for a distribution of disks with random inclination.

\subsubsection{Constraints on the other model parameters}
\label{other_parameters}

Since inner velocity gradients reaching a plateau ($V_t$) within less than one pixel could not be resolved with our spatial resolution which is higher than four pixels, the turnover radius $r_t$ was constrained to be at least one pixel during the fitting process. However, when the model converged toward this value, the statistical error becomes null.
On the other hand, when the plateau is not reached according to the $\chi^2$ minimization, the error on $r_t$ becomes large and so does the error on $V_t$ because the models mainly constrain the slope. Therefore the errors on these parameters are difficult to interpret and to propagate to compute the error on the inner slope of the velocity gradient.
A solution would consist in using a model described by the slope $\alpha$ and the turnover radius $r_t$ to estimate the error on the slope. This test has been performed but did not give more realistic errors in many cases.
Because this solution was not convincing, we used the original model. For these galaxies, no error is indicated for $r_t$ and $V_t$ in Table \ref{kinem_table}.
Although $r_t$ is not well constrained for small galaxies, \citet{Epinat:2010} showed with 137 galaxies projected at $z\sim 1.7$ that leaving $r_t$ as a free parameter statistically gives a good estimate of the shape of the rotation curve. It is therefore necessary to let $r_t$ be a free parameter, in particular for the largest galaxies.

\subsubsection{Parameters deduced from the kinematic maps and models}
\noindent
\textbf{Radius of the H$\alpha$ extent $R_{last}$}
\\
This radius is computed from the cleaned maps. The center and position angle from the kinematic best-fit models were used to derive the radius of each pixel. $R_{last}$ is the radius that both sides of the galaxy reach.
\\
\\
\textbf{Velocity shear $V_{shear}$}
\label{vshear}
\\
$V_{shear}$ was computed as the total shear observed in the modeled velocity field shown in Appendix \ref{kin_maps} (i.e. with an S/N threshold of 3) which takes into account the uncertainty on the observed velocity field. It is a projected velocity along the line of sight and is not inclination-corrected and consequently makes no assumption on the geometry of the galaxy. The use of the model enables us to smooth the velocity field at the outskirts.
\\
\\
\textbf{Maximum rotation velocity $V_{max}$}
\\
This was computed according to the model at the $R_{last}$ radius.
Two sources of uncertainties were added in quadrature to compute the final uncertainty on $V_{max}$.

The first one is the uncertainty on the inclination (cf. Section \ref{inclination}). To propagate uncertainties from inclination to rotation velocity, we used a Monte Carlo method assuming a Gaussian distribution for the inclination. The uncertainty on $V_{max}$ was the standard deviation of the resulting distribution on this parameter.

The second source of uncertainty is related to the modeling. Because the uncertainty on $V_t$ is a statistical one, it can be fairly small. Instead of using an approach based on these errors, the GHASP sample \citep{Epinat:2008_2,Epinat:2008_1} was exploited to compute model uncertainties related to the size of the galaxies with respect to the seeing. Using the same method as \citet{Epinat:2010}, 136 GHASP galaxies were projected at $z=1.33$ under typical seeing and sampling of SINFONI observations of the MASSIV sample and the S/N of the simulations was adjusted to match the H$\alpha$ fluxes from the MASSIV sample.
%  (see Section \ref{simulations_ghasp}). 
Figure 13 of \citet{Epinat:2010} was reproduced with these new simulations. This figure displays the relative error on the maximum velocity determination with respect to the beam-smearing parameter defined as $B={D_{25}/2}{s}$, where $D_{25}$ is the optical diameter and $s$ is the seeing FWHM. A linear regression was performed to estimate the evolution of the accuracy of the fit with galaxy size. The best fit gives

\begin{equation}
\delta V_{max}=V_{max}\times \frac{27.5-5.8 B}{100}\;.
\label{vmaxuncertainty}
\end{equation}

The percentage of galaxies below this relation is about 60\%.
Assuming a Gaussian distribution, 1-$\sigma$ corresponds to a confidence level of 68\%. We therefore simply approximated that $\delta V_{max}$ is the 1-$\sigma$ uncertainty.
For the MASSIV sample, $B$ was estimated from half-light radii ($R_e$) determined by GALFIT on I-band images (see Section \ref{model_morph}). Indeed, assuming an exponential disk, the following relation can be written: $D_{25}/2 = 1.9 R_e$. The model uncertainty was consequently determined following Equation \ref{vmaxuncertainty}. A minimum uncertainty of 10 km/s was imposed, however. This approach is well-adapted for rotators because it was built from a control sample of rotators.
\\
\\
\textbf{Local velocity dispersion $\sigma$}
\\
The velocity dispersion was computed as in \citet{Epinat:2009}: this is the average of the velocity dispersion map corrected for beam-smearing effects (see Section \ref{model_kin}) and spectral PSF. A weight proportional to the inverse of the uncertainty on the velocity dispersion (estimated before correcting for beam-smearing and spectral resolution) was attributed to each pixel to compute the average. The uncertainty on this parameter was computed as the weighted standard deviation.

\section{Kinematics classification}
\label{classification}

\begin{table*}
\caption{Kinematics and close environment classification of MASSIV ``first epoch'' sample galaxies}
\begin{center}
\begin{tabular}{cccccccccc}
\hline
VVDS ID & S/N & $\frac{R_{last}}{Seeing}$ & $\Delta PA$ [\degr] & $\frac{Res}{V_{shear}/2}$ & $V_{max}/\sigma$ & Shear & Rotator & Isolated & Isolation flag \\
 (1) & (2) & (3) & (4) & (5) & (6) & (7) & (8) & (9) & (10) \\
\hline
020106882 &   7.6 &   1.2 & $   23\pm    4$ &     0.07 &   3.2 & High & Yes & Yes & B \\
020116027 &   8.2 &   1.3 & $   23\pm    6$ &     0.27 &   0.6 & Low & No & No & A \\
020126402 & $<3$ & $-$ & $-$ & $-$ & $-$ & $-$ & $-$ & $-$ & $-$ \\
020147106 &  12.2 &   1.4 & $    7\pm    3$ &     0.18 &   0.3 & Low & Yes & Yes & B \\
020149061 &   6.2 &   0.7 & $   34\pm   90$ &     0.17 &   1.5 & Low & Yes & Yes & B \\
020164388 &  11.0 &   1.2 & $   65\pm    5$ &     0.11 &   1.5 & Low & No & Yes & B \\
020167131 &   3.7 &   0.3 & $   87\pm   11$ &     0.22 &   5.0 & Low & No & No & A \\
020182331 &   6.1 &   0.9 & $   34\pm    6$ &     0.09 &   2.0 & High & No & Yes & B \\
020193070 &   4.9 &   0.8 & $   31\pm    3$ &     0.10 &   3.6 & High & No & Yes & B \\
020208482 &   4.1 &   0.3 & $   20\pm   12$ &     0.18 &  22.9 & Low & Yes & Yes & B \\
020214655 &   7.5 &   0.8 & $   51\pm    5$ &     0.17 &   0.8 & Low & No & Yes & B \\
020217890 & $<3$ & $-$ & $-$ & $-$ & $-$ & $-$ & $-$ & $-$ & $-$ \\
020239133 &   5.3 &   0.8 & $   14\pm    6$ &     0.11 &   2.0 & High & Yes & Yes & B \\
020240675 &   5.5 &   0.5 & $    7\pm   18$ &     0.40 &   1.5 & Low & No & Yes & B \\
020255799 &   4.5 &   0.6 & $   75\pm   77$ &     1.04 &   0.2 & Low & No & Yes & B \\
020261328 &   7.8 &   1.0 & $    8\pm    8$ &     0.07 &   2.4 & High & Yes & Yes & B \\
020278667 &   4.1 &   0.2 & $   35\pm   14$ &     0.34 &   1.5 & Low & No & Yes & C \\
020283083 &   7.0 &   0.9 & $   58\pm    5$ &     0.18 &   1.5 & Low & No & No & B \\
020283830 &   4.8 &   1.2 & $   14\pm    2$ &     0.08 &  11.3 & High & Yes & No & B \\
020294045$^1$ &   7.0 &   1.2 & $    2\pm    4$ &     0.12 &   3.9 & Low & No & No & B \\
020306817 & $<3$ & $-$ & $-$ & $-$ & $-$ & $-$ & $-$ & $-$ & $-$ \\
020363717 &  11.4 &   1.1 & $   49\pm    8$ &     0.19 &   0.5 & Low & No & Yes & B \\
020370467 &   5.5 &   0.9 & $   31\pm   27$ &     0.45 &   0.6 & Low & No & Yes & B \\
020386743 &   8.8 &   0.9 & $   67\pm    8$ &     0.26 &   0.8 & Low & No & No & A \\
020461235 &   5.8 &   1.1 & $   19\pm    4$ &     0.10 &   3.5 & High & Yes & No & B \\
020461893 &   7.0 &   1.3 & $    4\pm    5$ &     0.10 &   0.9 & Low & Yes & Yes & B \\
020465775 &   6.8 &   0.7 & $   31\pm    5$ &     0.24 &   0.8 & Low & No & No & B \\
140083410 &   6.0 &   1.0 & $   88\pm   12$ &     0.35 &   0.4 & Low & No & Yes & A \\
140096645 &  10.3 &   1.0 & $   20\pm   90$ &     0.07 &   3.9 & High & Yes & No & B \\
140123568 &   3.5 &   0.1 & $   43\pm   90$ &     0.44 &   0.7 & Low & No & Yes & B \\
140137235 &   3.2 &   0.1 & $    8\pm   29$ &     0.11 &   3.5 & Low & Yes & Yes & B \\
140217425 &   6.8 &   1.9 & $    2\pm    1$ &     0.04 &   7.1 & High & Yes & Yes & B \\
140258511 &   8.1 &   1.3 & $   52\pm   42$ &     0.12 &   5.1 & High & Yes & Yes & A \\
140262766 &   6.5 &   1.0 & $   32\pm   28$ &     0.07 &   3.1 & High & Yes & Yes & B \\
140545062 &   8.5 &   1.3 & $   14\pm   41$ &     0.06 &   3.1 & High & Yes & Yes & B \\
220014252 &  11.0 &   1.7 & $    5\pm    2$ &     0.08 &   1.4 & High & Yes & Yes & B \\
220015726 &  10.4 &   1.0 & $    8\pm   12$ &     0.04 &   3.7 & High & Yes & Yes & B \\
220071601 & $<3$ & $-$ & $-$ & $-$ & $-$ & $-$ & $-$ & $-$ & $-$ \\
220148046 &   4.0 &   0.4 & $   25\pm   18$ &     0.39 &   0.9 & Low & No & Yes & C \\
220376206 &  12.6 &   2.4 & $   10\pm    1$ &     0.05 &   2.8 & High & Yes & No & B \\
220386469 &   4.5 &   1.1 & $   17\pm   12$ &     0.42 &   0.9 & Low & No & Yes & B \\
220397579 &  15.7 &   2.0 & $   31\pm   10$ &     0.68 &   0.2 & Low & No & No & A \\
220397579s &   5.4 &   1.2 & $   11\pm    3$ &     0.06 &   8.3 & High & Yes & No & A \\
220544103 &  10.4 &   1.2 & $   11\pm   55$ &     0.07 &   1.9 & High & Yes & No & B \\
220544394 &  10.5 &   1.1 & $   54\pm    7$ &     0.08 &   1.1 & Low & No & No & A \\
% 220544394s &   5.3 &   1.0 & $   35\pm   11$ &     0.15 &   2.9 & High & No & $-$ & $-$ \\
220576226 &  11.6 &   1.3 & $   45\pm   15$ &     0.15 &   0.6 & Low & No & Yes & B \\
220578040 &   5.9 &   1.4 & $   10\pm   14$ &     0.08 &   4.9 & High & Yes & Yes & C \\
220584167 &  13.8 &   2.1 & $   15\pm    2$ &     0.04 &   4.8 & High & Yes & Yes & B \\
220596913 &   5.7 &   6.2 & $    8\pm    1$ &     0.09 &   3.7 & High & Yes & Yes & B \\
910193711 &   6.0 &   1.8 & $   30\pm    4$ &     0.22 &   0.8 & Low & No & Yes & C \\
910279515 &   3.5 &   1.5 & $   23\pm    8$ &     0.11 &   3.9 & High & Yes & Yes & C \\
\hline
\end{tabular}
\end{center}
\footnotesize{
Column (2) gives the average S/N of the S/N map. The size of the galaxy with respect to the seeing is given in column (3). The criteria used to determine if galaxies are in rotation or not (8) are the mismatch between morphological and kinematic position angles (4) and the residuals normalized by the velocity shear (5). The ratio of the rotation velocity over the local velocity dispersion is given in column (6).
Galaxies are classified according the observed velocity shear in column (7): galaxies with $V_{shear}<100$\kms\ or $V_{shear}<100$\kms\ are respectively classified as low and high shear). The close environment classification and its associated quality flag are given in (9) and (10).\\
The suffix ``s'' refers to secondary objects which have been detected both in H$\alpha$ and in the I-band image.\\
$^1$ This galaxy is classified as non-rotating even if it fulfill the criteria because the kinematics seem to indicate that this system is composed of two close companions (cf. Appendix \ref{comments}).}
\label{classification_table}
\end{table*}

To distinguish between galaxy formation scenarii, it is necessary to know the dynamical state of galaxies for large samples and at various redshifts.

\subsection{Previous classification schemes of large samples}

The first kinematics classification of distant galaxies observed using integral field unit techniques was provided by \citet{Flores:2006} in the frame of the IMAGES sample ($0.4<z<0.75$) which contains 68 classified galaxies \citep{Neichel:2008}. Their classification is a visual one that relies on both HST optical images and FLAMES/GIRAFFE data. The galaxies were split into three classes: (i) rotating disks whose morphological and kinematic position angles match well and with a velocity dispersion peak in the center, (ii) perturbed rotators whose morphological and kinematic position angles match well but that have a peak in the velocity dispersion offset from the center and (iii) complex kinematics systems with both a disagreement between morphological and kinematic position angles and a peak in the velocity dispersion offset from the center. The IMAGES survey has a majority of galaxies with complex kinematics (44\%), and a nearly equivalent fraction of rotating disks (29.5\%) and of perturbed rotators (26.5\%) \citep{Yang:2008,Neichel:2008}. From these numbers, they concluded that at that epoch merging is still more active than in the local Universe.

The classification of the SINS sample ($z\sim 2.2$) relies on the asymmetries measured on both the velocity and velocity dispersion fields to distinguish rotator-like from merger-like galaxies. These asymmetries were derived either from a kinemetry analysis \citep{Shapiro:2008} or from a qualitative assessment \citep{Forster-Schreiber:2009}. Then, comparing the mean local velocity dispersion and the rotational velocity, these galaxies (both mergers and rotators) were classified either as rotation-dominated ($V_{max}/\sigma_0>1$) or dispersion-dominated ($V_{max}/\sigma_0<1$). For galaxies with small size or low S/N, they instead compared the full velocity shear $v_{obs}$ with the integrated line width $\sigma_{int}$ and used a threshold of $v_{obs}/(2\sigma_{int})\sim 0.4$. On the one hand, \citet{Forster-Schreiber:2009} found that one third of the 62 galaxies in the \ha\ SINS sample contains mergers. On the the other hand, they also found that one third of this sample contains rotation-dominated systems, another third corresponds to dispersion-dominated systems, and the last third are not classified. The authors interpret the large amount of dispersion-dominated disks as an evidence for cold gas accretion along cosmic web filaments.

For the LSD/AMAZE sample, the classification only distinguishes rotator-like from perturbed galaxies because galaxies have
% are mainly small and with
a fairly low S/N. This classification relies on the modeling  of the velocity field as an inclined X-Y plane \citep{Gnerucci:2011}. The criterion is based on the $\chi^2$ of the fit by the plane and on a constraint on the reliability of the inclination of the plane.

In the pilot run study of MASSIV \citep{Epinat:2009}, which was based on a visual kinematic classification, nine galaxies with $1.2<z<1.6$ were classified into three groups: (i) three galaxies are mergers, showing disturbed kinematics and possibly several components, (ii) two objects are classified as rotation-dominated disks, with clear signs of rotation and with a maximum rotation velocity higher than the mean local velocity dispersion and (iii) four galaxies are dispersion-dominated disks, with clear signs of rotation but with a maximum rotation velocity lower than the mean local velocity dispersion.
It was concluded that about one third of these galaxies are observed during some merging event, whereas one third are rotation-dominated and the other third are dispersion-dominated.

In this paper, we present a new kinematic classification for the MASSIV sample to put constraints on the dynamical state of galaxies at $z\sim1.2$.
The classification scheme was developed in several steps. First, eight people of the collaboration independently defined their own criteria.
These criteria were then reconciled and discussed during a common session and led to a unique classification based both on the close environment and on the velocity shear strength.
The visual classification helped in defining measurable criteria to build an automatic, thus reproducible, classification (this led to minor changes in the final classes). The final classification was also refined to describe the dynamical state and support of the galaxies.

\subsection{Galaxy small-scale environment}

The study of galaxy close environment is crucial for inferring a merger rate on the population probed by the MASSIV sample at $z\sim 1.2$. Using SINFONI data and CFHT imaging, galaxies could be classified as interacting or isolated. Flags have also been attributed to this classification to qualify its reliability. Flags ``A'', ``B'' and ``C'' mean secure classification ($> 90$\% probability), confident ($\sim 75$\% probability) classification and poor reliability ($\sim 50$\% probability), respectively.

The SINFONI field-of-view was explored to detect emission lines that might be attributed to companions. The same field was also explored systematically in the I-band images. Owing to the nodding strategy, the field-of-view is larger for seeing-limited observations. In that case, the shape of the field results from the superposition of two 8\arcsec\ square fields overlapped on their opposite corners (4\arcsec\ quadrants). Thus, the field-of-view of the combined data around each target can be as large as 6\arcsec\ (corresponding to $\sim50$ kpc at $z\sim1.2$). In the non-overlapping regions, the exposure time is half the total on-source exposure time but is sufficient to at least detect emission lines of objects. The explored field is not a full 12\arcsec$\times$ 12\arcsec\ square field but only 78\% of this area. For AO observations, the field of view is restricted to 3\arcsec\ (corresponding to $\sim25$ kpc), which is insufficient for studying the environment and therefore only the I-band image was explored to see if a possible companion was observed in the 6\arcsec\ around the main object.

Galaxies were classified as interacting with a confidence flag ``A'' when the following conditions were met: (i) the average S/N (see Table \ref{classification_table}) has to be higher than three in an area larger than the seeing, (ii) the systemic velocity difference between the two components has to be lower than 1000 \kms\ and (iii) an optical counterpart has to be observed within the CFHT I-band image. If the companion was extended but slightly smaller than the seeing, we assigned a flag ``B''.
When no counterpart was observed in I-band or if some galaxies were observed in the neighborhood in I-band but not in the SINFONI datacube, the object was considered as isolated with a flag ``B''.
A flag ``C'' was assigned for AO observations unless absolutely no other galaxy was observed in I-band around the object. This is detailed in Appendix \ref{comments}.

In some cases, the morphology is elongated and several \ha\ blobs can be detected along the elongation, sometimes also in I-band images (see comments in Appendix \ref{comments}). It is not straightforward to asses if these blobs are small objects that are accreted and not star forming regions within one unique galaxy.
A monotonic velocity field along the direction of the major axis is likely evidence for rotation. However, if the elongation is the result of the projection of the two components and if the two objects are small (no gradient detectable in each component), this could mimic a monotonic velocity field caused by the difference of radial velocity between the components.
Some additional features can also support a merger hypothesis:
(i) the angular separation is larger than 3\arcsec\ (ie $\sim25$ kpc), because galaxies this large are not expected at high redshift, nor are they observed in the local Universe;
(ii) the radial velocity gap is larger than $\sim600$\kms, because this rotational velocity would imply an unrealistic dynamical mass for a unique object.
If one of these features was seen, the galaxies were classified as interacting but with a flag ``B''. If none of them was observed, the galaxies were classified as isolated but a flag ``B'' was assigned.
For the difficult cases, we were able to use the morphology traced by old stars (mainly observable in near-infrared bands): the old star population may not follow the star forming regions in case of unique objects. The S/N of these J- or H-band continuum morphologies recovered from SINFONI data is fairly low, however. It is therefore not straightforward to arrive at definite conclusions.

When components are very close, i.e. about to merge, the situation is more complex. However, perturbed line profiles (e.g. VVDS020294045, cf. Appendix \ref{comments})
% \ref{intslow})
in the region between the two blobs favor a strong discontinuity in the velocity field, which points toward a two-component hypothesis. These cases received a flag ``B''.

For ongoing mergers one expects to observe peculiar kinematic signatures. Therefore, we classified some galaxies as interacting because their velocity fields, velocity dispersion maps and morphologies were presenting perturbations in the same positions, suggesting for instance the presence of tidal arms (e.g. VVDS020283083, cf. Appendix \ref{comments}).
% \ref{intfast}).
Since these signatures are more subjective, these galaxies usually have a flag ``B'' for the isolation criterion.

Other galaxies were classified as isolated. However, when there were some perturbations in the velocity field or a peak in the velocity dispersion map, this could be interpreted as signs for merger remnants, but because it concerns the galaxy outskirts, it is more likely noise and we assigned a flag ``B''.

\subsection{Velocity shear strength}

For the kinematic classification, we used a first criterion as simple and objective as possible based on the velocity shear $V_{shear}$ (see Section \ref{vshear}). The sample was accordingly divided into low-velocity shear galaxies ($V_{shear} < 100$\kms) and high-velocity shear galaxies ($V_{shear} > 100$\kms). This gives a rough idea about the dynamical state of a galaxy without taking into account more complex motions.
For VVDS220397579 and VVDS220544394 it was possible to study the velocity shear of the companion as well.

This criterion has the advantage that it can be easily measured.
However, the value of $V_{shear}$ for a given galaxy may vary with the orientation. This is particularly true for galaxies in ordered rotation ($V_{shear}$ is lower when observed face-on).
However, the distribution of the inclinations, as seen in Figure \ref{histo_incl}, shows that only eight galaxies according to the theoretical distribution and only three according to the observed distribution may need a correction larger than a factor of two (corresponding to an inclination of 30\degr) for the measurement of the velocity shear.

In galaxies with a low-velocity shear there might be several classes of objects: (i) very low-mass objects, (ii) face-on rotating galaxies, (iii) ongoing mergers in a transient state, (iv) spheroids, if one expects that the gas in these objects follows the distribution of stars, and (v) galaxies with a non/slowly rotating gaseous component.

\begin{figure}
\includegraphics[width=8.5cm]{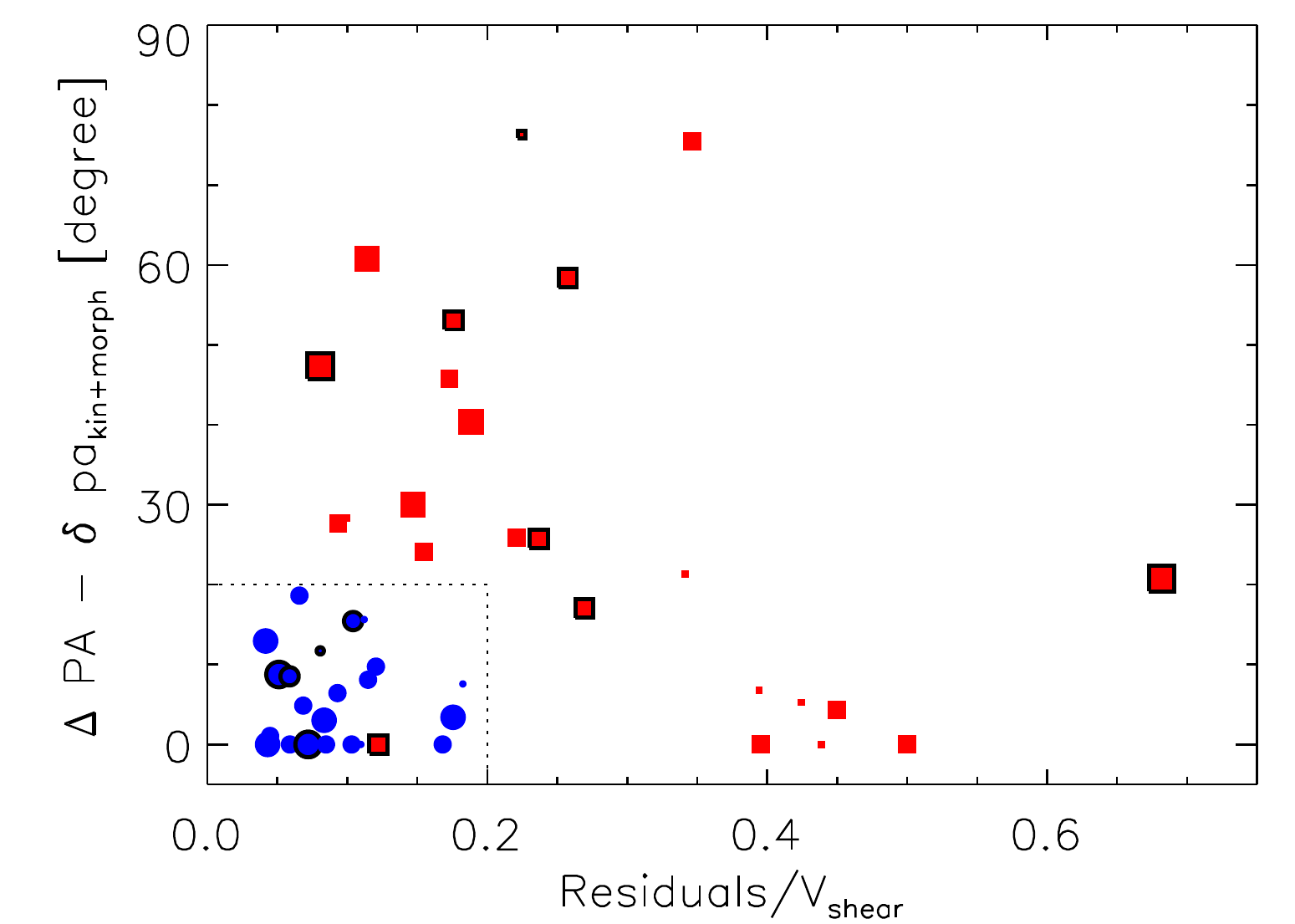}
\caption{
Disagreement between morphological and kinematic position angles (within the errors) as a function of the velocity field residuals normalized by the velocity shear.
Blue dots and red squares represent rotating and non-rotating galaxies. Symbols with black contours mark interacting galaxies. The symbol size is related to the S/N of the SINFONI data: small, medium and large symbols refer to galaxies with $S/N<5$, $5<S/N<10$ and $S/N>10$, respectively.
One galaxy is not in the expected region (VVDS020294045, cf. Appendix \ref{comments}).
}
\label{classification_res_dpa}
\end{figure}

\subsection{Rotating and non-rotating galaxies}
\label{class_rot_nrot}

For a specific analysis (e.g. Tully-Fisher relation, see \citeauthor{Vergani:2012} submitted), it is essential to be able to define a robust sample of rotating galaxies.
We expect a rough agreement of the morphological and kinematic position angles for rotators. They can be slightly different due to perturbations like bars or strong spiral arms. We also expect for these galaxies that the rotation motions dominate over perturbations. To distinguish rotators from non-rotating galaxies, we made a diagram in Figure \ref{classification_res_dpa} with two quantities that quantify these two arguments: the disagreement between morphological and kinematic position angles derived from the models presented in sections \ref{model_morph} and \ref{model_kin} within the uncertainties, $|PA_{m}-PA_{k}| - \sqrt{\delta PA_{m}^2+\delta PA_{k}^2}$, and the mean weighted velocity field residuals normalized by the velocity shear.
Rotators are therefore defined as galaxies with a position angle difference lower than 20\degr\ and velocity field perturbations lower than 20\%. These thresholds were chosen from the distribution seen in Figure \ref{classification_res_dpa} because they enable us to isolate a clear cloud of points near the origin (the rotators). We note that these values are also representative of relatively small deviations from rotational motion and position angles.
There is one exception in this diagram: VVDS020294045 was classified visually as non-rotating although it lies in the rotator's region of the diagram. Indeed, the velocity shear visible in its velocity field is probably caused by a very close companion that thus mimics a rotating disk velocity field. The shape of the line suggests a merger. If we were to exclude the companion, which is overlapping with the main galaxy, the resulting velocity field would probably be classified as non-rotating (see Appendix \ref{comments}).

We checked the agreement between this classification and the classification based on kinemetry used in the SINS sample \citep{Shapiro:2008}.
The agreement is reasonable but we find that galaxies that would be classified as mergers according to their criteria are mainly those that we have classified as non-rotating.
Indeed, the classification of \citet{Shapiro:2008} relies on the position in a diagram in which galaxies are placed according to their asymmetric velocity and velocity dispersion components. Galaxies show a strong correlation between these two quantities. This correlation arises because asymmetries are normalized by the global velocity shear. We are able to show that the normalized residuals of the model are well-correlated with the normalized asymmetries in the velocity field. This is true for large and small galaxies. Thus, our classification adds the agreement between the position angles in morphology and kinematics as a new criterion.
One advantage of this criterion is that it is not restricted to galaxies observed with a good S/N and with many resolution elements.

\subsection{Dynamical support}

The last criterion is related to the dynamical support of galaxies. Indeed, it has been shown that at high redshift ($z>2$), galaxies are on average more dispersion-dominated than in the intermediate ($z\sim 0.5$) and local Universe.
One of our goals is to determine the evolution of the fraction of dispersion-dominated systems with redshift since MASSIV might probe a transitional redshift range.

To quantify the dynamical support, we assumed that all galaxies have some rotation and that this rotation is adequately described by the rotating-disk model. Then, we additionally assumed that the local velocity dispersion of the gas is representative of the random motions in the galaxy. Thus, the dynamical support is estimated by the ratio of the maximum rotation velocity over the local velocity dispersion $V_{max}/\sigma$. This description is intrinsically better justified for rotators, however.

The various classes and related parameters are summarized in Table \ref{classification_table}.

\section{Discussion}
\label{discussion}

Using the classification scheme described above, we were able to classify as rotating or non-rotating the 46 galaxies (including one companion) in the redshift range $0.9<z<1.6$ for which some emission line was detected in the SINFONI datacubes. For part of the discussion below we limit the sample to the 36 galaxies with an average S/N (defined in Section \ref{maps_extraction} and reported in Table \ref{classification_table}) higher than 5 for which the kinematic classification is more robust (we note that these low S/N galaxies are the majority of galaxies with $R_{last}/seeing<1$, which means that the kinematic model might be less robust).
Accordingly among an initial sample of 46 secure galaxies with $0.9<z<1.6$, 22\% are not robustly classified or not classified from their dynamical properties. This proportion is on the same order for the LSD/AMAZE (\citealp{Gnerucci:2011}). In the MASSIV sample, these galaxies have on average lower stellar masses and star formation rates (see Table \ref{class_properties}), the latter explaining the lower S/N of these observations. We also emphasize that the high dispersions are caused by the two galaxies observed with AO, which have larger stellar masses and SFR than the rest of the unclassified galaxies. For those two galaxies the low S/N is due to the small pixel size rather than an intrinsically low SFR.
Concerning the close environment criterion, the 41 systems classified with flags ``A'' and ``B'' are considered.

\begin{table*}
\caption{Physical properties of MASSIV subclasses}
\begin{center}
\begin{tabular}{c|c|ccc|ccc|ccc|ccc}
Class & N & \multicolumn{3}{c|}{$M_{star}$} & \multicolumn{3}{c|}{$SFR$} & \multicolumn{3}{c|}{$R_e$} & \multicolumn{3}{c}{$\sigma$} \\
      &        & \multicolumn{3}{c|}{[$10^{10} \msun$]} & \multicolumn{3}{c|}{[\msunyr]} & \multicolumn{3}{c|}{[kpc]} & \multicolumn{3}{c}{[km~s$^{-1}$]} \\
 &  & Med & Mean & Dev & Med & Mean & Dev & Med & Mean & Dev & Med & Mean & Dev \\
\hline
\hline
Rotating     & 19 &  3.9 &  4.0 &  3.7 &  46 &  60 &  44 &  3.1 &  3.8 &  2.4 &  62 &  58 &  19 \\
Non-Rotating & 16 &  1.2 &  1.6 &  1.2 &  30 &  39 &  29 &  2.7 &  2.7 &  1.2 &  60 &  61 &  17 \\
Detected but unclassified &10 &  1.5 &  2.3 &  2.1 &  24 &  34 &  30 &  2.3 &  2.3 &  1.0 &  43 &  39 &  24 \\
\hline
Isolated    & 28 &  1.4 &  3.0 &  3.4 &  37 &  47 &  41 &  2.7 &  3.1 &  2.2 &  62 &  56 &  22 \\
Interacting & 13 &  1.7 &  2.1 &  1.4 &  30 &  43 &  32 &  4.0 &  3.9 &  1.4 &  53 &  52 &  21 \\
\hline
Rotating + Isolated    & 14 &  3.9 &  4.1 &  4.3 &  58 &  63 &  49 &  2.9 &  3.8 &  2.7 &  62 &  57 &  19 \\
Rotating + Interacting &  4 &  4.6 &  3.6 &  1.4 &  64 &  55 &  33 &  5.3 &  4.2 &  1.7 &  73 &  61 &  25 \\
Non-Rotating + Isolated    &  8 &  1.4 &  2.0 &  1.7 &  42 &  37 &  22 &  1.9 &  2.0 &  1.1 &  66 &  64 &  19 \\
Non-Rotating + Interacting &  7 &  1.2 &  1.3 &  0.5 &  26 &  35 &  33 &  3.4 &  3.5 &  0.7 &  53 &  56 &  15 \\
\hline
\hline
\end{tabular}
\end{center}
\footnotesize{N is the number of galaxies in each subclass; med: median value; mean: average; dev: standard deviation. Only the main objects are considered.}
\label{class_properties}
\end{table*}

\subsection{Rotating disks vs. non rotating galaxies}

Rotating disks represent at least 44\% (20 out of 46 detected galaxies) of the MASSIV ``first epoch'' sample and about 56\% of the high S/N sample.
We find a lower percentage of secure non-rotating systems (35\% or about 44\% of the high S/N sample) that have no observed rotation in the gaseous component or have very disturbed kinematics with respect to their broad band morphology.
The physical properties of these two classes of galaxies are on average different.
We have performed Kolmogorov-Smirnov tests \citep{Press:1992} on stellar mass, SFR, half-light radius and velocity dispersion distributions.
The two populations look different in terms of stellar mass and SFR at a level slightly higher than $1\sigma$ (the probability to follow the same distribution is lower than 0.2). Distributions for size and velocity dispersion are fully compatible.
Rotating objects are on average more massive ($M_{star}=4.0\times 10^{10}\msun$), more star-forming ($SFR=60$\msunyr) and have larger radii ($R_e=3.8$ kpc) than non-rotating ones ($M_{star}=1.6\times 10^{10}\msun$, $SFR=39$\msunyr\ and $R_e=2.7$ kpc).
On the other hand, these two types of galaxies have very similar velocity dispersions around 60\kms\ (see Table \ref{class_properties}).
At $z\sim 2.2$, \citet{Forster-Schreiber:2009} already noticed a similar trend from the SINS sample: dispersion-dominated systems are on average smaller than rotation-dominated ones.

The percentage of rotating systems is higher in MASSIV at $z\sim 1.2$ than in the LSD/AMAZE sample at $z\sim3.3$. Indeed, \citet{Gnerucci:2011} found a lower limit of 34\% of rotators in their sample (11 out of 32 detected galaxies that they were able to classify).
Their classification is closest to ours since it relies both on velocity field modeling and on the agreement between morphological and kinematic position angle of the major axis. However, owing to their small statistics and their different selection function it cannot be excluded that these proportions are compatible.
At $z\sim 2.2$, the comparison with the SINS sample is more difficult because of the selection function, as for LSD/AMAZE, but also because of the classification scheme, which mainly relies on a kinemetry analysis and on a visual inspection \citep{Forster-Schreiber:2009}. \citeauthor{Forster-Schreiber:2009} were able to unambiguously identify 18 rotating systems (discussed in \citealp{Cresci:2009}), i.e. a lower limit of 35\% (18 out of 52 detected systems). These authors were also able to identify 14 dispersion-dominated systems (27\% of the SINS sample) based on $V_{max}/\sigma$ ratio. Considering only the reliably classified systems, SINS and MASSIV percentages are similar.
At lower redshift, the results of the IMAGES sample ($z\sim 0.6$) show that 63\% of the sample show signs of rotation (both rotating disks and perturbed rotators classes, \citealp{Puech:2008}). This percentage is higher than for MASSIV and favors an interpretation in which gas in star-forming systems is stabilizing into disks while the Universe evolves. This conclusion has to be balanced with the various selection functions and classification methods, in the same way as for the comparison with the other surveys discussed previously. Concerning the selection, the various authors claim that they observed representative sets of star-forming galaxies in each redshift range.

\begin{figure}
\includegraphics[width=8.5cm]{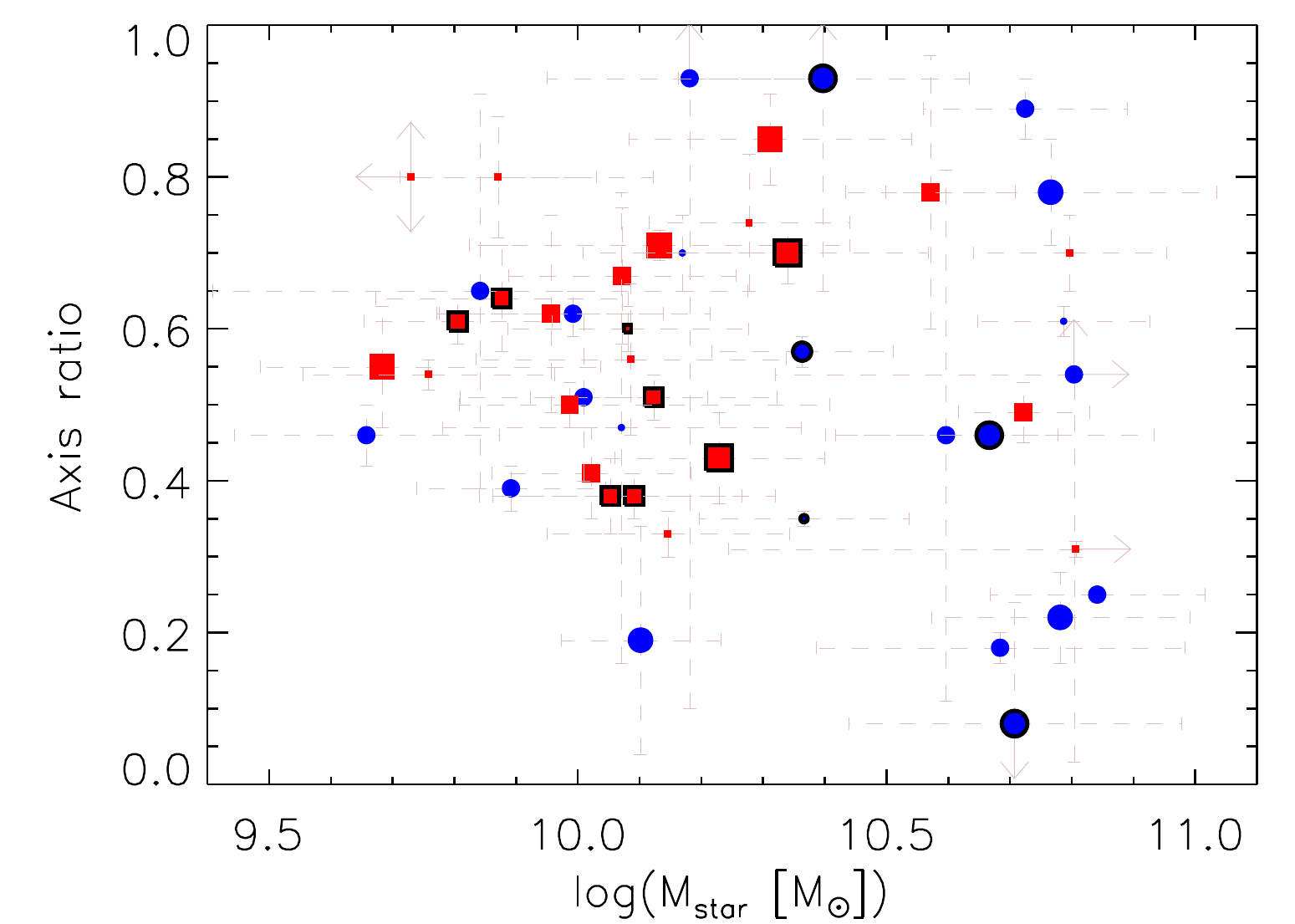}\\
\includegraphics[width=8.5cm]{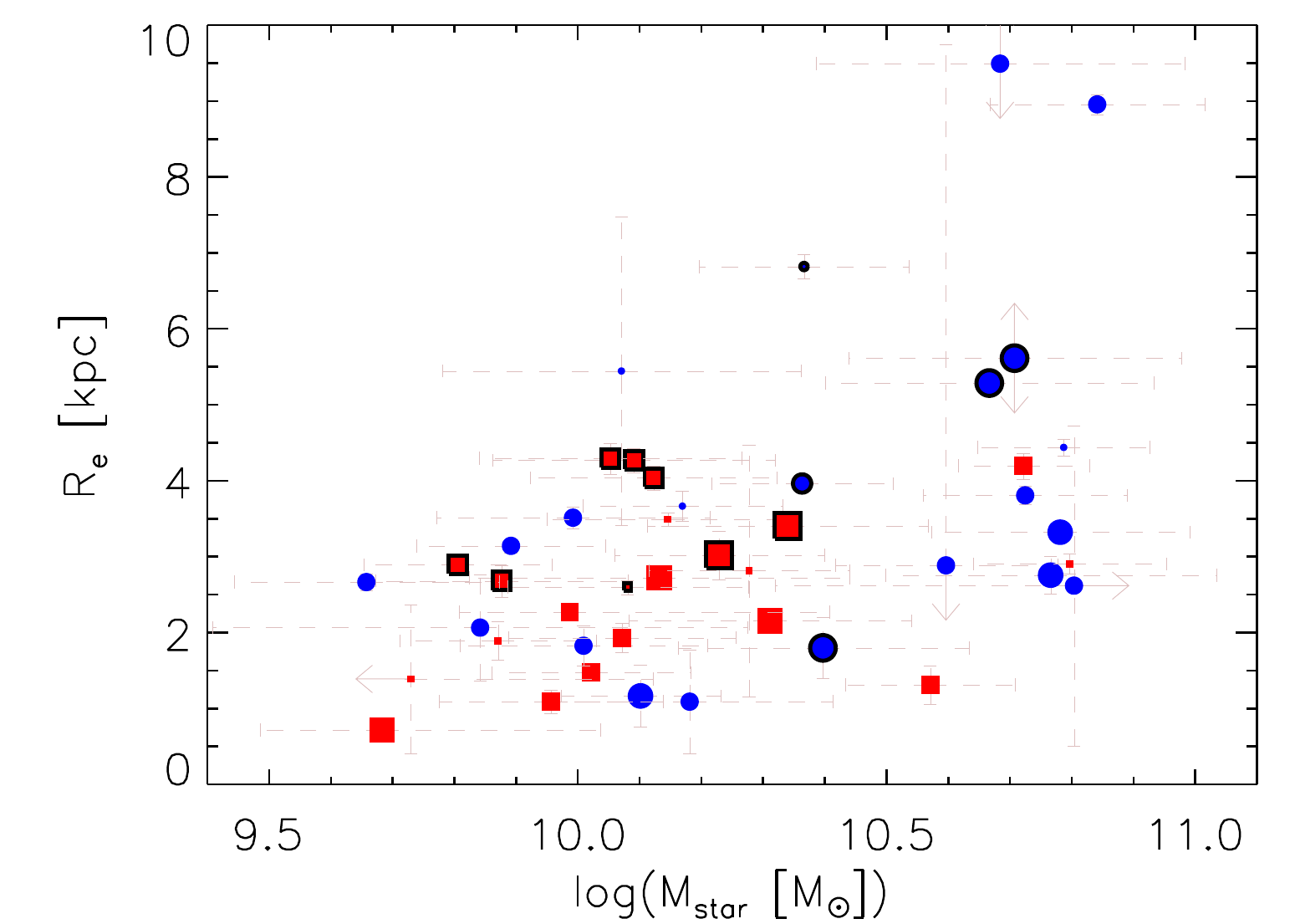}
\caption{
\textbf{Top:} Axis ratio as a function of the stellar mass.
\textbf{Bottom:} Half-light radius as a function of the stellar mass.
Same symbols as Figure \ref{classification_res_dpa}.
% Blue dots and red squares represent rotating and non-rotating galaxies. Symbols with black contours mark interacting galaxies. The symbol size is related to the S/N of the SINFONI data: small, medium and large symbols refer to galaxies with $S/N<5$, $5<S/N<10$ and $S/N>10$, respectively.
Arrows indicate that uncertainties are exceeding the displayed range.
}
\label{mstar_plots}
\end{figure}

In the MASSIV sample, the most massive galaxies ($\log{M_{star}} > 10.5$) are mainly identified as disks in rotation. However, this is true for the gaseous phase and does not necessarily imply that the stars are settled into a disk. The best spheroid candidates may be the most compact and roundest massive galaxies. Indeed, elliptical galaxies can be flattened, too, but should be rounder than disks on average. There are four potential candidates even if only one of them has an axis ratio fully compatible with zero (see Figure \ref{mstar_plots}). These may also be nearly face-on disks if observed velocity shear is very low.
In addition, given our definition of rotators, it is highly probable that gas and stars share a common disk since both morphological and kinematic major axis agree well.
The morphology obtained from the CFHT imaging used for MASSIV has an insufficient low spatial resolution to allow addressing this problem unambiguously but it is worth noticing that in the local Universe stellar disks are observed in elliptical galaxies (e.g. \citealp{Kuntschner:2010}). Usually they are quite young ($\sim 1$ Gyr) but they can also be as old as $\sim 10$ Gyr, which would be compatible with our observations.

The intermediate mass galaxies  ($\log{M_{star}} < 10.5$) are almost equally divided between rotating and non-rotating systems but the smallest ones are principally non-rotating systems (see Figure \ref{mstar_plots}). Since both rotating and non-rotating systems are observed with small masses and radii, we can tentatively conclude that non-rotating systems are not caused by observational artifacts. However, we cannot exclude the possibility that
for some of the latter systems, when $R_{last}/Seeing\sim 0.5$,
the absence of observed rotation is due to an episode of star formation in single non-resolved regions.
The fact that we observe large non-rotating galaxies incompatible with face-on systems is also very intriguing. In addition, those with $R_e > 2.8$ kpc all interact. This property is probably related to the nature itself of these systems.

\subsection{Dynamical support of disks}

\begin{figure}
\includegraphics[width=8.5cm]{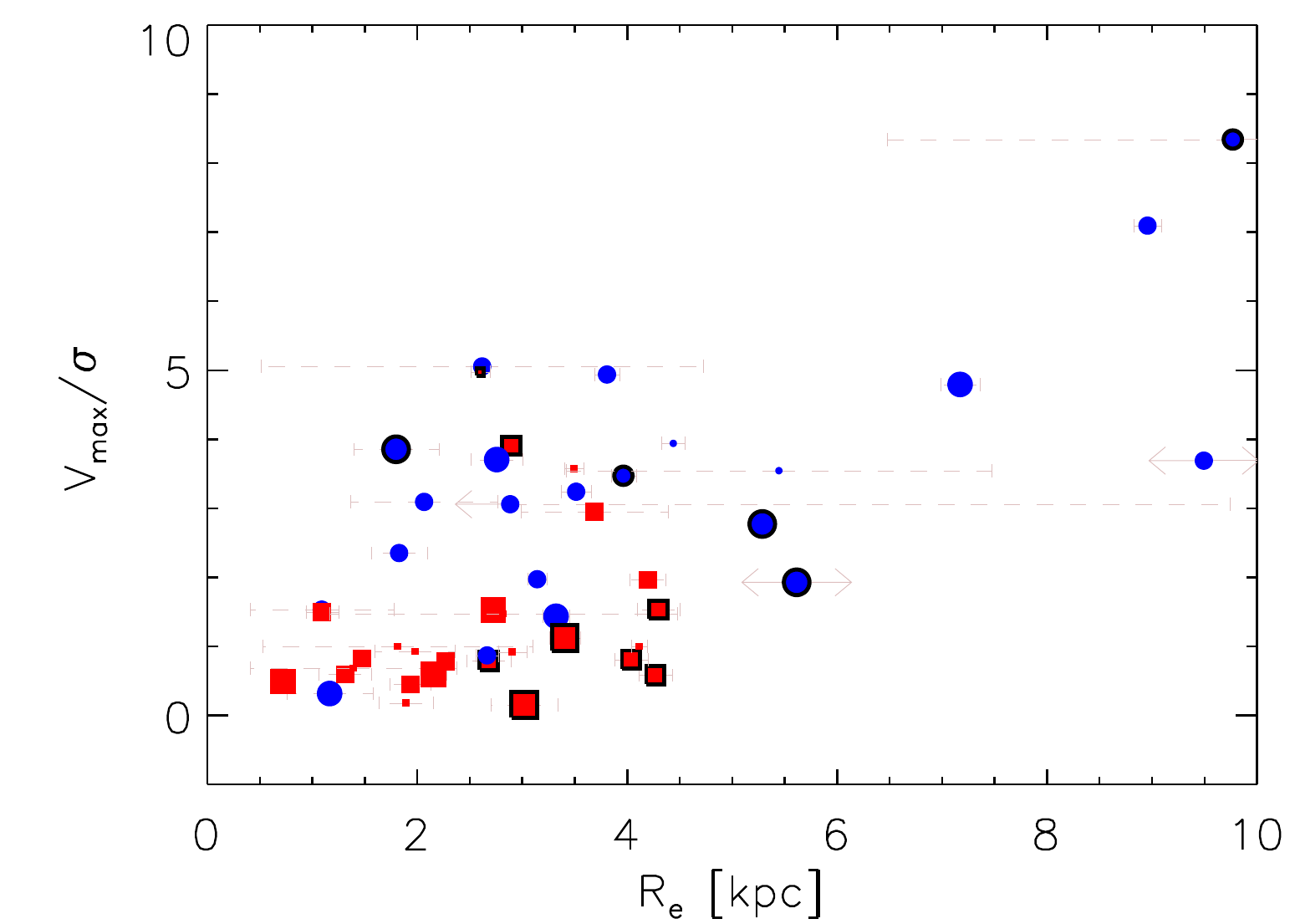}
\caption{
$V_{max}/\sigma$ as a function of the half light radius.
Same symbols as Figure \ref{mstar_plots}.
% Blue dots and red squares represent respectively rotating and non-rotating galaxies. Symbols with black contours mark interacting galaxies. Symbol size is related to the S/N of the SINFONI data: small, medium and large symbols refers to galaxies with $S/N<5$, $5<S/N<10$ and $S/N>10$ respectively. Arrows indicate that uncertainties are exceeding the displayed range.
}
\label{vs_plots}
\end{figure}

\begin{figure}
\includegraphics[width=8.5cm]{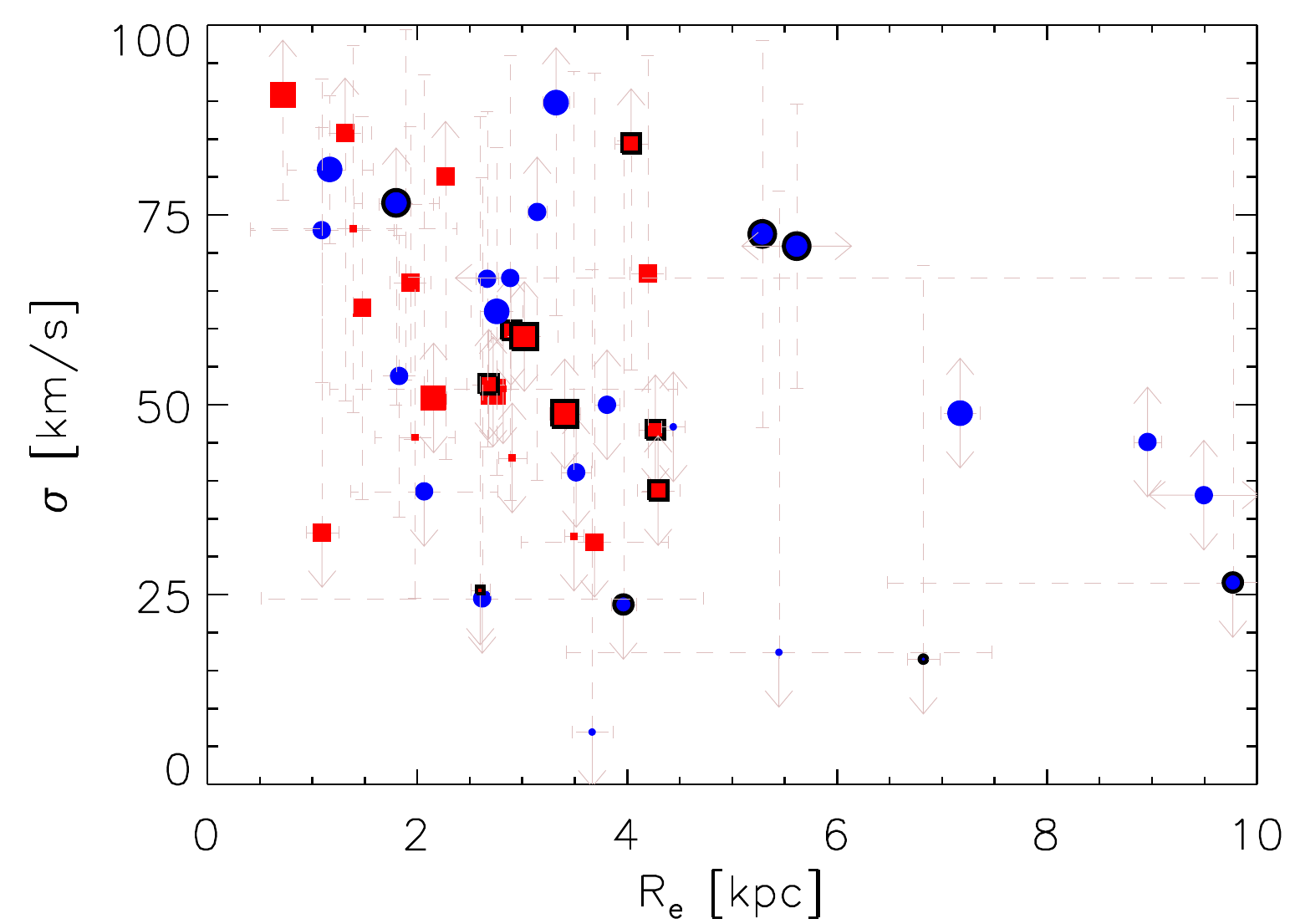}
\caption{Velocity dispersion as a function of the effective radius.
Same symbols as Figure \ref{mstar_plots}.
% Blue dots and red squares represent respectively rotating and non-rotating galaxies. Symbols with black contours mark interacting galaxies. Symbol size is related to the S/N of the SINFONI data: small, medium and large symbols refers to galaxies with $S/N<5$, $5<S/N<10$ and $S/N>10$ respectively. Arrows indicate that uncertainties are exceeding the displayed range.
}
\label{sigma_re}
\end{figure}

To quantify the dynamical support, it is common to study the ratio of the rotation velocity over the local velocity dispersion. In Figure \ref{vs_plots}, we show this ratio as a function of the half-light radius. Only two out of the 20 secure rotators in MASSIV have a ratio lower than unity and are consequently dispersion-dominated. Only five out of these 20 secure rotators have $V_{max}/\sigma < 2$, which indicates that the majority of rotating disks are
clearly rotation-supported at $z<1.5$. Moreover, there seems to exist a trend that the largest galaxies have the highest $V_{max}/\sigma$ ratio.
As seen in Table \ref{class_properties}, the median velocity dispersion is $\sigma=62$\kms\ for rotators. In addition, the median rotation velocity $V_{max}=141$\kms\ and the median $V_{max}/\sigma$ ratio is 3.2.
By restricting the analysis on rotating systems with $R_e > 3.2$ kpc (median for all rotators), we end up with ten objects with a median velocity dispersion $\sigma=49$\kms, a median rotation velocity $V=201$\kms\ and a median $V_{max}/\sigma = 3.7$, which clearly indicates that the largest disks are more stable.
This seems to be supported in Figure \ref{sigma_re} where we clearly see that the velocity dispersion is low for the largest rotators ($R_e > 6$ kpc).

At higher redshift, rotators in the LSD/AMAZE sample ($z\sim 3.3$) have a typical gaseous velocity dispersion of $\sigma=90$\kms\ and $V_{max}/\sigma\sim 1.6$ \citep{Gnerucci:2011}, whereas in the SINS sample ($z\sim 2.2$), \citet{Cresci:2009} found $\sigma=52$\kms\ and $V_{max}/\sigma\sim 4.5$ as median values for their unambiguous rotators. The comparison with SINS is not straightforward since they have made these measurements only for a fraction of these systems (12 out of 18).
For their sample of rotating systems, the median rotation velocity is $V_{max}\sim240$\kms\ and the median half-light radius (using the conversion between half-light radius and disk scale length $R_e=1.68 R_d$ valid for an exponential disk distribution) is $R_e=3.4$ kpc. These median values are higher than those we obtain for our sample of 20 rotating galaxies ($V_{max}\sim140$\kms\ and $R_e=3.2$ kpc), which indicates that while their sample of rotation-dominated galaxies probes the massive disks population at $z\sim 2.2$, our sample of rotating systems spans a wider range in terms of dynamical mass.
At lower redshift, from the IMAGES sample ($z\sim 0.6$), the typical gaseous velocity dispersion is $\sigma \sim 45$\kms\ (Puech, private communication) and $V_{max}/\sigma \sim 4$ considering both rotating disks and perturbed rotators, with a trend for higher $V_{max}/\sigma$ for rotating disks \citep{Puech:2007}.
The GHASP sample is the largest 2D kinematics sample of local spiral galaxies with star-formation rates typical of the local Universe. We consider here the subsample of 136 galaxies that we projected at $z\sim1.33$ with similar spatial resolution conditions as the MASSIV sample (see section \ref{other_parameters}). These galaxies have a median optical radius of 8.5 kpc. Since $R_{opt}\sim 1.9 R_e$ for an exponential disk distribution, this gives a median half-light radius of 4.5 kpc. The median rotational velocity is 164\kms, the median velocity dispersion is 24\kms\ and the median $V_{max}/\sigma$ is $\sim 7$.
Despite the small differences observed from $z\sim 2.2$ to $z\sim 0.6$, observations of various samples suggest a trend of decreasing velocity dispersion with decreasing redshift. Associated to local and $z\sim 3$ observations, this supports the idea that the gaseous phase of galaxy disks becomes less turbulent with cosmic time.

\begin{figure}
\includegraphics[width=8.5cm]{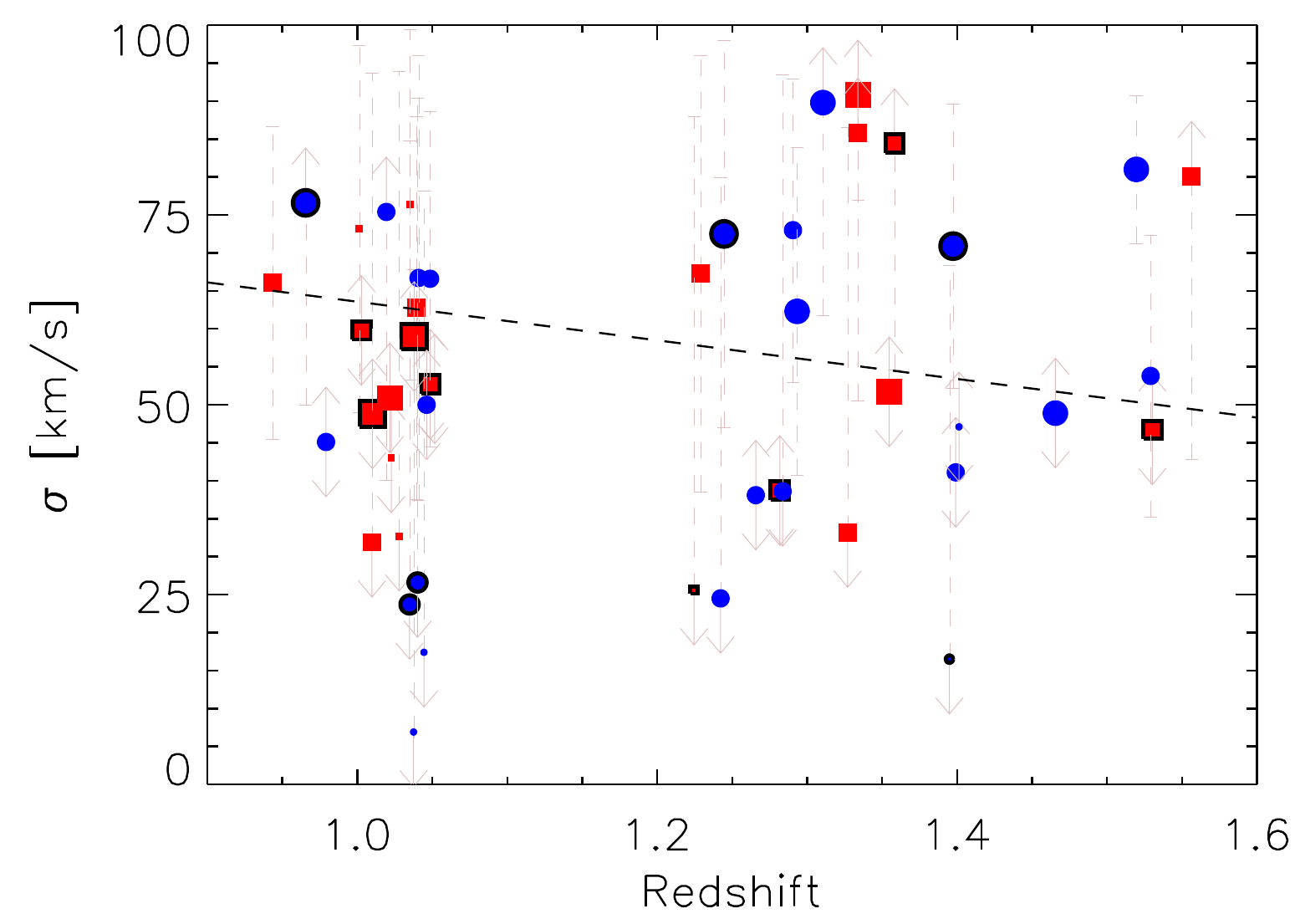}
\caption{Velocity dispersion as a function of the redshift.
Same symbols as Figure \ref{mstar_plots}.
% Blue dots and red squares represent respectively rotating and non-rotating galaxies. Symbols with black contours mark interacting galaxies. Symbol size is related to the S/N of the SINFONI data: small, medium and large symbols refers to galaxies with $S/N<5$, $5<S/N<10$ and $S/N>10$ respectively.
The dotted line represents the velocity dispersion corresponding to the spectral PSF of SINFONI.
}
\label{sigma_z}
\end{figure}

We also observe that about half of the sample of rotators have a gaseous velocity dispersion unambiguously higher than 60\kms.
These may be interpreted as clumpy disks.
Indeed, it has been claimed that a high-velocity dispersion is expected for clumpy disks created from a smooth cold gas accretion (e.g. \citealp{Bournaud:2009}). This mode of accretion is believed to be more efficient at $z>2$ \citep{Keres:2009} and it is therefore not surprising to also observe a significant percentage of galaxies in MASSIV that resemble stable rotating disks with velocity dispersions compatible with $\sim 20 - 50$\kms (cf. figure \ref{sigma_z}).

We also point out that the gaseous velocity dispersion is not significantly different when considering non-rotating systems, which raises the question of the origin of this gaseous velocity dispersion. If these non-rotating systems are indeed transient mergers of star-forming disks in an unstable phase, then a high-velocity dispersion is expected \citep{Bournaud:2011}, but then it is more difficult to explain cases of large non-rotating galaxies with relatively low gaseous velocity dispersion.

As pointed out by e.g. \citet{Lehnert:2009}, \citet{Green:2010}, and \citet{Letiran:2011}, star formation might be responsible for the gaseous turbulence. In particular, star formation intensity might be correlated with gaseous velocity dispersion.
Within this hypothesis, the lowering of the velocity dispersion from high to low redshift might be explained by the fact that the samples discussed in this paper have on average a decreasing star formation rate from $z\sim 3$ to $z=0$ (see \citealp{Contini:2011}). This behavior is due to cosmological surface brightness dimming but also to galaxy evolution itself: on average, galaxies have a larger size and form less stars at $z=0$ than at $z\sim2$.
The relations between star formation and velocity dispersion will be addressed with the full MASSIV sample in a forthcoming paper.

\subsection{Importance of merging at $z\sim1.2$}

The proportion of interacting galaxies in our sample is at least 29\% (13 interacting systems with flag ``A'' or ``B'' out of the 45 detected systems with $0.9<z<1.6$) and is 32\% if we only consider the 41 systems with a flag ``A'' or ``B'' in the redshift range $0.9<z<1.6$. Because of the way we determined which systems are in interaction, this proportion is a lower limit. Indeed, on the one hand, it could be that a fraction of non-rotating galaxies are ongoing mergers or merger remnants. On the other hand, close mergers could mimic rotators (elongation and velocity gradient along a common axis), as is the case for VVDS020294045 (see Appendix \ref{comments}). In addition, owing to the non-circular final field-of-view of our SINFONI data we only cover a fraction of the area at a given radius around our targets. There is also a substantial percentage of objects for which close galaxies are observed in I-band but not in SINFONI datacubes (see Appendix \ref{comments}). It is possible that a fraction of these objects are passive galaxies at the same redshift as the MASSIV targets.
It is interesting to notice that on average, galaxies in interaction are larger than isolated galaxies and have a lower gaseous velocity dispersion (see Table \ref{class_properties}).

The proportion of interacting systems in MASSIV is comparable to the percentage of mergers in the SINS sample. However, SINS mergers (around one third of the sample) are mainly identified using kinemetry, a technique based on the degree of perturbation observed in the kinematic maps, whereas the 29\% of interacting galaxies in MASSIV are mainly based on the detection of several components. On the one hand, only three pairs out of 52 detected systems were identified in SINS (and three pairs out of 29 systems in the LSD/AMAZE sample), which is much lower than the pair fraction observed in MASSIV. On the other hand, SINS mergers found from kinemetry may mainly correspond to our non-rotating galaxies (see Section \ref{class_rot_nrot}).
That means that using this latter method, we would probably find that some MASSIV isolated galaxies are likely mergers.
This is additionally supported by the fact that the percentage of non-rotating systems is larger in interacting galaxies:
among the 11 systems with interactions in the sample (with both S/N$>$5 and isolation flag ``A'' or ``B''), seven have at least one non-rotating component (64\%), whereas among the 22 isolated systems (with both S/N$>$5 and isolation flag ``A'' or ``B''), there are only eight non-rotating galaxies (36\%).
This could indicate that some isolated non-rotating galaxies are in fact ongoing mergers or merger remnants.
This seems to be also supported by the number of interacting galaxies, which is larger at $z\sim 1$ (e.g. Figure \ref{sigma_z}).
The conclusion is therefore that the percentage of mergers at $z\sim1.2$ is higher than at $z\sim 2.2$ as observed in SINS.

At lower redshift, 44\% of galaxies in the IMAGES sample have complex kinematics that could be explained by ongoing mergers of star-forming galaxies \citep{Yang:2008}. This is more than the proportion of interacting galaxies in MASSIV. However, the definition of galaxies with complex kinematics is probably closer to our definition of non-rotating galaxies (as is the case for the comparison with SINS) and the proportions of these two classes in each sample are fully comparable. It is probable that for IMAGES, a larger field of view would have led to the detection of close companions. Consequently, a direct comparison of merging with MASSIV is not straightforward.

The number of mergers deduced from the pair number is higher in MASSIV than in SINS. Our results are still consistent with a peak in the merger activity  at  $1 < z < 2$ if we consider systems with complex kinematics.
However, it is not yet clear if these kinematic signatures of mergers are related to minor or major mergers, whereas from the MASSIV sample we have a way to infer the mass ratio between pair members.
A dedicated analysis on the merger rate from the observed pairs will be performed for the entire MASSIV sample (L\'{o}pez-Sanjuan et al., in prep.) and will probably enable us to infer the rate of observed ongoing mergers.

\subsection{Nature of non-rotating galaxies}
\label{nrgals}

Clearly, non-rotating systems are mainly galaxies classified as low-velocity shear galaxies plus some interacting galaxies.
Among the 16 (and eight additional with $S/N<5$) low-velocity shear galaxies only three (and two additional that have very poor S/N) could be considered as rotating. Two of these galaxies are compatible with nearly face-on systems.
The exact nature of these non-rotating objects is still unclear. Such a population of galaxies has already been observed at higher redshift both in the SINS sample at $z\sim 2.2$ and in the LSD/AMAZE sample at $z\sim 3.3$.
These objects are smaller on average than rotators and are often associated to interacting systems (see Table \ref{class_properties}).

An unexpected trend is observed for these objects in the MASSIV sample (Figure \ref{sigma_re}): there exists an anti-correlation between the mean velocity dispersion of the gas and the effective radius of the stellar component.
This correlation is in contrast to what would be expected for elliptical galaxies located in the fundamental plane \citep{Dressler:1987,Djorgovski:1987}. However, the fundamental plane applies for the central stellar velocity dispersion, which is a quantity that cannot currently be measured in high-redshift galaxies from spectroscopic data.
Alternatively, this correlation could reflect a beam-smearing effect since an unresolved velocity gradient could reproduce this trend. However, the fact that we also observe velocity gradients in other galaxies with similar sizes contradicts this interpretation but it is closely linked to the spatial distribution of the ionized gas. For face-on disks (possible only for low-velocity shear galaxies), it would be expected to recover the same trend as for rotators (i.e. lower velocity dispersion for large disks). This is very unlikely, however, because velocity gradients may be observed for the largest galaxies and also because we do not expect to observe more than four galaxies with an inclination lower than 20\degr\ (see Figure \ref{histo_incl}).
If the gas were more concentrated, though, it might collapse more violently to form stars at a higher density if we consider that star formation can drive the gaseous velocity dispersion as suggested by e.g. \citet{Green:2010}.

As was been proposed in the previous sections, these non-rotating objects can be interpreted as
\begin{itemize}
 \item Merger remnants in a transient unstable state for the gaseous phase (e.g. collapse of gas due to interaction, anti-spin mergers, etc.).
 \item Rotators with a special gaseous distribution: the most plausible being a unique highly concentrated star-forming region when galaxies are small.
 \item Galaxies with an unstable gaseous phase. This could be supported by the relatively low mass (thus a low potential well) of these systems.
 \item Nearly face-on disks (for low-velocity shear galaxies only): this would explain the similar velocity dispersion as for the rotators but the number of these galaxies is statistically low.
 \item Spheroids: however, (i) these galaxies may be the most massive ones, (ii) there is no reason to think that the gas is not in a disk as observed at lower redshift (e.g. \citealp{Peletier:2007} in early-type spirals; e.g. \citealp{Bournaud:2007} in numerical simulations of gas-rich mergers) and (iii) if the gas is randomly distributed, a higher velocity dispersion than for rotators would be expected.
\end{itemize}

\section{Summary and conclusions}

We presented the data and their processing for a subsample of 50 galaxies of the MASSIV sample (described in \citealp{Contini:2011}). We focused on the analysis of the dynamical state of galaxies using kinematic maps derived from SINFONI data-cubes. We classified this sample based on the modeling of these maps and based on a comparison between morphology and kinematics. This classification describes on the one hand the rotating or non-rotating nature of the objects and on the other hand the close environment of galaxies.

We separated rotating galaxies and non-rotating galaxies based on two arguments: (i) the agreement between morphological and kinematic major axis position angle and (ii) the accuracy of the description of a rotating disk model. We showed that about half of the sample displays rotation, whereas one third does not show a dominant ordered rotation (the remaining fraction is not classified). However, the ionized gas turbulence in these two classes is similar ($\sigma \sim 60$\kms), which marks a transition between higher redshift galaxies showing on average a higher velocity dispersion ($\sigma \sim 60-90$\kms), as in SINS at $z\sim 2.2$ or LSD/AMAZE at $z\sim 3.3$, and lower galaxies in IMAGES at $z\sim 0.6$ and GHASP at $z=0$, which are characterized by a lower velocity dispersion ($\sigma \sim 20-40$\kms). All these samples are 3D spectroscopic samples observing ionized gas emission lines. This transition of the gaseous velocity dispersion seen from these high- to low-redshift samples is partly responsible for the increase of the rotational support (deduced from the $V_{max}/\sigma$ ratio) when galaxies evolve.
It could be related to a common process that would induce a decrease of the star formation rate.
We observe that about half of our galaxies have a velocity dispersion compatible with that in local star-forming galaxies, whereas the other half is more compatible with galaxies at higher redshift.
At high redshift, this high-velocity dispersion seen in isolated disks is thought to be caused by a mass assembly driven by cold flows.
In this framework, our results suggest that at $z\sim 1.2$ cold gas accretion is less efficient than at higher redshift but more efficient than at lower redshift, consistent with cosmological simulations \citep{Keres:2005}.
When we compare our sample to these other high-redshift samples, it seems that the fraction of disks increases in star-forming galaxies while the Universe evolves.

By studying strong kinematic signatures of merging and detecting pairs in our data-cubes and broad band images, we have shown that the fraction of interacting galaxies is up to at least one third of the sample. The fraction of non-rotating objects in these systems is higher than in isolated ones. This suggests that a significant fraction of isolated non-rotating objects could be mergers in a transient state in which the gas is not dynamically stable. However, the nature of non-rotators is still unclear. Compared to higher and lower redshift 3D spectroscopic surveys, our findings seem to indicate that
at the peak of star formation activity, the fraction of star-forming galaxies in interaction is also at its maximum, corresponding to a peak in the merging activity \citep{Ryan:2008,Conselice:2008,Lopez-Sanjuan:2009}.

All these results favor a scenario in which the mass assembly of star-forming galaxies is progressively shifting from a predominance of smooth cold gas accretion to a predominance of merging as cosmic time evolves, with a transition epoch around a redshift $z\sim 1-1.5$.

\begin{acknowledgements}
We thank the referee for useful comments that helped in improving the paper.
This work has been partially supported by the CNRS-INSU and its Programme National Cosmologie-Galaxies (France) and by the French ANR grant ANR-07-JCJC-0009
\end{acknowledgements}

\bibliographystyle{aa}
\bibliography{biblio}

\begin{appendix}

\section{Individual comments}
\label{comments}
\noindent \textbf{VVDS020106882}
This galaxy has a regular velocity field. The velocity dispersion is higher on the southeast side but remains low. Some external pixels are detected on the northwest side. They seem to be related to regions associated to the galaxy due to concordant velocities.
Two objects (including a very faint one) are detected in I-band at less than 6\arcsec\ but are not detected in the SINFONI datacube.
\\
\\ \noindent \textbf{VVDS020116027}
This galaxy has a small companion at 4\arcsec\ to the north. The companion is detected in H$\alpha$ but with a low S/N and over a region of the size of the seeing. This companion is less exposed than the main galaxy. The main galaxy itself seems to have two components in the H$\alpha$ flux distribution and has a very low velocity shear.
\\
\\ \noindent \textbf{VVDS020147106}
Despite a low-velocity gradient and some perturbations in the low S/N regions, the kinematic position angle of this galaxy agrees well with the I-band morphology. This galaxy may therefore be a nearly face-on disk (at least the ionized gas) but with a high velocity dispersion. There are several objects detected in I-band but not in the SINFONI datacube at less than 6\arcsec.
\\
\\ \noindent \textbf{VVDS020149061}
The I-band image suggests small and extremely faint companions that are not detected in the SINFONI datacube, however .
The seeing is half the size of the H$\alpha$ emission. The velocity field is irregular on the edges where the S/N is lower and the velocity dispersion is higher than 70\kms\ everywhere in the galaxy (except at the edges).
\\
\\ \noindent \textbf{VVDS020164388}
Except for the edge, the velocity field and the velocity dispersion map are regular. Velocity dispersion is not higher than 60\kms. In the I-band image, the galaxy is quite round. Several faint close objects are also observed in this image but are not detected in H$\alpha$.
\\
\\ \noindent \textbf{VVDS020167131}
Some [O\textsc{iii}] is detected in the center but there are only a few pixels and the S/N is lower than 4.
In addition, some emission line is detected at the same redshift in the center of the galaxy on the east seen in the I-band. This indicates that these two galaxy are in interaction.
\\
\\ \noindent \textbf{VVDS020182331}
The velocity gradient is clear even if we exclude low S/N regions (bluest and reddest velocities). However, the velocity field and velocity dispersion map look perturbed, probably because of the low S/N of the observation.
A very close companion is detected in the I-band image that is undetected in H$\alpha$.
\\
\\ \noindent \textbf{VVDS020193070}
Despite a generally low S/N, the velocity field is quite regular except in the outskirts. The velocity dispersion is low but slightly asymmetric.
The morphology is elongated both in I-band and in H$\alpha$ (but with a difference of $\sim 10$\degr\ between morphological and kinematic major axis position angles).
Three very faint objects are detected at less than 5\arcsec\ from the galaxy but are not detected in H$\alpha$.
\\
\\ \noindent \textbf{VVDS020208482}
The S/N is fairly low but there is a clear velocity shear.
The I-band image shows a close galaxy and other objects, farther away, that are not detected in H$\alpha$.
\\
\\ \noindent \textbf{VVDS020214655}
This galaxy displays a fairly smooth low velocity gradient, but not aligned with the morphology, which is quite round. There are some perturbations in the low S/N regions of the velocity field. There are also small perturbations in the velocity dispersion map, which is fairly flat and has low values.
There is a close galaxy in the I-band (at $\sim2$\arcsec) that is not detected in H$\alpha$.
\\
\\ \noindent \textbf{VVDS020239133}
The morphology from I-band is clearly elongated and is more extended than that in H$\alpha$. The velocity field is asymmetric but shows a clear gradient. The velocity dispersion map is also perturbed and has a high-velocity dispersion in the south that could be explained by a sky line residual. Two objects are observed in I-band but not detected in H$\alpha$ within the SINFONI field-of-view.
\\
\\ \noindent \textbf{VVDS020240675}
The velocity field is perturbed, with a low-velocity gradient. The velocity dispersion is fairly low ($\sim 40$\kms). In addition, the size of the H$\alpha$ emission is not large compared to the seeing. There is some emission detected at 2\arcsec\ in the east of the main galaxy, but nothing is detected in the I-band at this location. There are also several objects in I-band within 6\arcsec\ that are not detected in H$\alpha$.
\\
\\ \noindent \textbf{VVDS020255799}
The signal in H$\alpha$ for this galaxy is not extended and with a mean S/N lower than 5.
There is no clear regular velocity gradient. One object is detected at 5\arcsec\ in I-band but not in H$\alpha$.
\\
\\ \noindent \textbf{VVDS020261328}
The velocity field is not completely regular in the blue and red sides. There is also a peak in the velocity dispersion where the S/N is lower. Several objects are detected in I-band but not in H$\alpha$.
\\
\\ \noindent \textbf{VVDS020278667}
Some H$\alpha$ is detected, but the extent is lower than the seeing, and the S/N is lower than 5. Therefore its classification is very doubtful. There is a close galaxy ($\sim1.5$\arcsec) in the I-band but it is not detected in H$\alpha$.
\\
\\ \noindent \textbf{VVDS020283083}
This galaxy looks asymmetric in the I-band and this coincides with the H$\alpha$ distribution and with the main perturbation of the velocity field. This could be related to a tidal tail. That is why this galaxy is classified as non-isolated.
\\
\\ \noindent \textbf{VVDS020283830}
In the H$\alpha$ flux map there is a faint detection smaller than the seeing that coincides with a detection in the I-band (North-East). For the main object, the distribution is asymmetric. This is also true in the H$\alpha$ distribution, which is clumpy. These clumps may be either star-forming regions in one single rotating galaxy or two small objects in the process of merging.
\\
\\ \noindent \textbf{VVDS020294045}
The total velocity shear is larger than 50\kms. However, both the I-band and the H$\alpha$ images are consistent with a system involving two components: one main component with a low-velocity gradient (south) and one small component with a size comparable to the seeing (north). This interpretation is supported by the high-velocity dispersion between the two components due to beam smearing between the two components at different radial velocities. There is a sky line residual at the velocity of the small component, but the flux detected is noticeable. It is therefore not possible to classify the small component. This system is classified as interacting with a flag B since it could also be a single object but with strong perturbations.
\\
\\ \noindent \textbf{VVDS020363717}
The I-band shows a close ($\sim 1$\arcsec), very diffuse and faint object that is not detected in the H$\alpha$ (southwest).
The velocity shear is low. The velocity field is perturbed in the low S/N regions. The velocity dispersion map is also perturbed and the velocity dispersion is high ($\sim 90$\kms) after beam smearing correction.
\\
\\ \noindent \textbf{VVDS020370467}
Both velocity field and velocity dispersion map are perturbed. The velocity gradient of the model is low and the velocity dispersion is quite high (higher than 70\kms) in particular at the north edge (higher than 150\kms). This could be a sign for a strong interaction due to merging,
but there is also a sky line residual that could induce the line width. In addition some objects are observed in the I-band image at less than 6\arcsec\ but are not detected in H$\alpha$.
\\
\\ \noindent \textbf{VVDS020386743}
The H$\alpha$ distribution is very elongated for this galaxy, but no velocity gradient is visible. Both velocity field and velocity dispersion are smooth.
The I-band image shows a distorted galaxy compatible with the H$\alpha$ distribution. In the north, the distortion is compatible with a faint detection in H$\alpha$ that could be related to a minor close companion.
\\
\\ \noindent \textbf{VVDS020461235}
The velocity field of this galaxy is perturbed.
There is a blob detected in H$\alpha$ in the southwest side. This blob is also suggested in the I-band morphology but is less clear. This blob has velocities compatible with the rotation of the main component but could as well be a minor companion in the process of merging. This system is therefore classified as interacting with a flag B.
\\
\\ \noindent \textbf{VVDS020461893}
The velocity field is regular but the velocity dispersion is less regular and high and could suggest a high-dispersion clump. An object is observed at 3\arcsec\ southward in the I-band image but is not detected in SINFONI data.
\\
\\ \noindent \textbf{VVDS020465775}
Both velocity field and velocity dispersion map are irregular. The gradient of the model is lower than 50\kms.
We claim that there is a companion in the northwest that induces broad and non-Gaussian (hence a lower S/N) lines interpreted as a sign of interaction that could also be responsible for the asymmetry in the I-band. Hence it is classified as interacting with a flag B since it could also be a single object but with strong perturbations.
\\
\\ \noindent \textbf{VVDS140083410}
The velocity field is perturbed and there is no clear overall velocity gradient. No galaxy is detected at less than 8\arcsec\ in the I-band image.
\\
\\ \noindent \textbf{VVDS140096645}
The velocity field is smooth but with some perturbations along the minor kinematic axis.
The velocity dispersion map is perturbed and larger on the southern side probably due partially to a sky line residual. Some emission is detected in the H$\alpha$ map outside the main component (north). This detection is smaller than the seeing and is not clearly confirmed in the I-band morphology (CFHT12k images are less deep than CFHTLS images).
It could be a minor companion but it is also compatible with the velocity of the main component and could be associated to it. Therefore it is classified as interacting with a flag B.
This galaxy also has an AGN (see \citealp{Queyrel:2011}).
\\
\\ \noindent \textbf{VVDS140123568}
Some H$\alpha$ is detected, but there are only a few pixels and the S/N is lower than 4. Therefore its classification is very doubtful. There is no detected object in the I-band image at less than 6\arcsec\ around the galaxy. Since the kinematics does not allow to investigate a possible ongoing merger, the isolation has a flag B.
\\
\\ \noindent \textbf{VVDS140137235}
Some H$\alpha$ is detected, but there are only a few pixels and the S/N is lower than 4. Therefore its classification is very doubtful. There is no detected object in the I-band image at less than 6\arcsec\ around the galaxy. Since the kinematics does not allow to investigate a possible ongoing merger, the isolation has a flag B.
\\
\\ \noindent \textbf{VVDS140217425}
The velocity field is regular and has the highest shear ($\sim 600$\kms). The velocity dispersion map shows two regions with high-velocity dispersions ($> 150$\kms). These regions are associated with a lower S/N. This is because there are double profiles in these regions that induce a poor fit. These double profiles are caused by the beam-smearing that mixes regions with different velocities. This can be interpreted as the presence of star-forming clumps at the edge of the galaxy and a large clump in the center. The decomposition of the profile using two Gaussian scales down the velocity dispersion to usual values, and shows that the two external clumps have no velocity gradient, whereas the central one does have one. This picture is compatible with a rotating disk in which the rotational velocity reaches a plateau before the external clumps.
Using the double profile decomposition instead of the kinematics model to derive the maximum velocity, we find $V_{max}\sim 312\pm \sim 20$\kms, which is much more realistic than the value found from the model. The velocities of the plateau are $\sim +313$ and $\sim -312$, which shows that the rotation curve is fairly symmetric and therefore favors a rotating disk hypothesis. Correcting for the inclination, we obtain $V_{max}=322$\kms.
However, the I-band image is asymmetric (which is not incompatible with the previous statement),
and it cannot be ruled out that the external clumps are instead objects about to merge with the main component.
\\
\\ \noindent \textbf{VVDS140258511}
The velocity field and velocity dispersion map look perturbed at the northeast edge but this may be due to a sky line residual.
Except at this location, the velocity dispersion is low (less than 40\kms).
\\
\\ \noindent \textbf{VVDS140262766}
The velocity field and the velocity dispersion map are not strongly perturbed and the velocity dispersion is about 50\kms. Several objects are seen at less than 5\arcsec\ in the I-band image but are not detected in H$\alpha$.
\\
\\ \noindent \textbf{VVDS140545062}
This object has a clear and smooth velocity shear compatible with the I-band and H$\alpha$ morphology. However, the velocity field is not very well reproduced by the rotating disk model: one side is faster than the other. The velocity dispersion map shows a peak in the southwest side. This peak is caused by a double profile that may be interpreted as a sign of merging (late stage or ongoing merger).
However, since the system as a whole is isolated (no galaxy at less than 7\arcsec\ in I-band), this galaxy was classified as isolated with a flag B.
\\
\\ \noindent \textbf{VVDS220014252}
Both velocity field and velocity dispersion map are perturbed. In addition, the morphology in I-band is asymmetric. This system has a high-velocity dispersion ($>80$\kms) and is peaked on some edges. This may be a sign of merging (ongoing or remnant). However, the system is isolated and there is no other evidence for some companion except one very faint object in I-band at $\sim5$\arcsec, which is not in the SINFONI field-of-view. So this system is classified as isolated.
\\
\\ \noindent \textbf{VVDS220015726}
The velocity field of this galaxy is well reproduced by a rotating disk model. The velocity dispersion map is peaked in the center, which agrees with the expected effect of beam smearing caused by the inner velocity gradient. The velocity field and velocity dispersion map are slightly perturbed. One very faint component is detected in I-band at around 4\arcsec\ but is not detected in H$\alpha$.
\\
\\ \noindent \textbf{VVDS220148046}
This galaxy has been observed using AO and is therefore less deep than seeing-limited observations. Some emission line was detected but with a low S/N ($\sim 4$). It appears that the redshift determined from the VVDS was wrong and that we observed [O\textsc{iii}] at $z=2.2442$. The emission is compact and therefore it is not possible to conclude about the kinematics.
\\
\\ \noindent \textbf{VVDS220376206}
A small component (smaller than the seeing, hence a flag B for the environment classification) is detected in H$\alpha$ and is confirmed in the I-band image (north). The residual velocity field shows signs of interaction. The main component has two blobs in H$\alpha$
and the I-band morphology is slightly asymmetric. The velocity dispersion map is also perturbed and peaked on the edges. This may be a sign of interaction or of the end of a merging event.
\\
\\ \noindent \textbf{VVDS220386469}
This observation was made with AO and is underexposed (owing to the small pixel size, cf. \citealp{Contini:2011}).
Even if the AO observation does not allow a fair comparison with other galaxies in the same conditions because of the smaller field of view, this system is classified as isolated with a flag B since the closest clear detection in the I-band image is farther than the 6\arcsec\ that would be reached using the seeing-limited observing strategy.
The velocity field is irregular and the velocity dispersion is very small but this may be due to the use of AO, which provides more details and a lower S/N than seeing-limited observations. The extent of the velocity field is small so it is difficult to argue that there is no velocity gradient at all in this galaxy.
\\
\\ \noindent \textbf{VVDS220397579}
This system is the clearest interacting system between two large galaxies.
The two objects are detected in H$\alpha$ and in the I-band. Both are elongated toward the same direction.
The main object in H$\alpha$ and in I-band has a smooth velocity field with no gradient except on the northern side, near the companion. This region coincides with a higher velocity dispersion and this is probably related to the interaction with the companion.
Since the galaxies are aligned and since the overall velocity gradient is monotonic, the whole system might be considered as a unique galaxy with two big clumps. However, such a system would be very much extended, a property which is physically unlikely at these redshifts.
% and this is physically difficult to believe this.
The companion shows a clear velocity shear and has been classified as well.
% can be classified as well. It as a clear velocity shear.
It is also worth noticing that the minor companion in I-band is the most luminous in the K-band (from UKIDSS Deep Extragalactic Survey, \citealp{Lawrence:2007}), therefore likely the most massive one.
\\
\\ \noindent \textbf{VVDS220544103}
This galaxy has a clear velocity gradient. The object is distorted in the I-band image as well as in H$\alpha$.
This can be explained by the presence of two components (the morphology was decomposed using this hypothesis). In that case, the southern component may be well described by a rotating galaxy and the brightest blob in H$\alpha$ (north) maybe a companion in a merging stage or a relic of merging. However, it cannot be ruled out that this is a single object with disturbed morphology and kinematics, hence the flag B.
\\
\\ \noindent \textbf{VVDS220544394}
The H$\alpha$ map clearly shows the presence of a main component and a faintest one in the north. Their velocity fields seem decorrelated enough to claim that these are two separated objects on a pre-merging stage. The small component is also clearly detected in the I-band.
The velocity field of the main component is perturbed but shows a velocity shear. Its velocity dispersion is about 50\kms. The small component is slightly larger than the seeing and may also be considered to have a velocity shear.
\\
\\ \noindent \textbf{VVDS220576226}
The velocity field is perturbed in the north where the S/N is the lowest. The velocity dispersion map is regular. Two objects are observed within 6\arcsec\ in the I-band image but are not detected in H$\alpha$.
\\
\\ \noindent \textbf{VVDS220578040}
The I-band image shows an asymmetric galaxy with a more diffuse emission in the east. The H$\alpha$ map shows two blobs. These can be interpreted either as two distinct galaxies or as two clumps in one single galaxy. The velocity field is perturbed and is not fully reproduced by the model. The velocity dispersion map is slightly perturbed. Since no strong kinematics perturbation is observed, this galaxy is classified as isolated (with a flag C) and therefore its kinematics is interpreted as a sign of rotation.
\\
\\ \noindent \textbf{VVDS220584167}
The I-band morphology is distorted and the H$\alpha$ distribution is asymmetric. The kinematic position angle, however, agrees very well with the morphology. The velocity field shows some perturbations such as a slight asymmetry. This could be due to the position of the kinematic center that would differ from the morphological one. The velocity dispersion map is peaked in the center, however, but the beam smearing may not account for the whole amplitude of the peak. Two objects are detected in I-band but not in H$\alpha$ at less than 6\arcsec.
\\
\\ \noindent \textbf{VVDS220596913}
This galaxy is a chain galaxy in the I-band. It has been observed both with and without AO (cf. \citealp{Contini:2011}). Several blobs are detected in H$\alpha$. They can be interpreted as several clumps in a single edge-on galaxy since the velocity field is monotonic. They can also be interpreted as several small galaxies in a merging stage. The continuum was extracted from the SINFONI data around H$\alpha$ and seems to be peaked in the center, between the H$\alpha$ blobs, which favors the edge-on hypothesis. Since no strong kinematics perturbation is observed, and since a previous seeing-limited observation \citep{Epinat:2009} revealed that no extra emission is detected, this galaxy is classified as isolated with a flag B and therefore its kinematics is interpreted as sign of rotation.
\\
\\ \noindent \textbf{VVDS910193711}
This observation was made with AO. The extent is larger than 1\arcsec. The velocity field is perturbed. The interpretation of the perturbations may not be comparable with seeing-limited observations because the S/N is lower and the scale of the perturbations smaller. The velocity dispersion is high (around 100\kms). In the close neighborhood that would be observed using the seeing-limited strategy, several objects are detected in the I-band image.
\\
\\ \noindent \textbf{VVDS910279515}
This galaxy was observed using AO and is therefore less deep than seeing-limited observations. This galaxy is unclassified since there are only a few pixels with a S/N larger than 3. There are several objects in the I-band image at less than 5\arcsec\ that would have been observed using a seeing-limited strategy.

\onecolumn

\section{Kinematic maps}
\label{kin_maps}
% \section{Kinematical maps}
% \label{appendix_kinem}
% 
% For each galaxy we show from left to right:\\

% (top) the I-band CFHT image (arbitrary scale), the H$\alpha$ flux map (arbitrary scale) and the S/N map,\\

% (middle) the observed velocity field, the rotating disk modeled velocity field, the residual velocity field, the uncertainty map on the velocity field,\\

% (bottom) the observed uncorrected velocity dispersion, the velocity dispersion map deduced from the velocity field model (beam smearing effect and spectral PSF), the beam smearing corrected velocity dispersion map and the uncertainty on the velocity dispersion.\\

% The VVDS ID and the redshift of each galaxy is indicated on the top-left.
% On each map, North is up and East is left. The center used for kinematics modeling is indicated as a double black and white cross, the position angle is indicated by the black line. This line ends at the effective radius. The seeing FWHM is indicated on the H$\alpha$ map as a circle.
% %  and the contours correspond to the H$\alpha$ flux distribution.
% For galaxies with detected companions, there are two or three set of maps: one for the whole system and one for each modeled component (the companion as a ``s'' suffix).\\

% % \textbf{arranger le cas avec les deux composantes pour enlever les modeles et residus + virer l'info sur le Ha flux}\\

This appendix shows the kinematic maps for all detected galaxies. There are several sets of maps for galaxies with detected companions: one for the whole system (without kinematics modeling) and one for each modeled component.

% \noindent
\begin{figure}[h]
\begin{center}
\includegraphics[width=16cm]{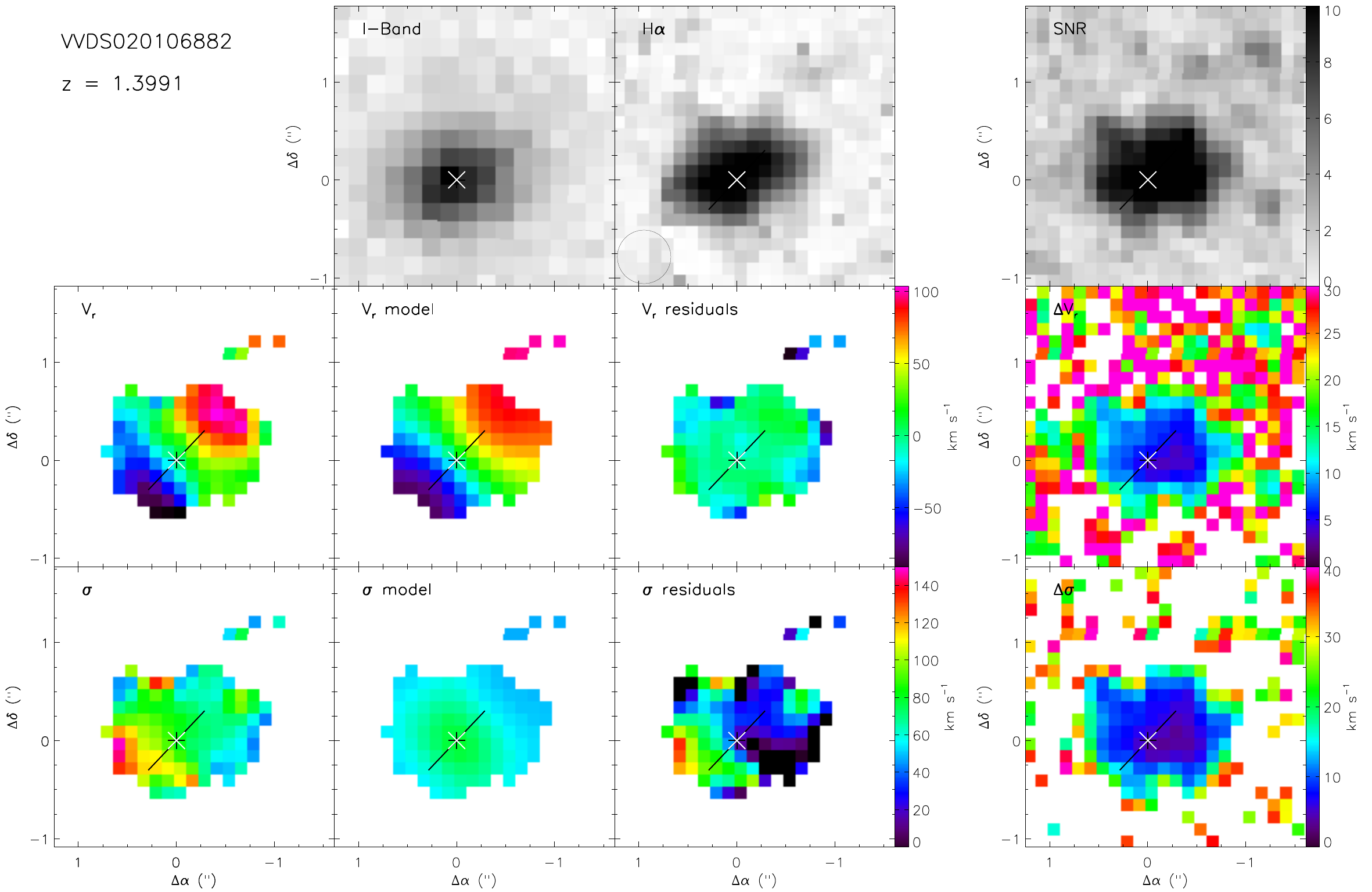}
\caption{Maps for VVDS020106882. From left to right:
(top) the I-band CFHT image (arbitrary scale), the H$\alpha$ flux map (arbitrary scale) and the S/N map,
(middle) the observed velocity field, the rotating-disk-modeled velocity field, the residual velocity field, the uncertainty map on the velocity field,
(bottom) the observed uncorrected velocity dispersion, the velocity dispersion map deduced from the velocity field model (beam-smearing effect and spectral PSF), the beam-smearing-corrected velocity dispersion map and the uncertainty on the velocity dispersion.
The redshift is indicated in the top-left.
In each map, north is up and east is left. The center used for kinematics modeling is indicated as a double black and white cross, the position angle is indicated by the black line. This line ends at the effective radius. The seeing FWHM is indicated on the H$\alpha$ map as a circle.}
%  and the contours correspond to the H$\alpha$ flux distribution.}
\label{maps_caption}
\end{center}
\end{figure}

\begin{figure}
\begin{center}
\includegraphics[width=16cm]{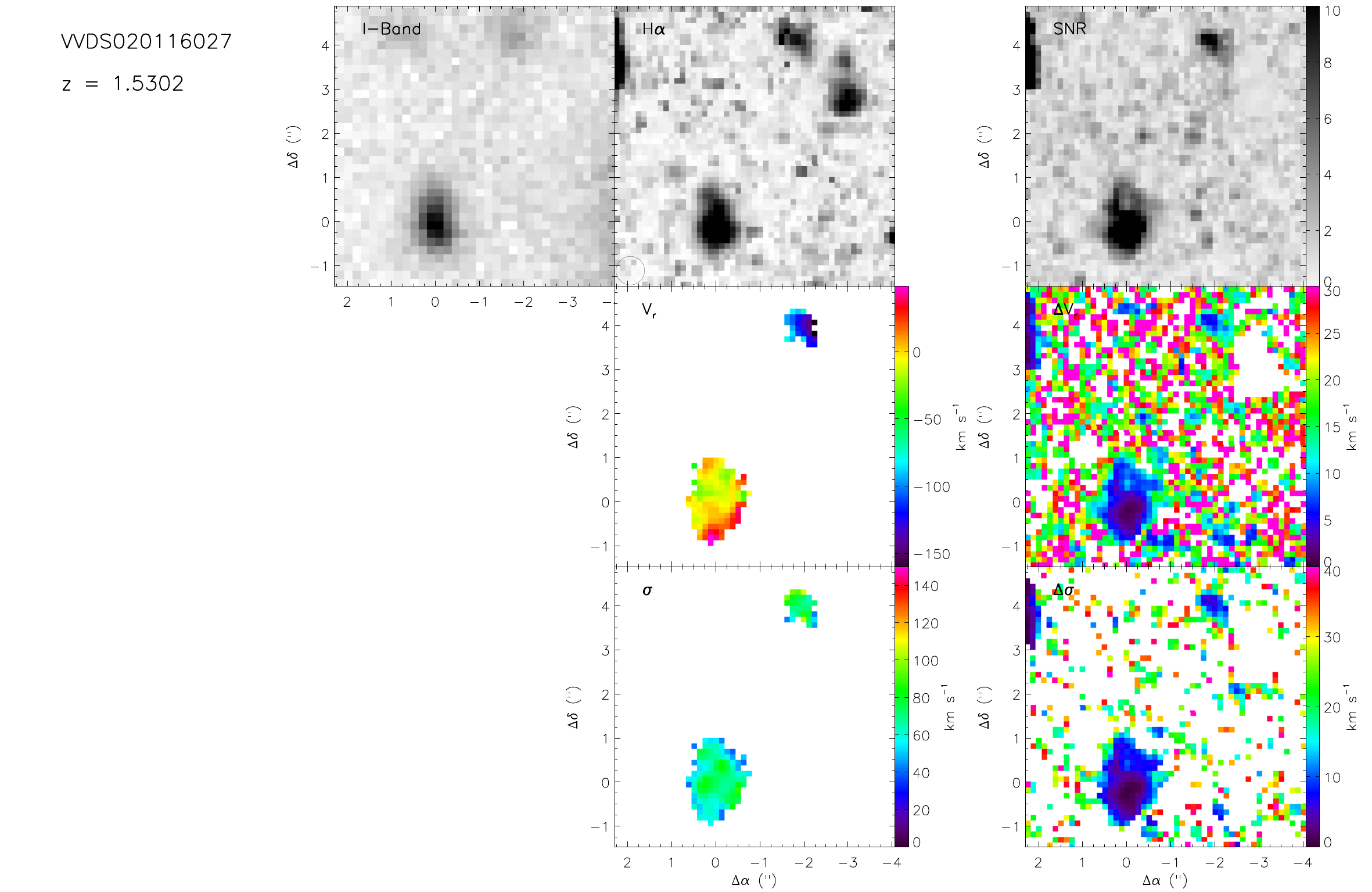}
\caption{Maps for VVDS020116027 system. From left to right:
(top) the I-band CFHT image (arbitrary scale), the H$\alpha$ flux map (arbitrary scale) and the S/N map,
(middle) the observed velocity field, the uncertainty map on the velocity field,
(bottom) the observed uncorrected velocity dispersion and the uncertainty on the velocity dispersion.
The redshift is indicated in the top-left.
In each map, north is up and east is left. The seeing FWHM is indicated on the H$\alpha$ map as a circle.}
\label{maps_caption2}
\end{center}
\end{figure}
\begin{figure}
\begin{center}
\includegraphics[width=16cm]{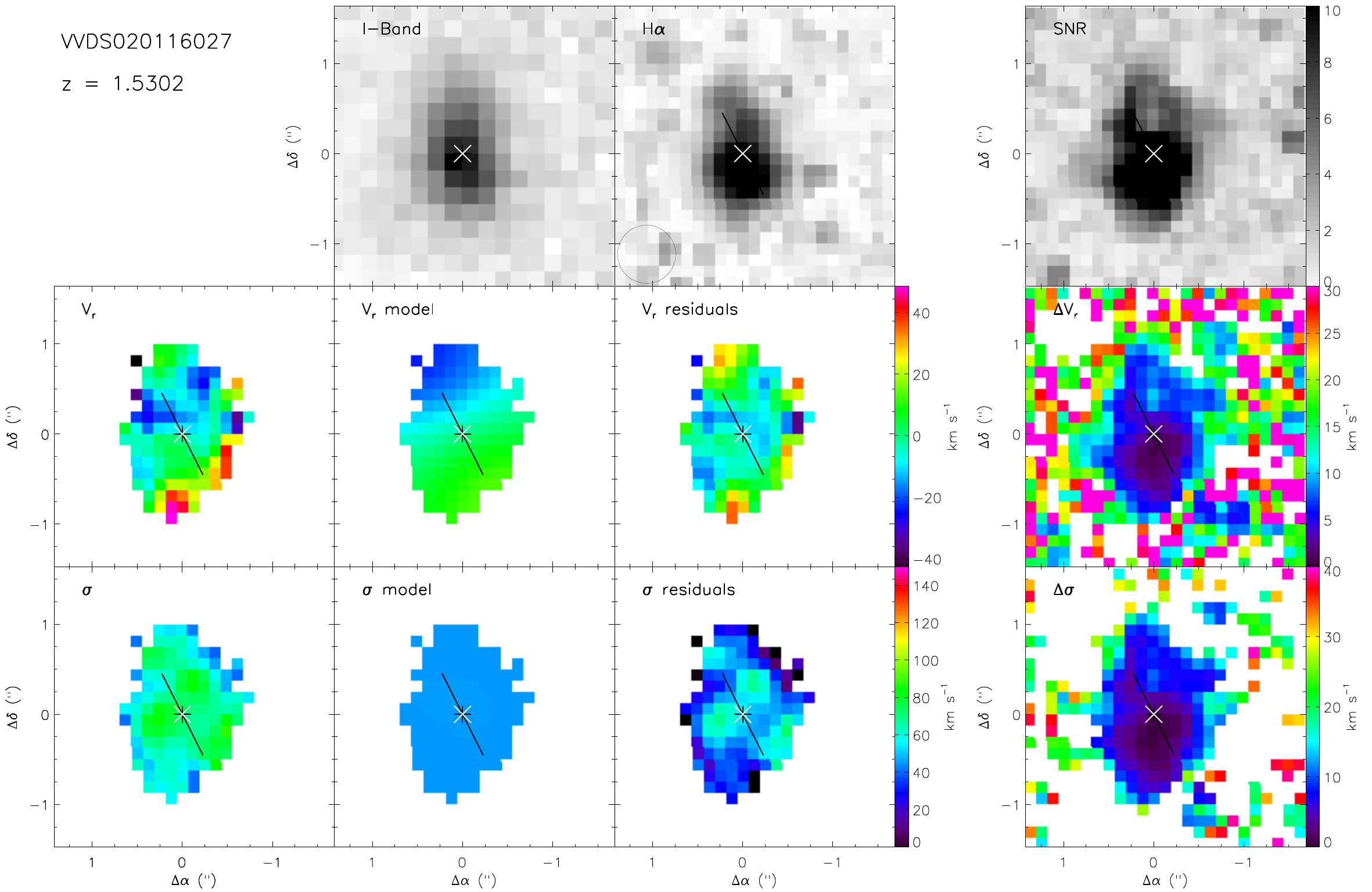}
\caption{Maps for VVDS020116027. Same caption as Figure \ref{maps_caption}.}
\end{center}
\end{figure}

\begin{figure}
\begin{center}
\includegraphics[width=16cm]{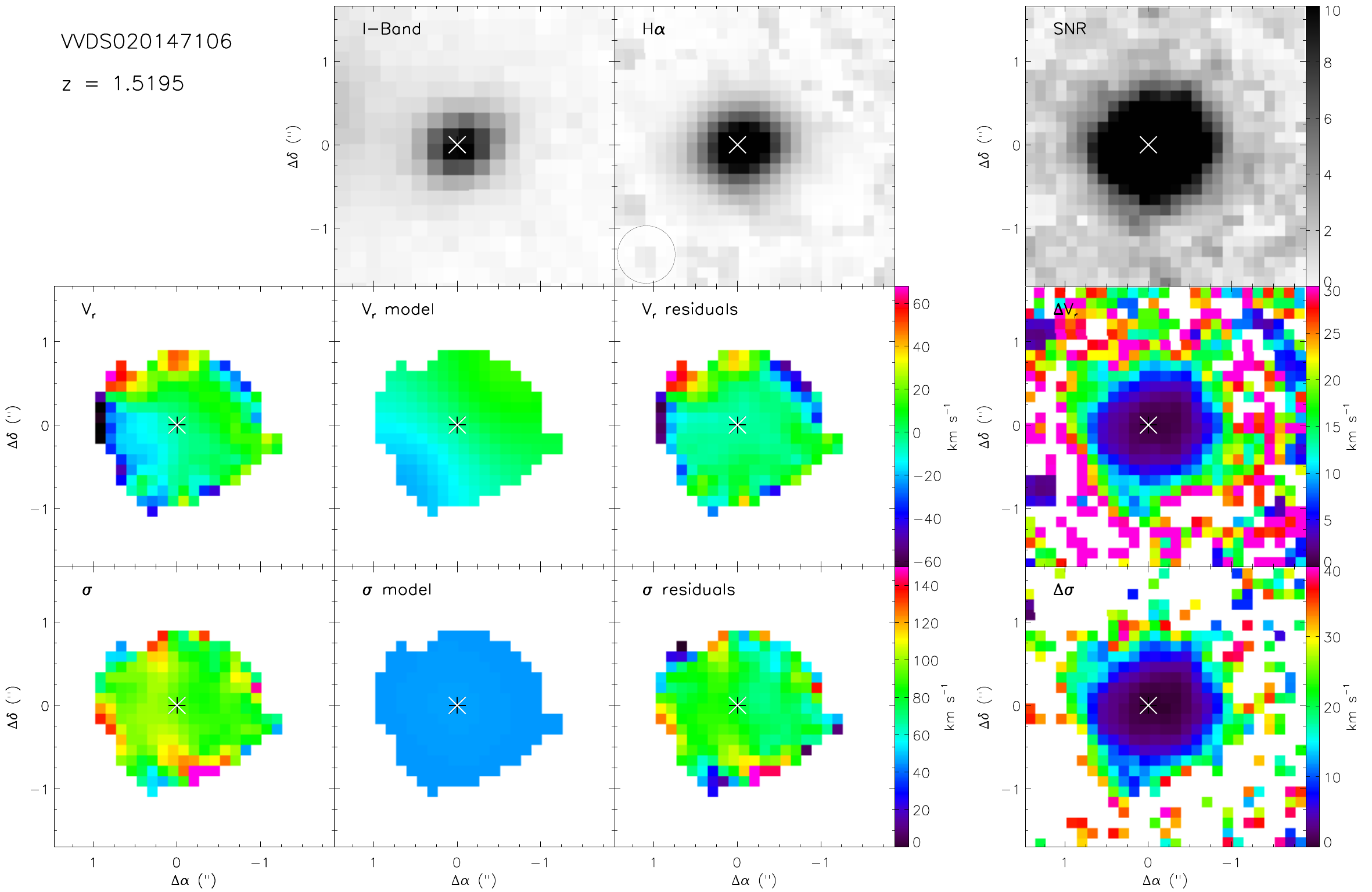}
\caption{Maps for VVDS020147106. Same caption as Figure \ref{maps_caption}.}
\end{center}
\end{figure}
\begin{figure}
\begin{center}
\includegraphics[width=16cm]{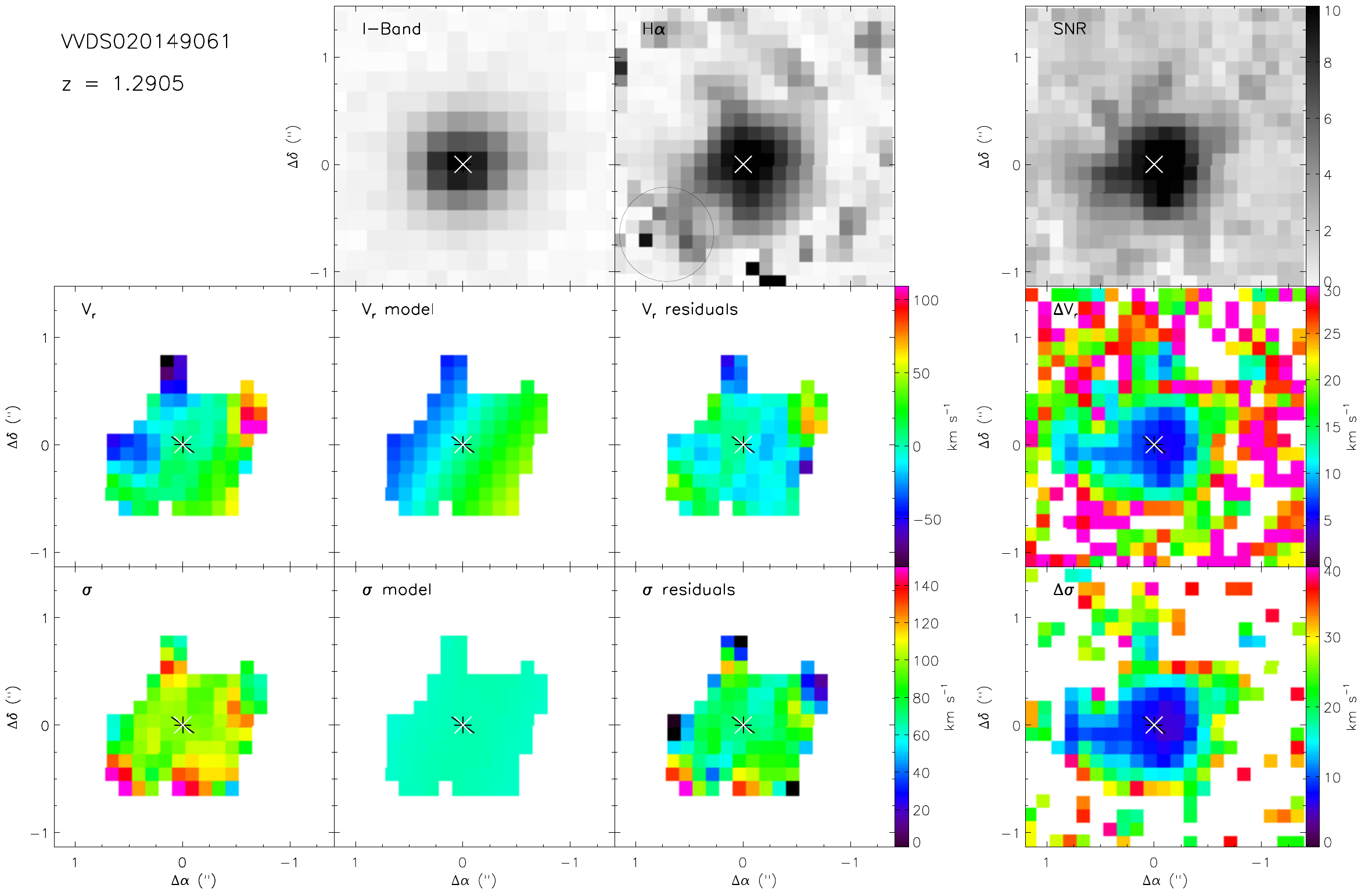}
\caption{Maps for VVDS020149061. Same caption as Figure \ref{maps_caption}.}
\end{center}
\end{figure}

\begin{figure}
\begin{center}
\includegraphics[width=16cm]{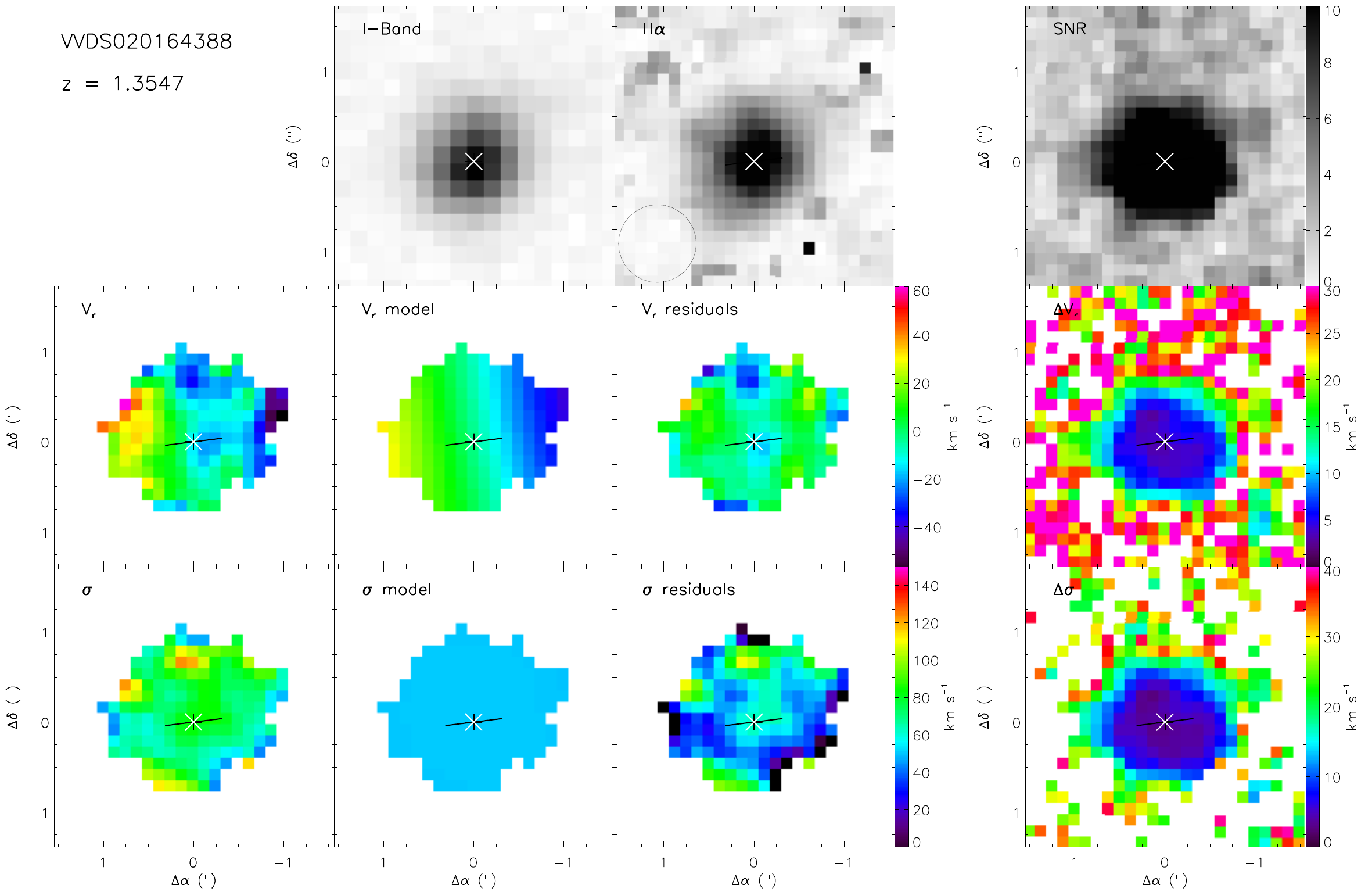}
\caption{Maps for VVDS020164388. Same caption as Figure \ref{maps_caption}.}
\end{center}
\end{figure}
\begin{figure}
\begin{center}
\includegraphics[width=16cm]{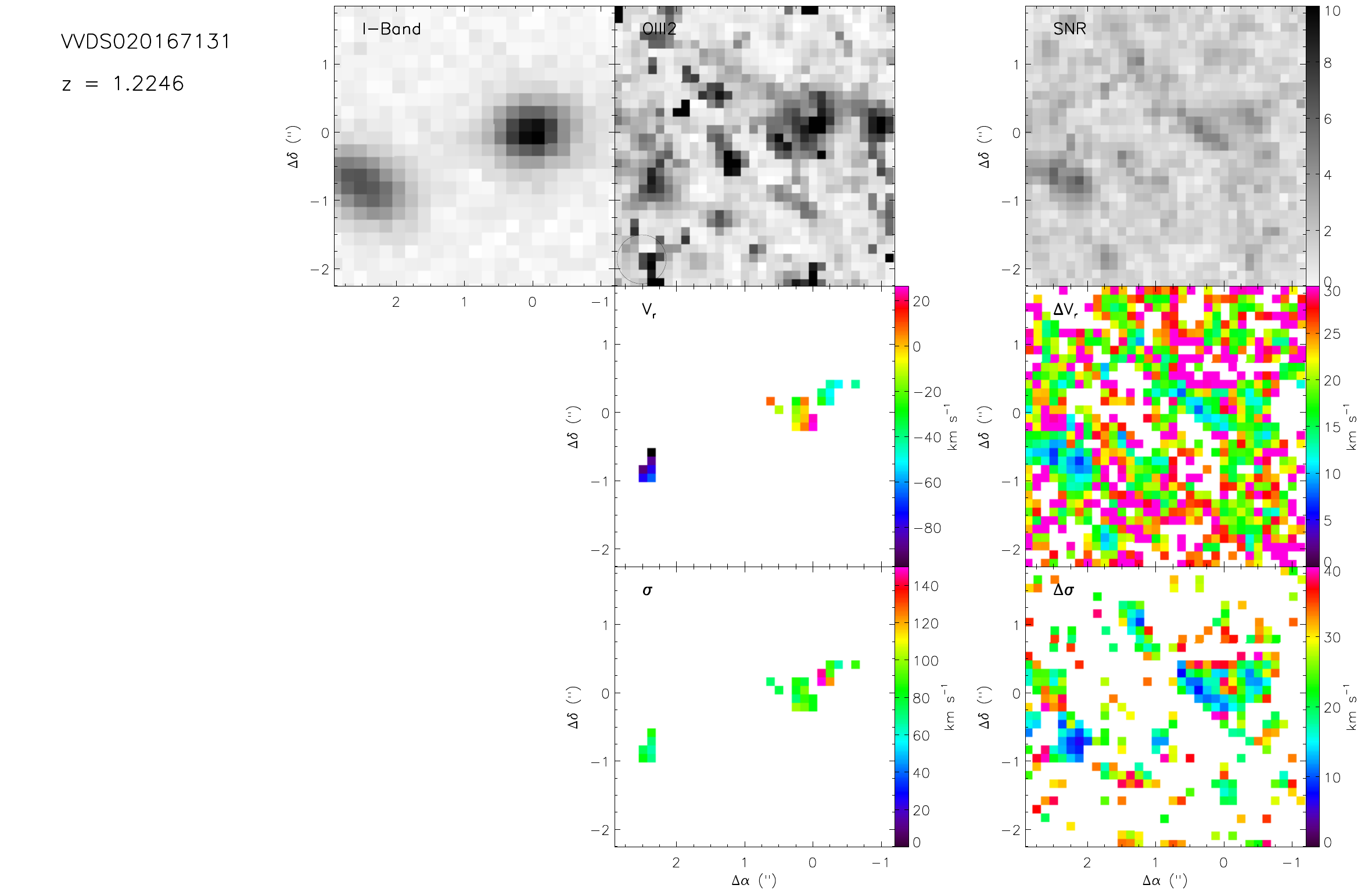}
\caption{Maps for VVDS020167131 system. Same caption as Figure \ref{maps_caption2}.}
\end{center}
\end{figure}

\begin{figure}
\begin{center}
\includegraphics[width=16cm]{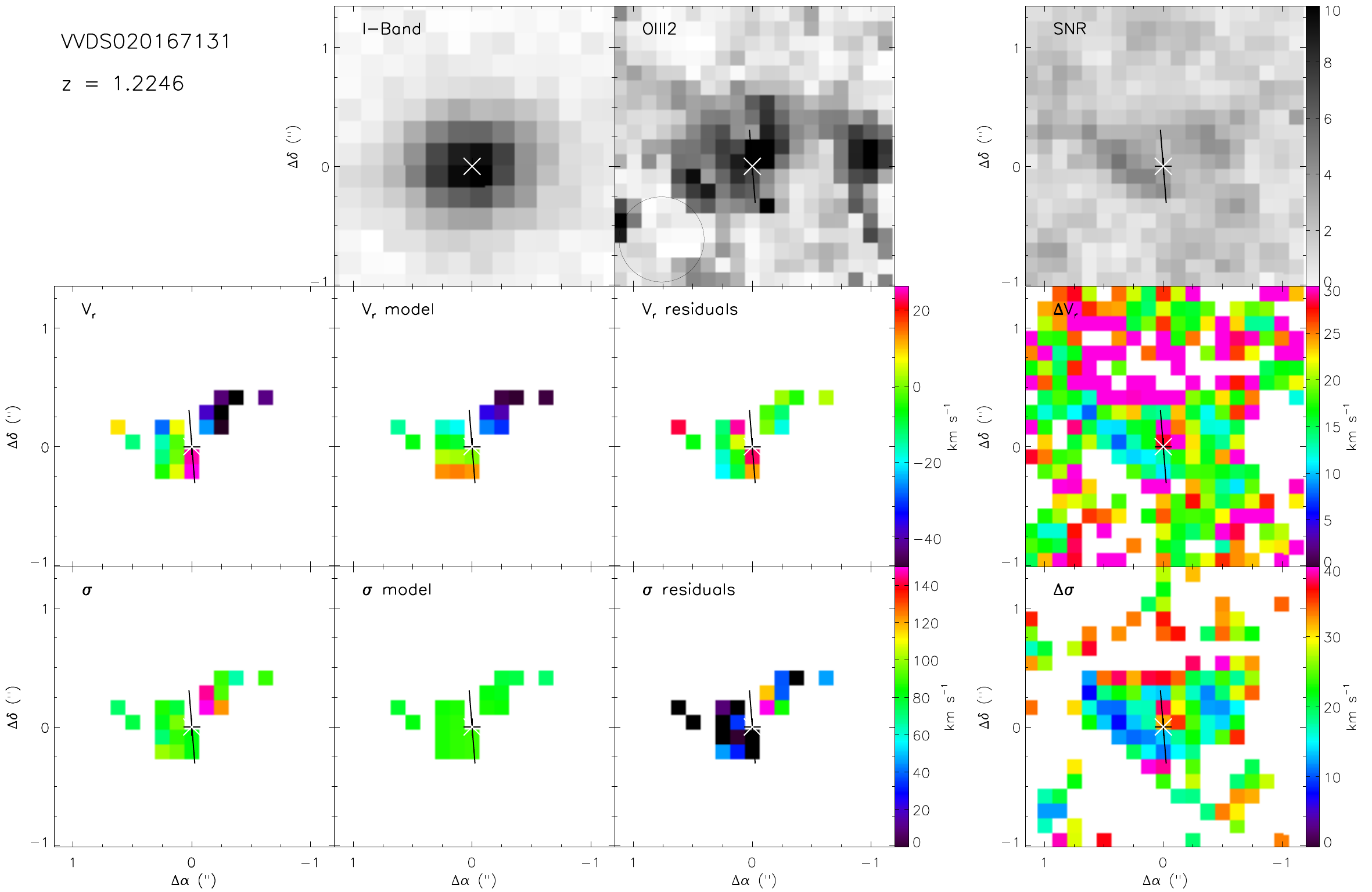}
\caption{Maps for VVDS020167131. Same caption as Figure \ref{maps_caption}.}
\end{center}
\end{figure}
\begin{figure}
\begin{center}
\includegraphics[width=16cm]{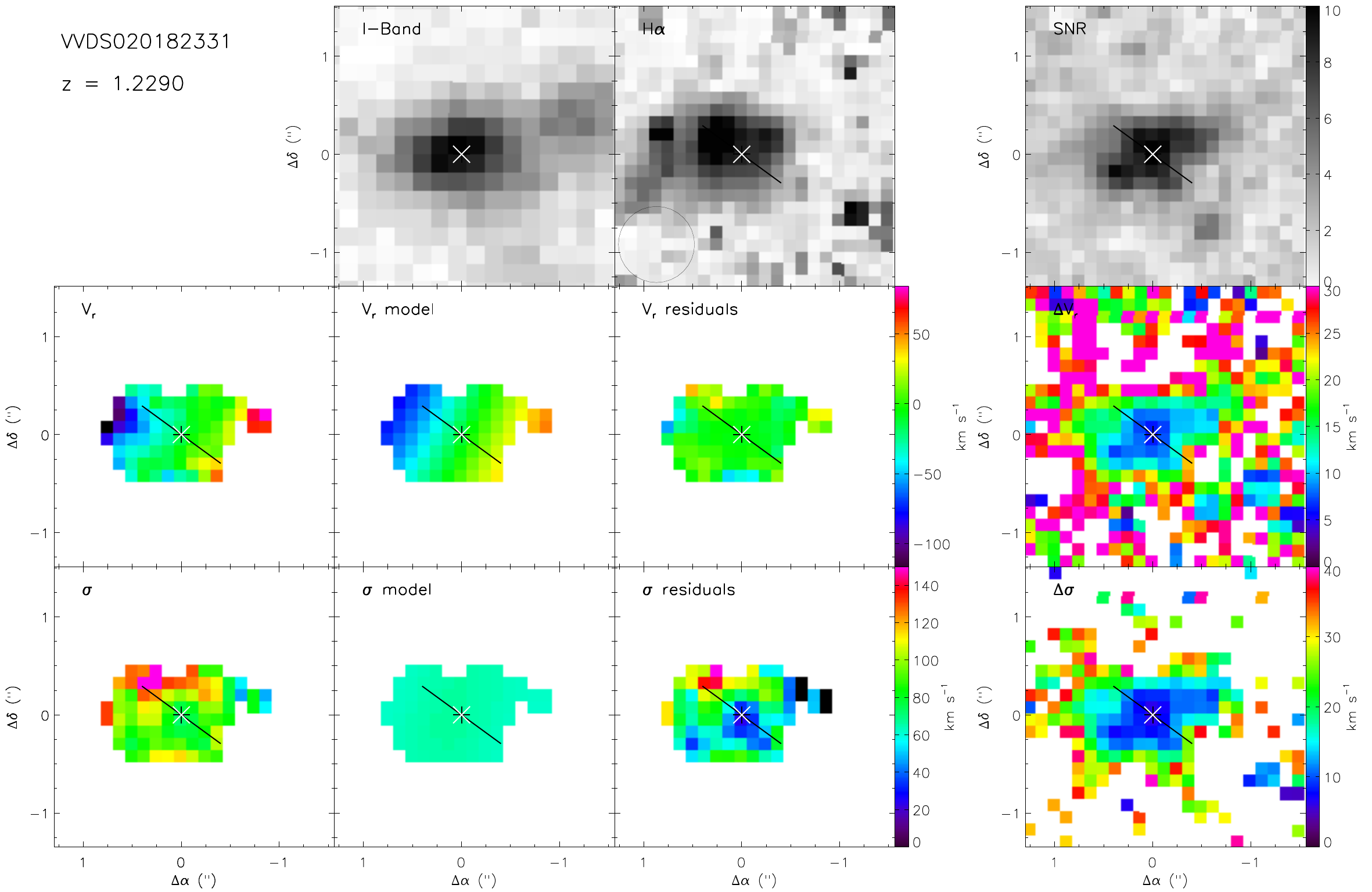}
\caption{Maps for VVDS020182331. Same caption as Figure \ref{maps_caption}.}
\end{center}
\end{figure}

\clearpage

\begin{figure}
\begin{center}
\includegraphics[width=16cm]{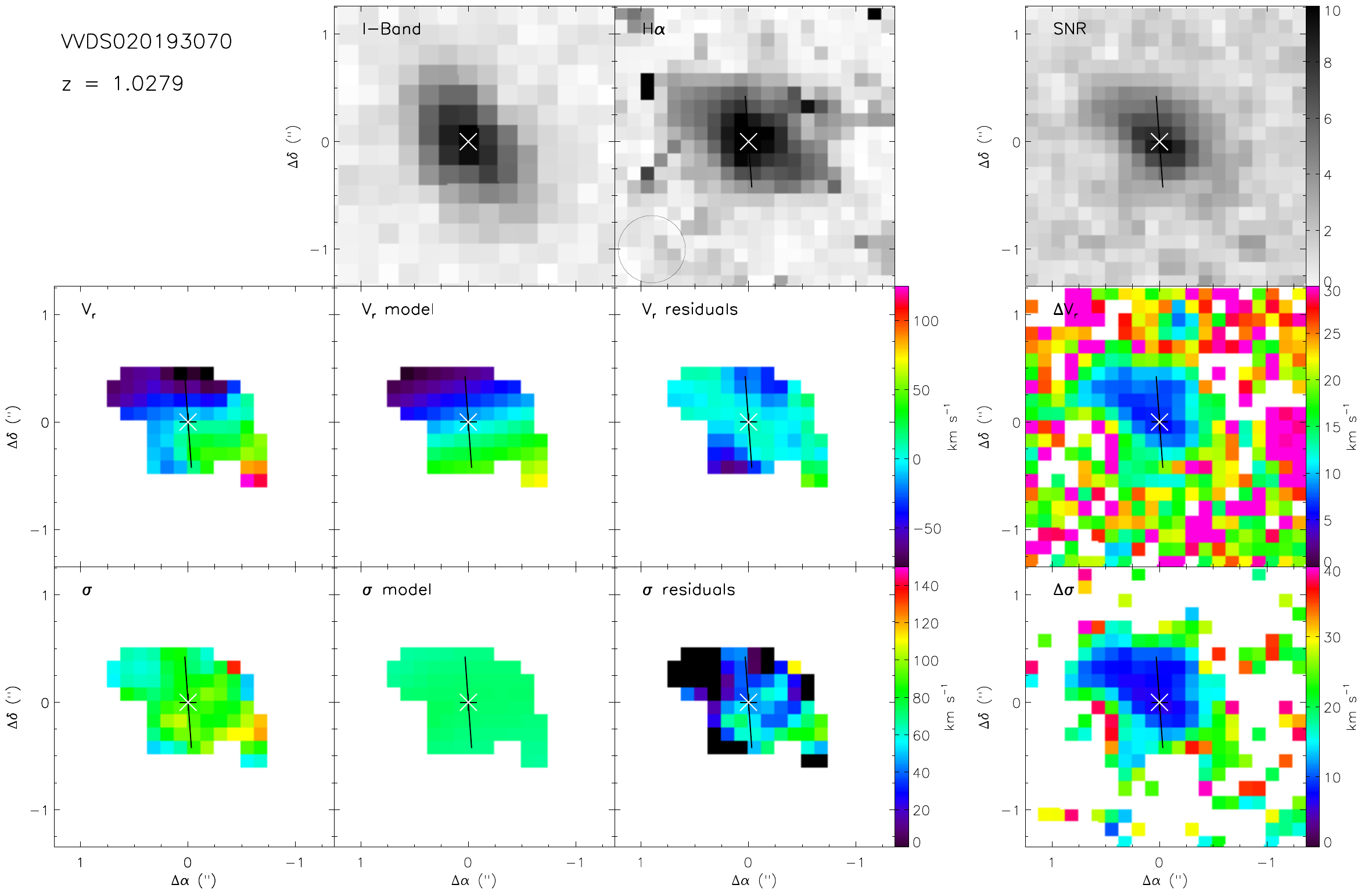}
\caption{Maps for VVDS020193070. Same caption as Figure \ref{maps_caption}.}
\end{center}
\end{figure}
\begin{figure}
\begin{center}
\includegraphics[width=16cm]{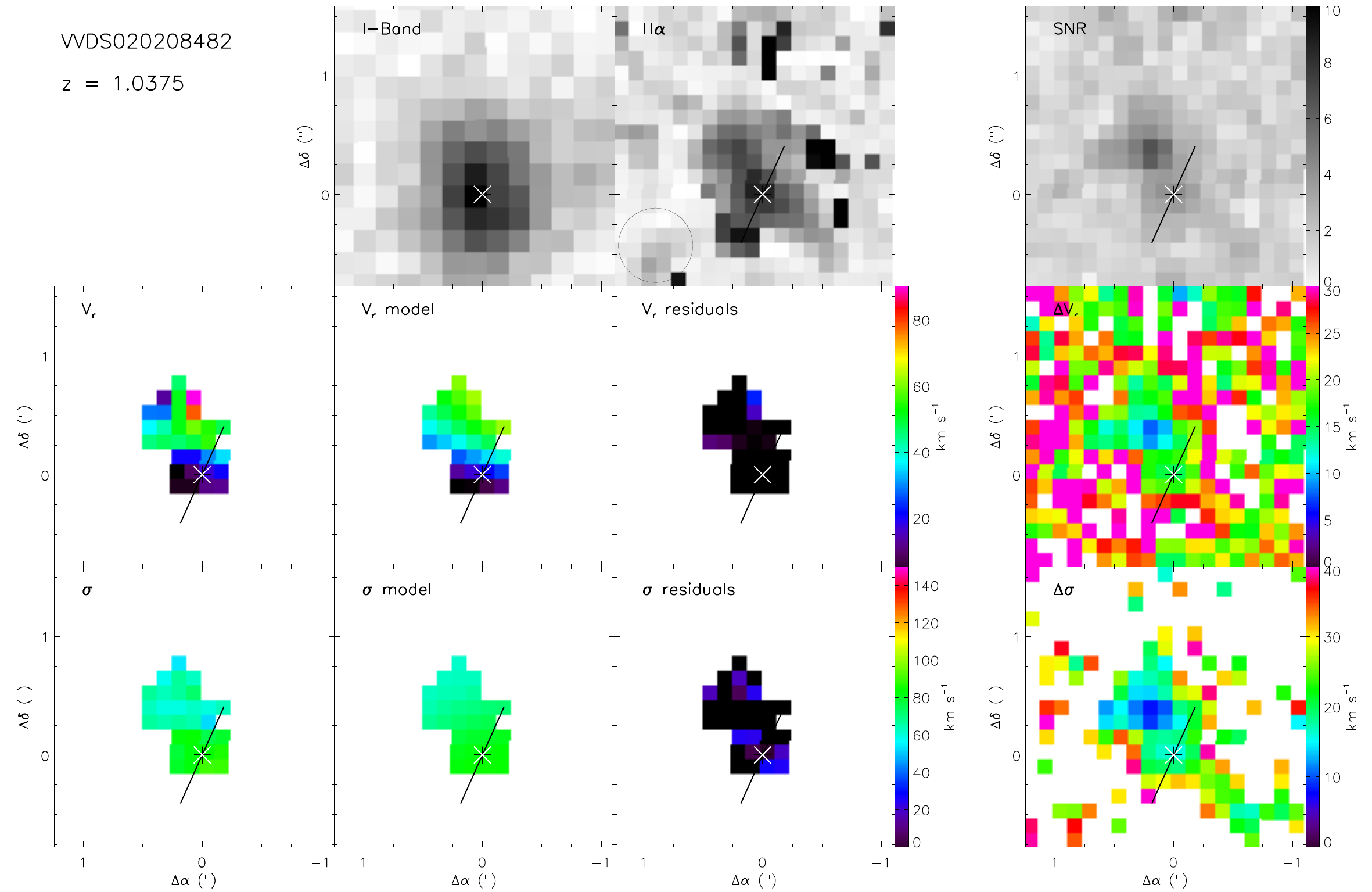}
\caption{Maps for VVDS020208482. Same caption as Figure \ref{maps_caption}.}
\end{center}
\end{figure}

\begin{figure}
\begin{center}
\includegraphics[width=16cm]{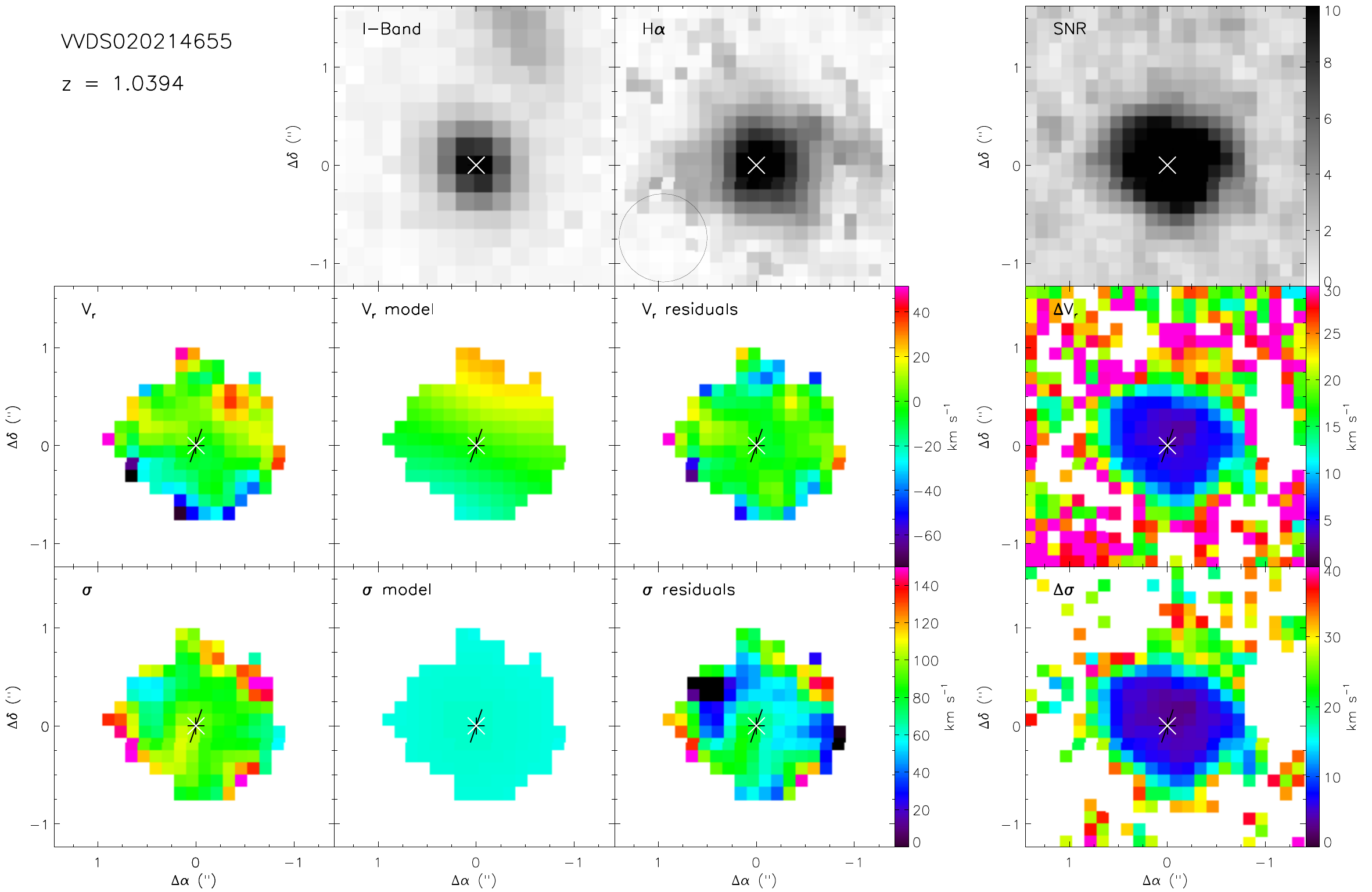}
\caption{Maps for VVDS020214655. Same caption as Figure \ref{maps_caption}.}
\end{center}
\end{figure}
\begin{figure}
\begin{center}
\includegraphics[width=16cm]{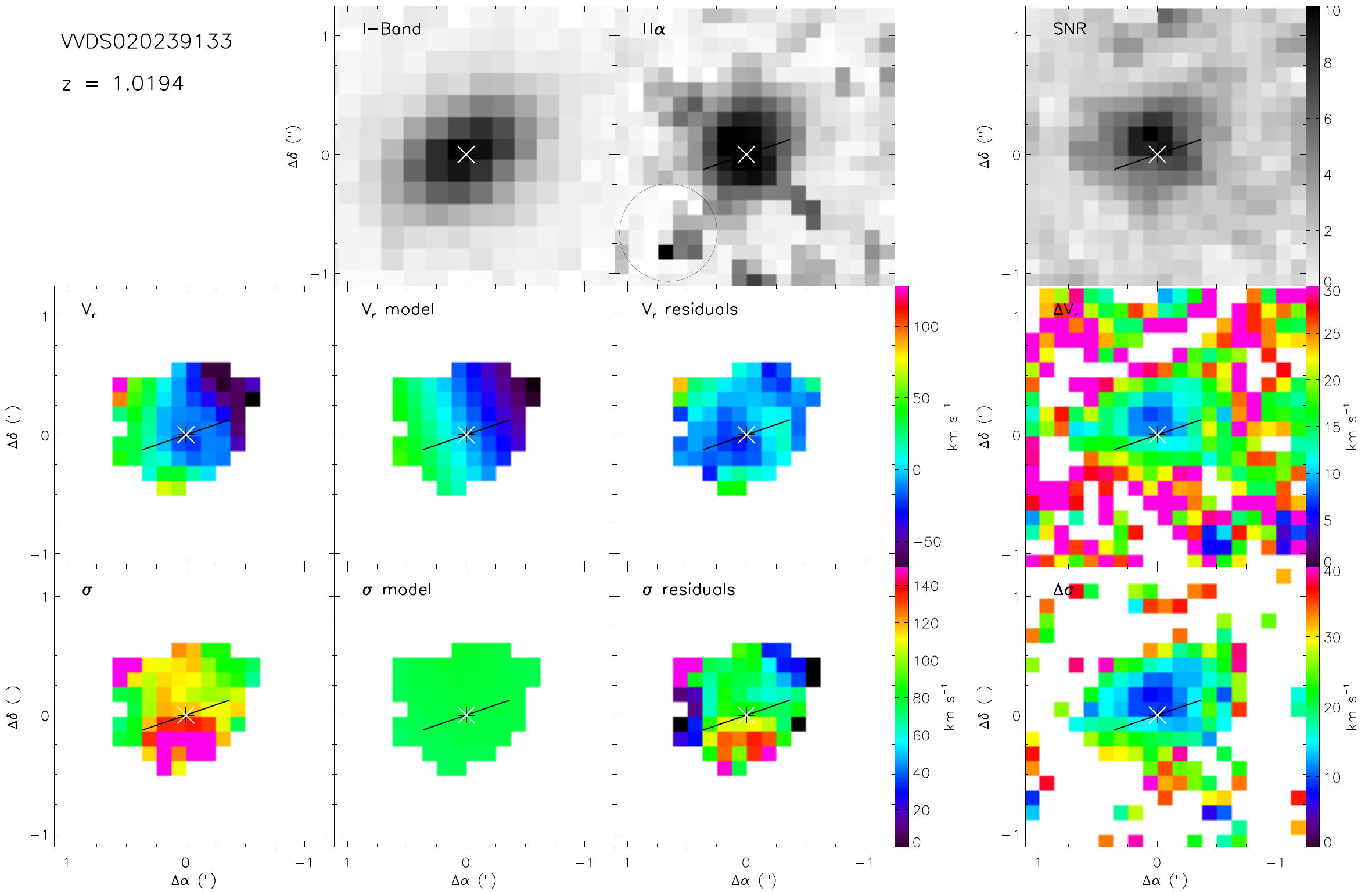}
\caption{Maps for VVDS020239133. Same caption as Figure \ref{maps_caption}.}
\end{center}
\end{figure}

\begin{figure}
\begin{center}
\includegraphics[width=16cm]{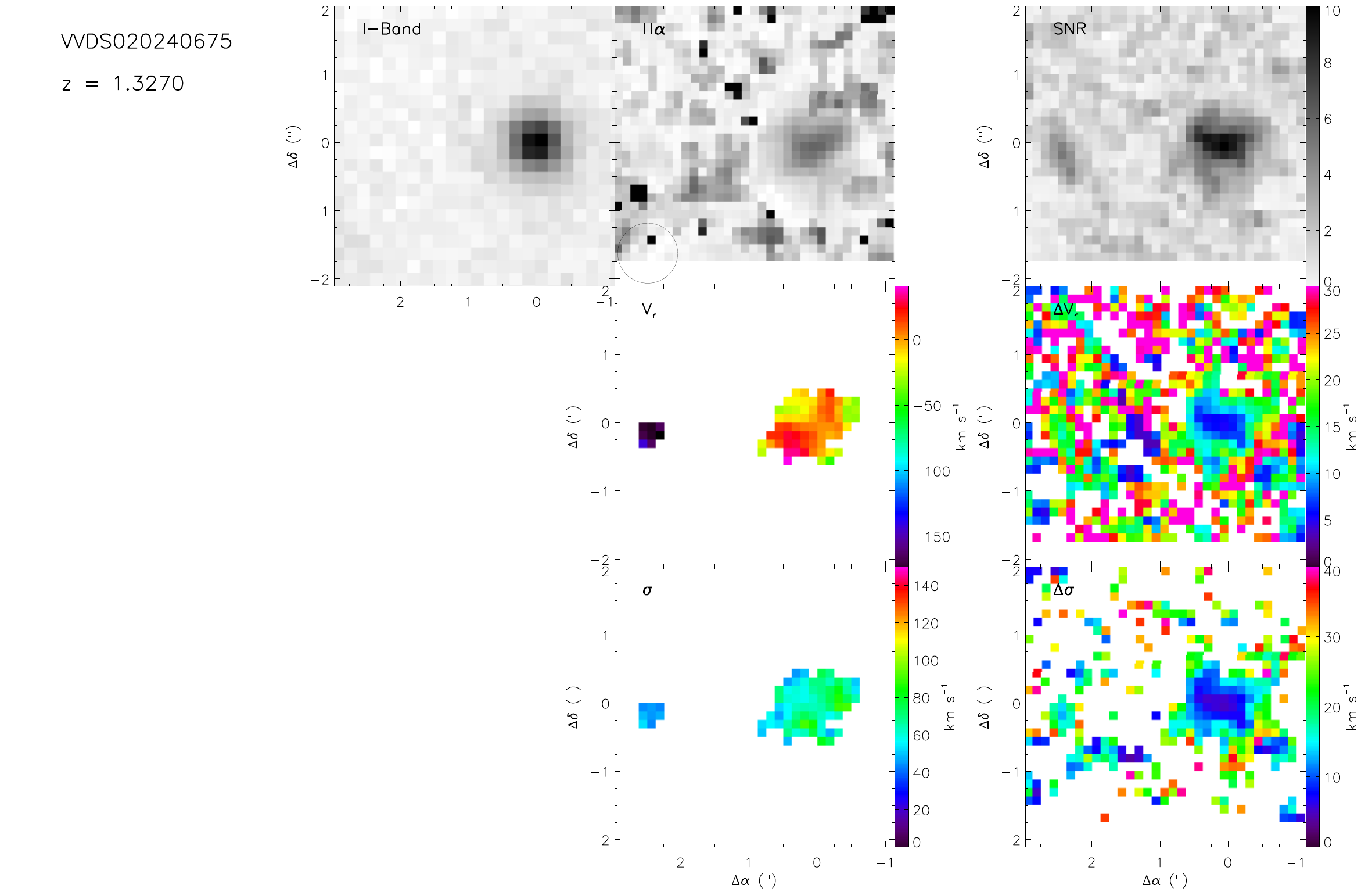}
\caption{Maps for VVDS020240675 system. Same caption as Figure \ref{maps_caption2}.}
\end{center}
\end{figure}
\begin{figure}
\begin{center}
\includegraphics[width=16cm]{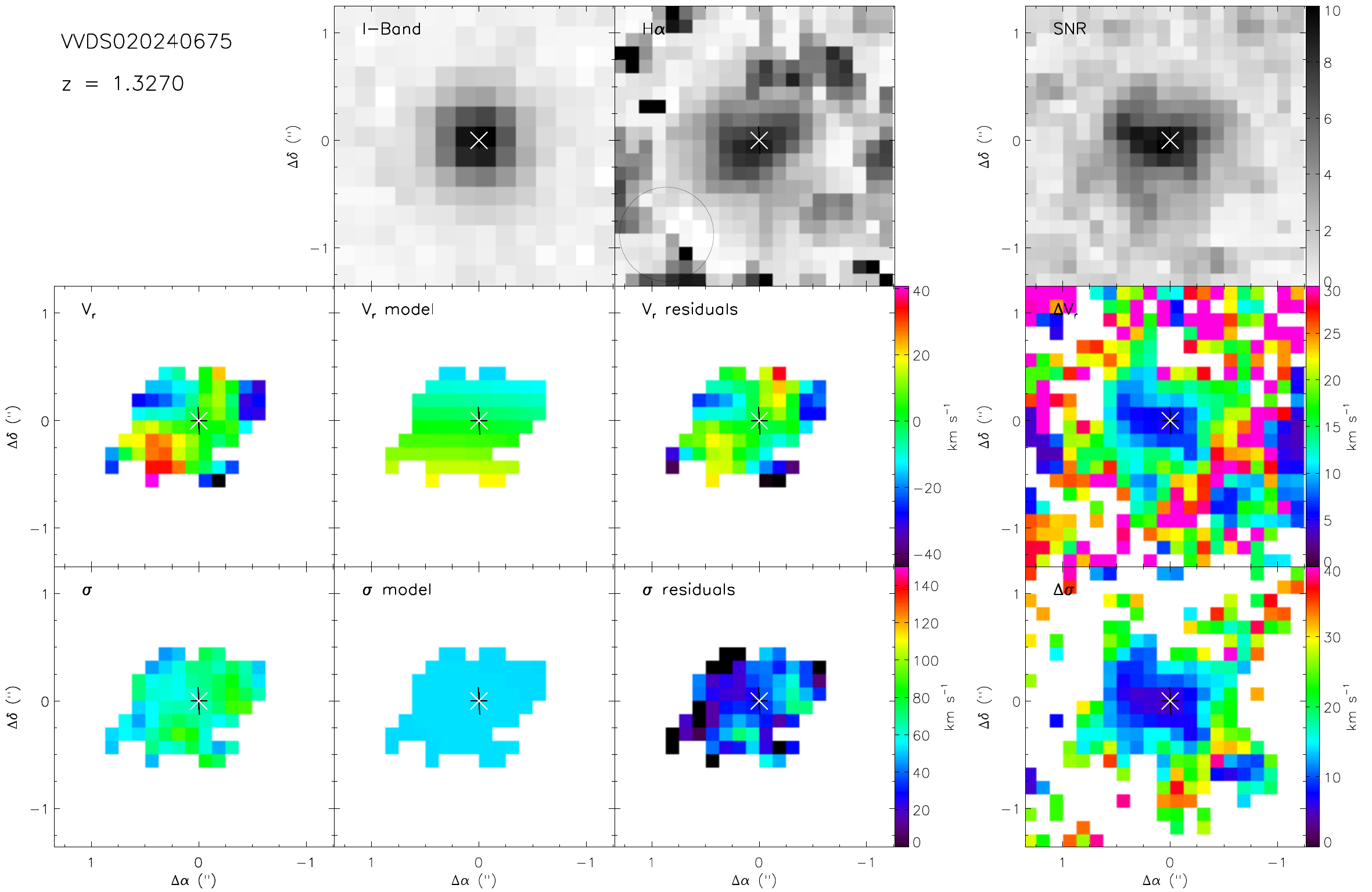}
\caption{Maps for VVDS020240675. Same caption as Figure \ref{maps_caption}.}
\end{center}
\end{figure}

\begin{figure}
\begin{center}
\includegraphics[width=16cm]{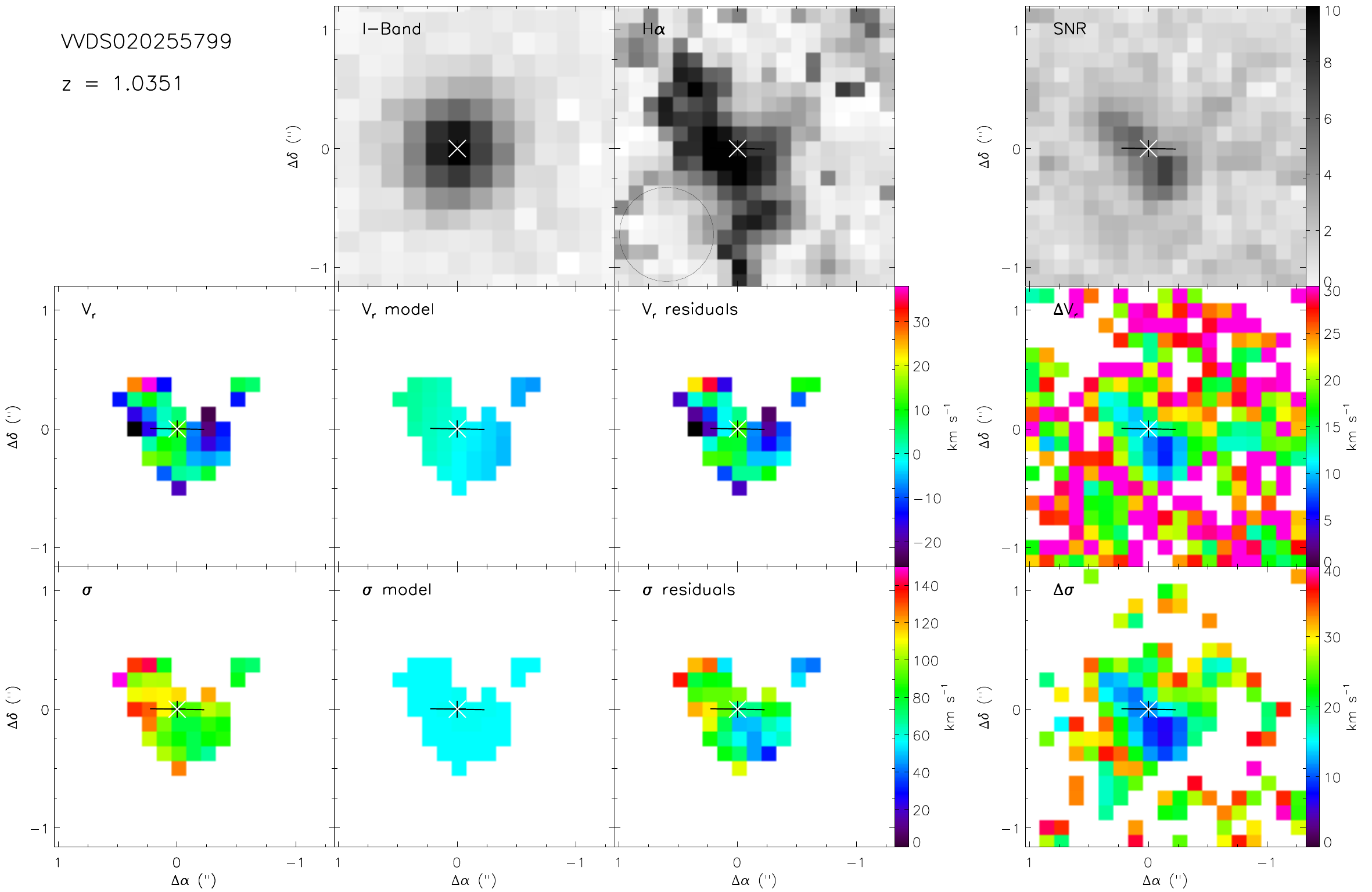}
\caption{Maps for VVDS020255799. Same caption as Figure \ref{maps_caption}.}
\end{center}
\end{figure}
\begin{figure}
\begin{center}
\includegraphics[width=16cm]{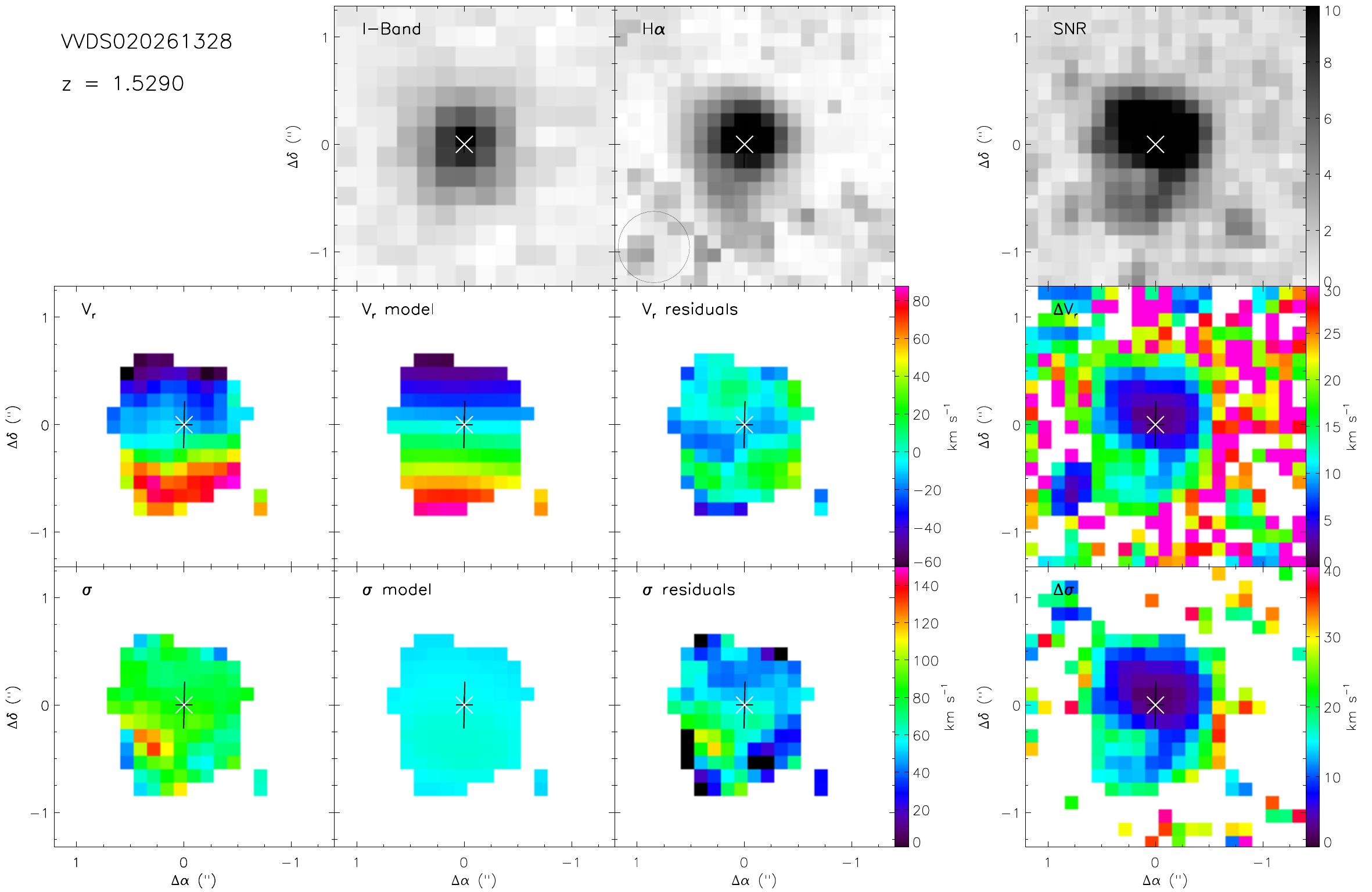}
\caption{Maps for VVDS020261328. Same caption as Figure \ref{maps_caption}.}
\end{center}
\end{figure}

\begin{figure}
\begin{center}
\includegraphics[width=16cm]{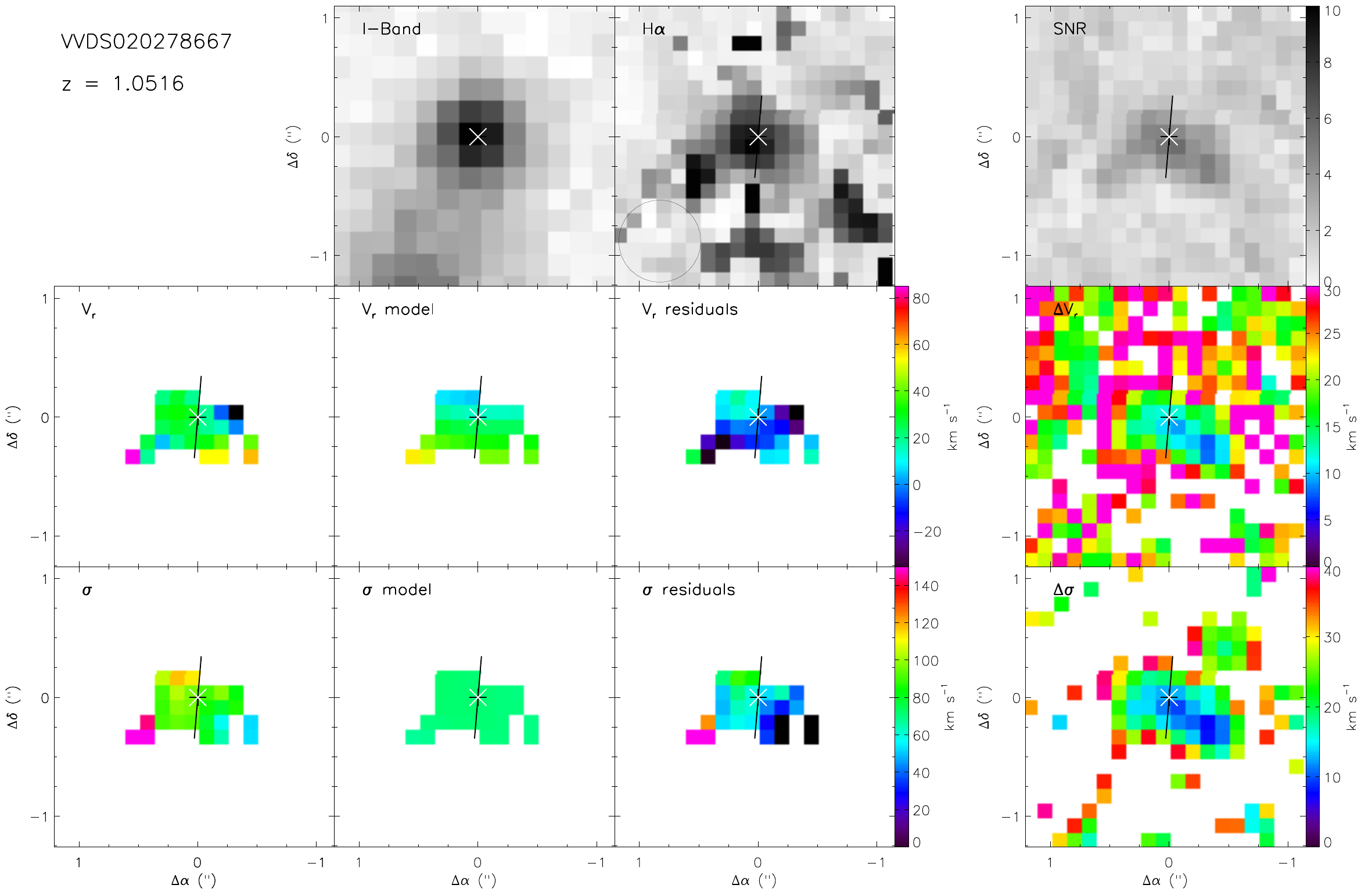}
\caption{Maps for VVDS020278667. Same caption as Figure \ref{maps_caption}.}
\end{center}
\end{figure}
\begin{figure}
\begin{center}
\includegraphics[width=16cm]{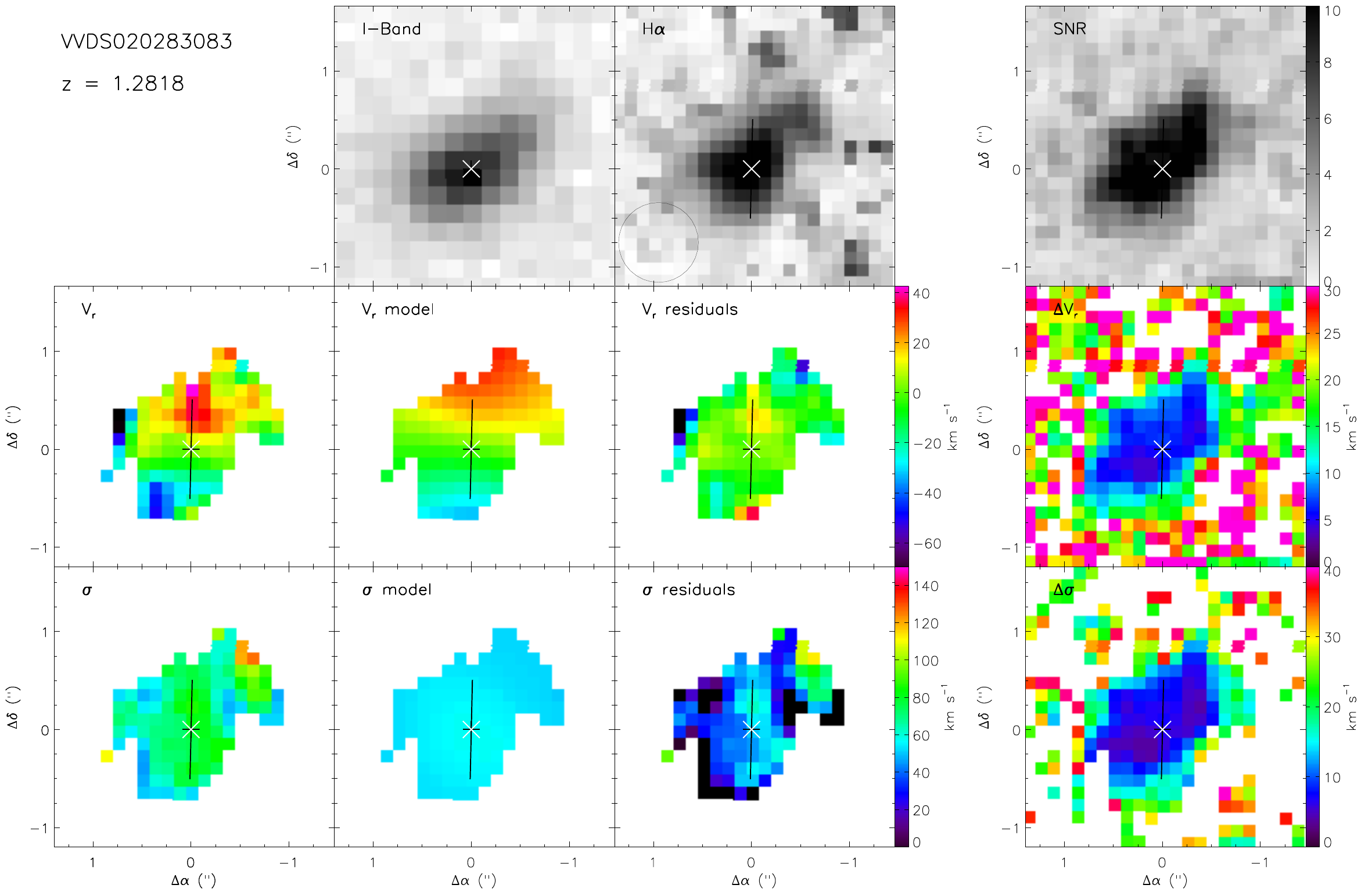}
\caption{Maps for VVDS020283083. Same caption as Figure \ref{maps_caption}.}
\end{center}
\end{figure}

\clearpage

\begin{figure}
\begin{center}
\includegraphics[width=16cm]{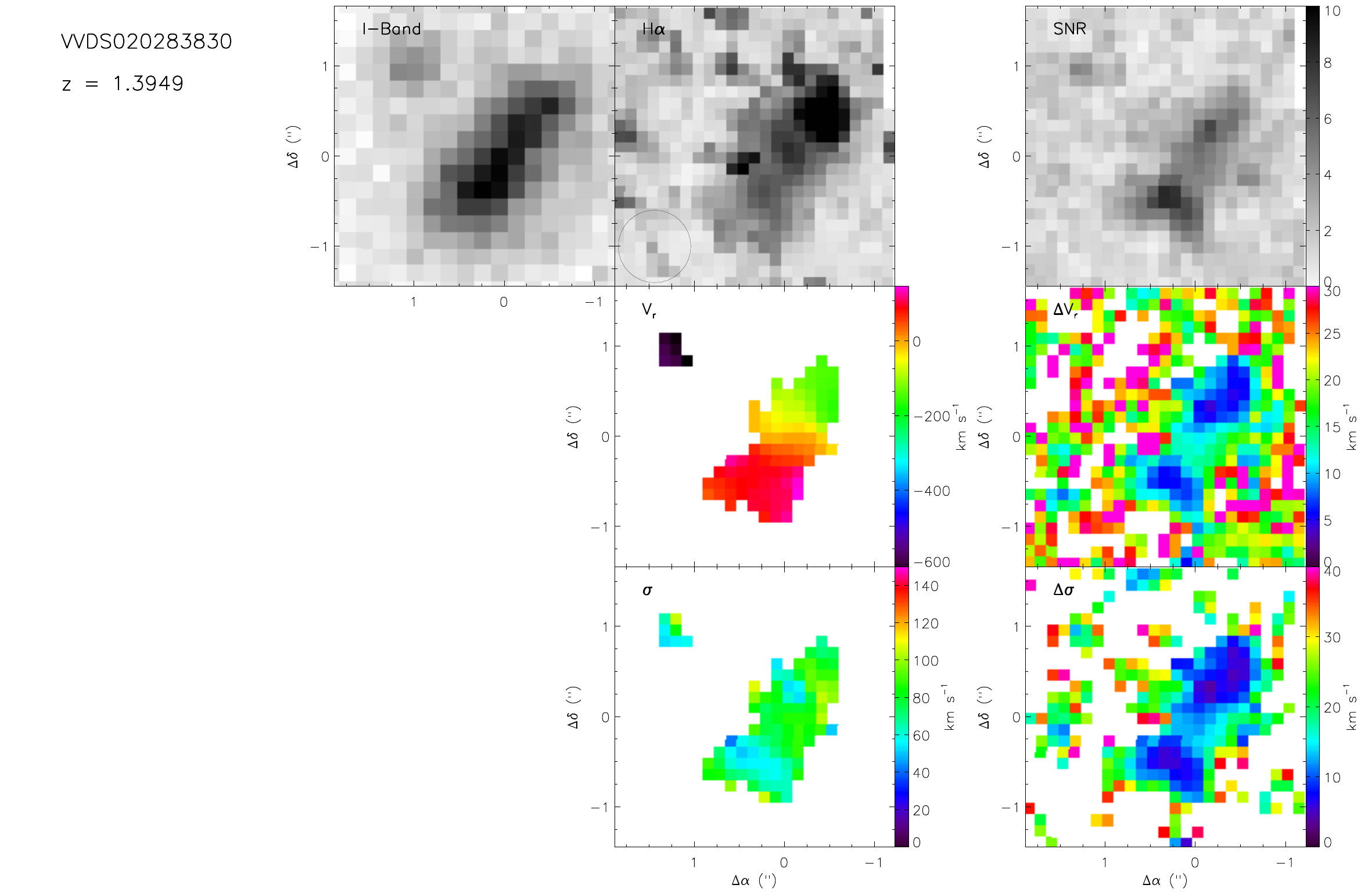}
\caption{Maps for VVDS020283830 system. Same caption as Figure \ref{maps_caption2}.}
\end{center}
\end{figure}
\begin{figure}
\begin{center}
\includegraphics[width=16cm]{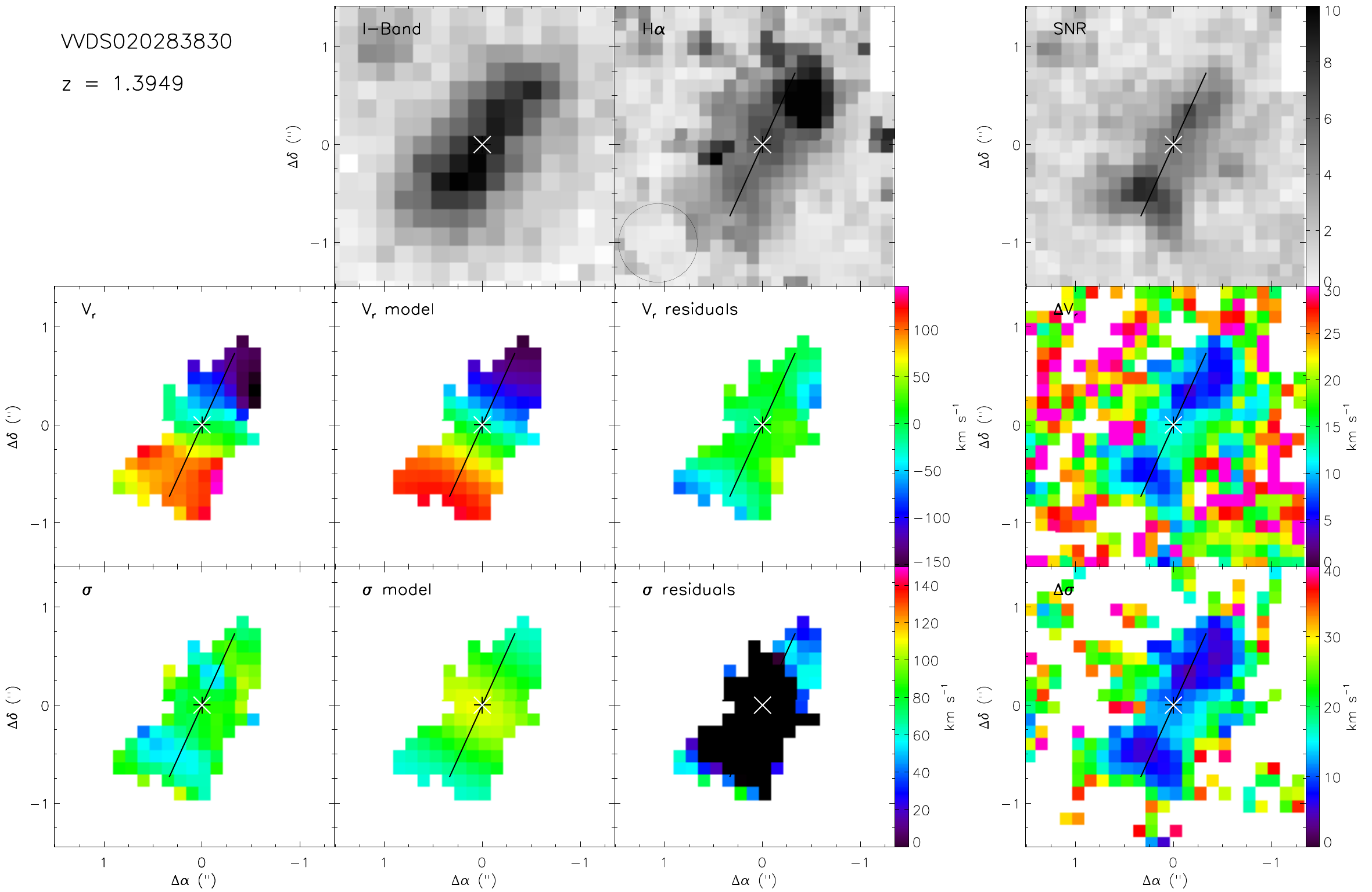}
\caption{Maps for VVDS020283830. Same caption as Figure \ref{maps_caption}.}
\end{center}
\end{figure}

\begin{figure}
\begin{center}
\includegraphics[width=16cm]{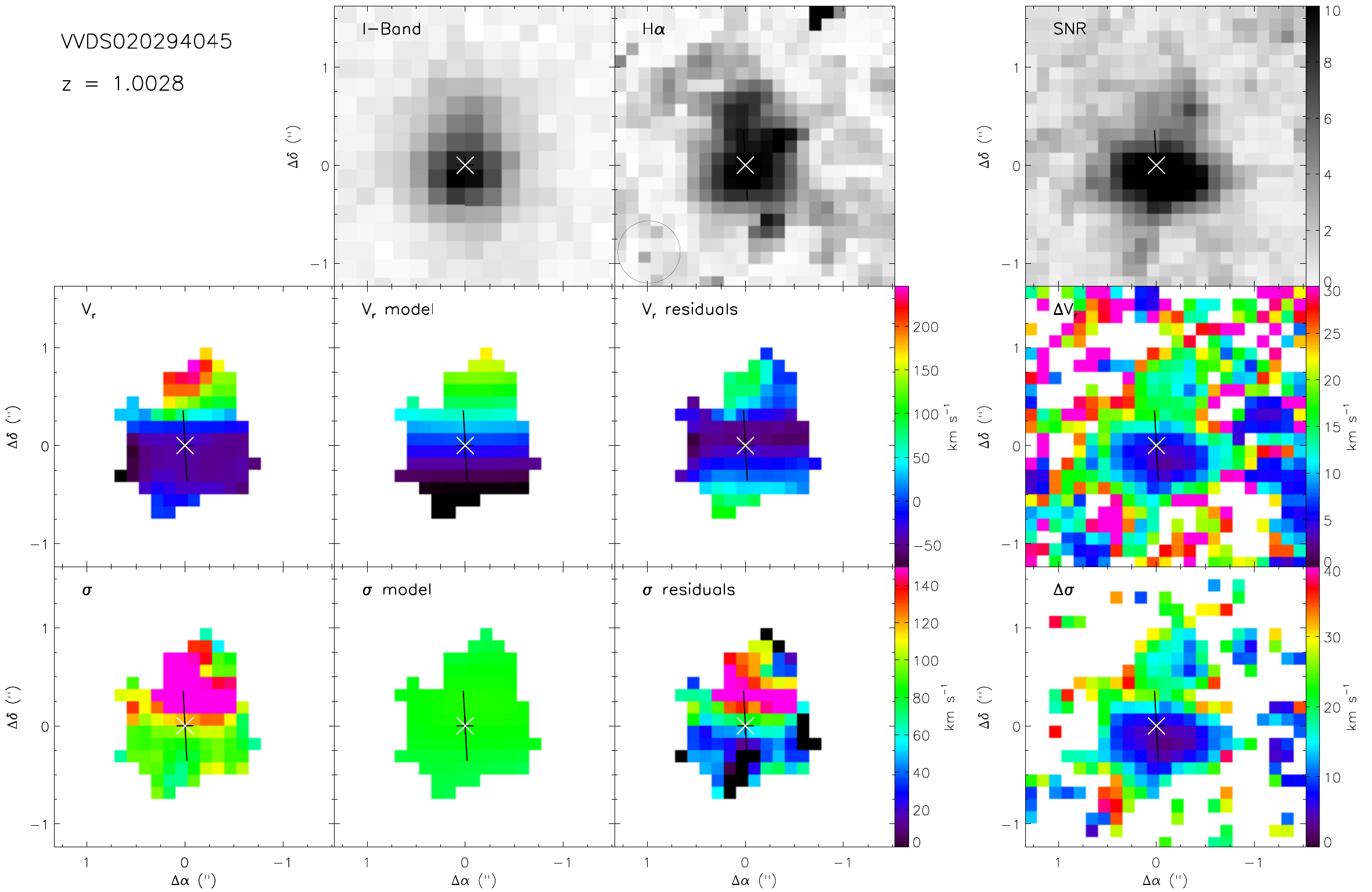}
\caption{Maps for VVDS020294045. Same caption as Figure \ref{maps_caption}.}
\end{center}
\end{figure}
\begin{figure}
\begin{center}
\includegraphics[width=16cm]{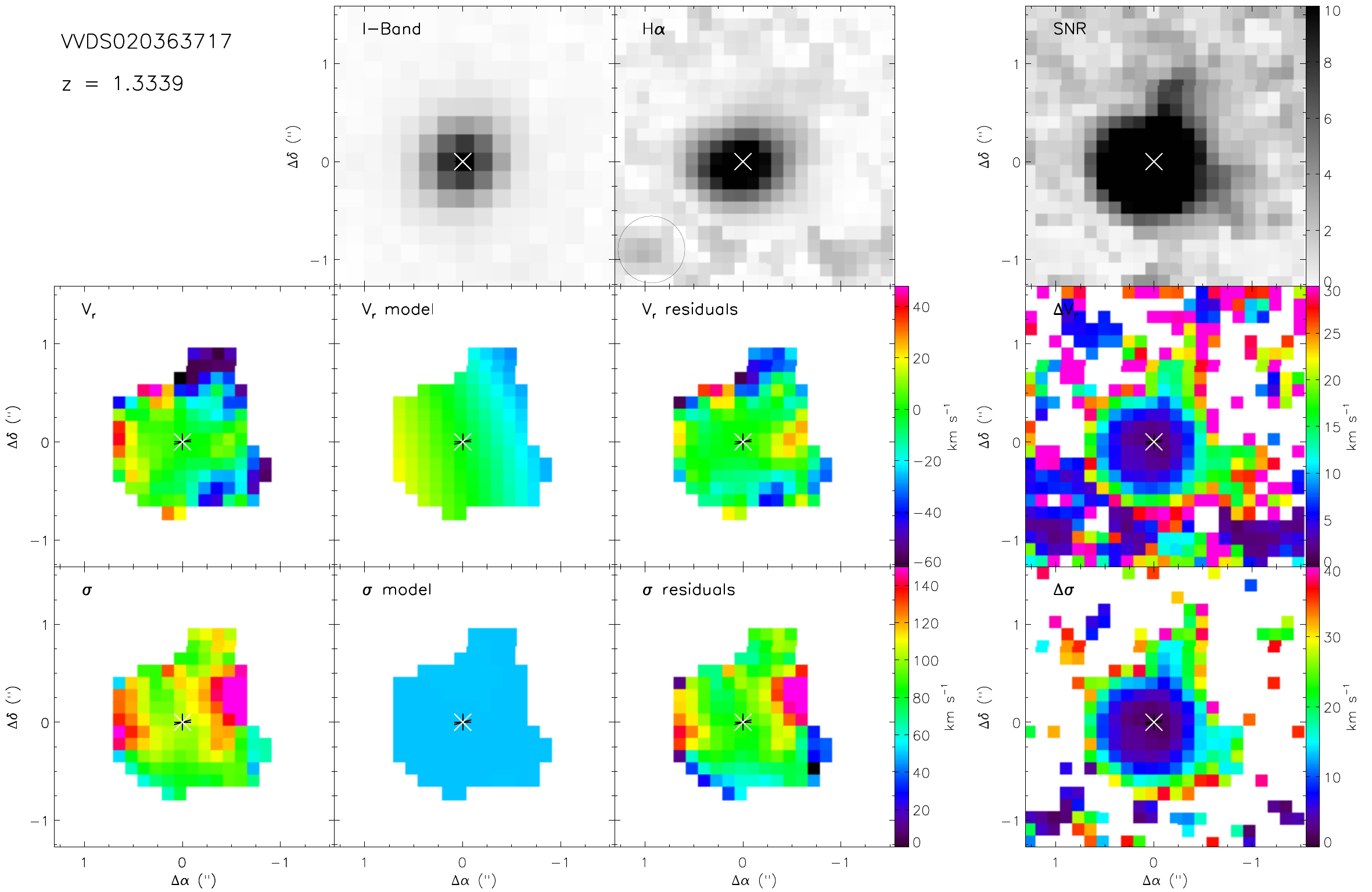}
\caption{Maps for VVDS020363717. Same caption as Figure \ref{maps_caption}.}
\end{center}
\end{figure}

\begin{figure}
\begin{center}
\includegraphics[width=16cm]{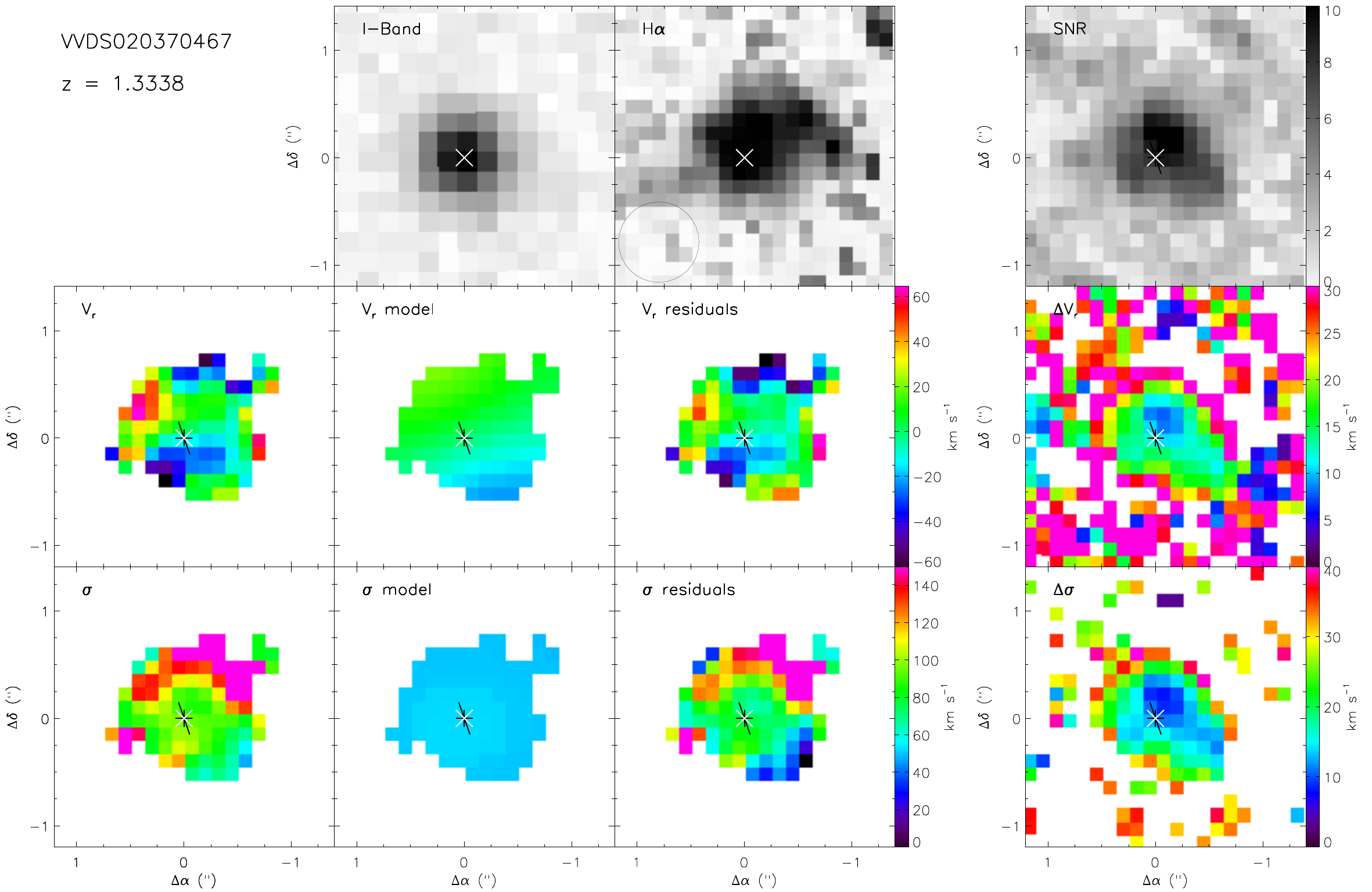}
\caption{Maps for VVDS020370467. Same caption as Figure \ref{maps_caption}.}
\end{center}
\end{figure}
\begin{figure}
\begin{center}
\includegraphics[width=16cm]{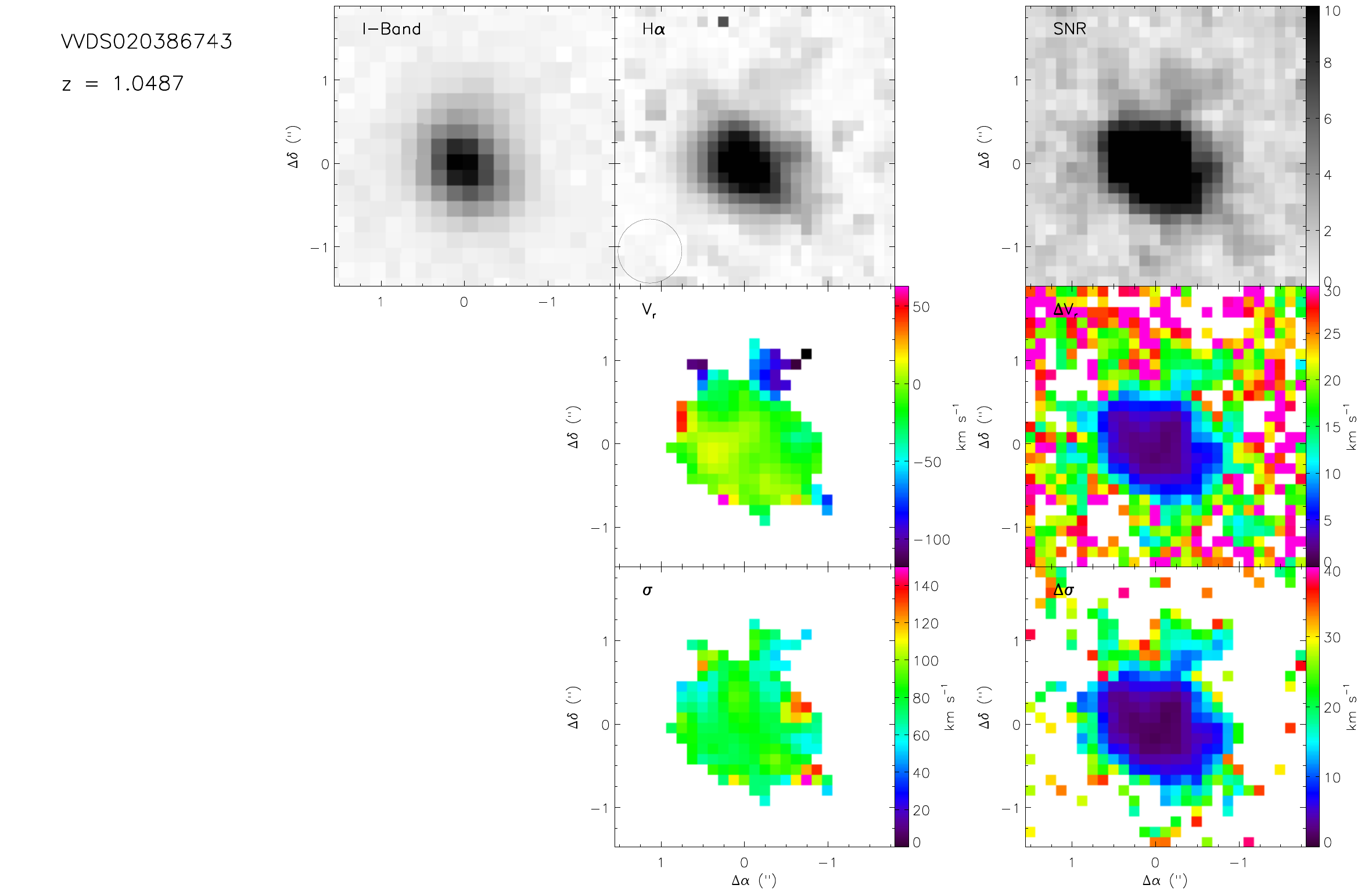}
\caption{Maps for VVDS020386743 system. Same caption as Figure \ref{maps_caption2}.}
\end{center}
\end{figure}

\begin{figure}
\begin{center}
\includegraphics[width=16cm]{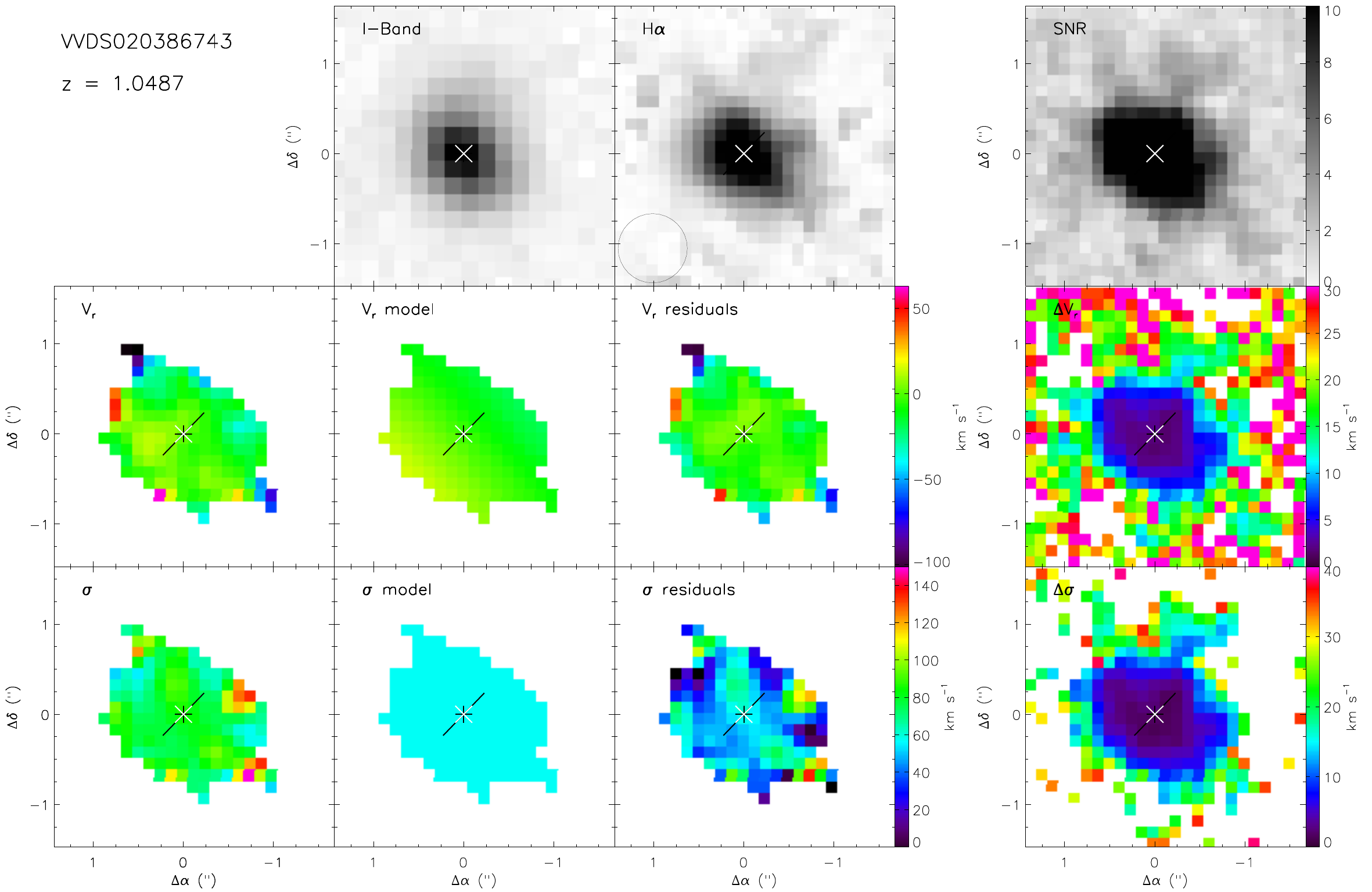}
\caption{Maps for VVDS020386743. Same caption as Figure \ref{maps_caption}.}
\end{center}
\end{figure}
\begin{figure}
\begin{center}
\includegraphics[width=16cm]{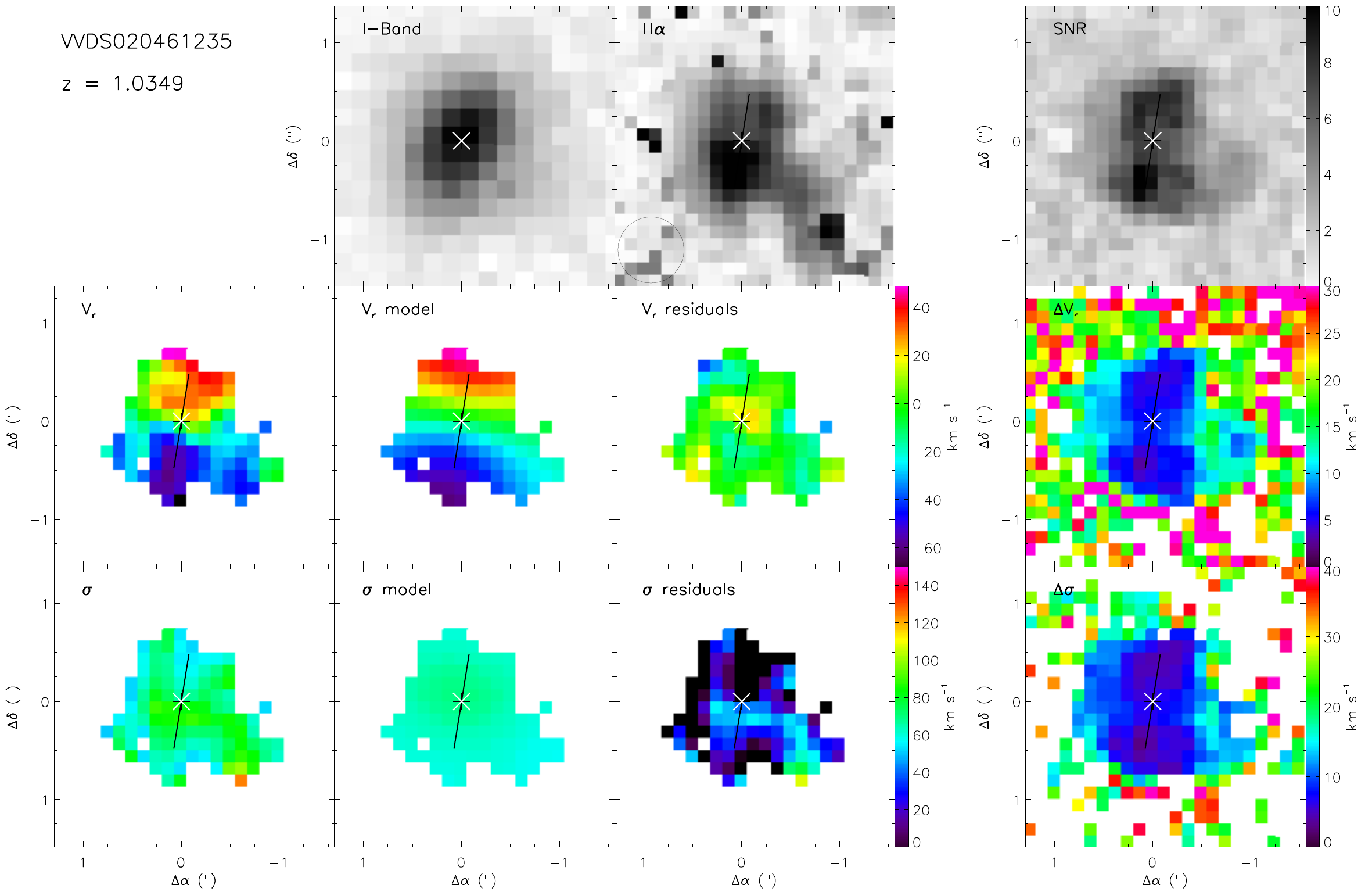}
\caption{Maps for VVDS020461235. Same caption as Figure \ref{maps_caption}.}
\end{center}
\end{figure}

\begin{figure}
\begin{center}
\includegraphics[width=16cm]{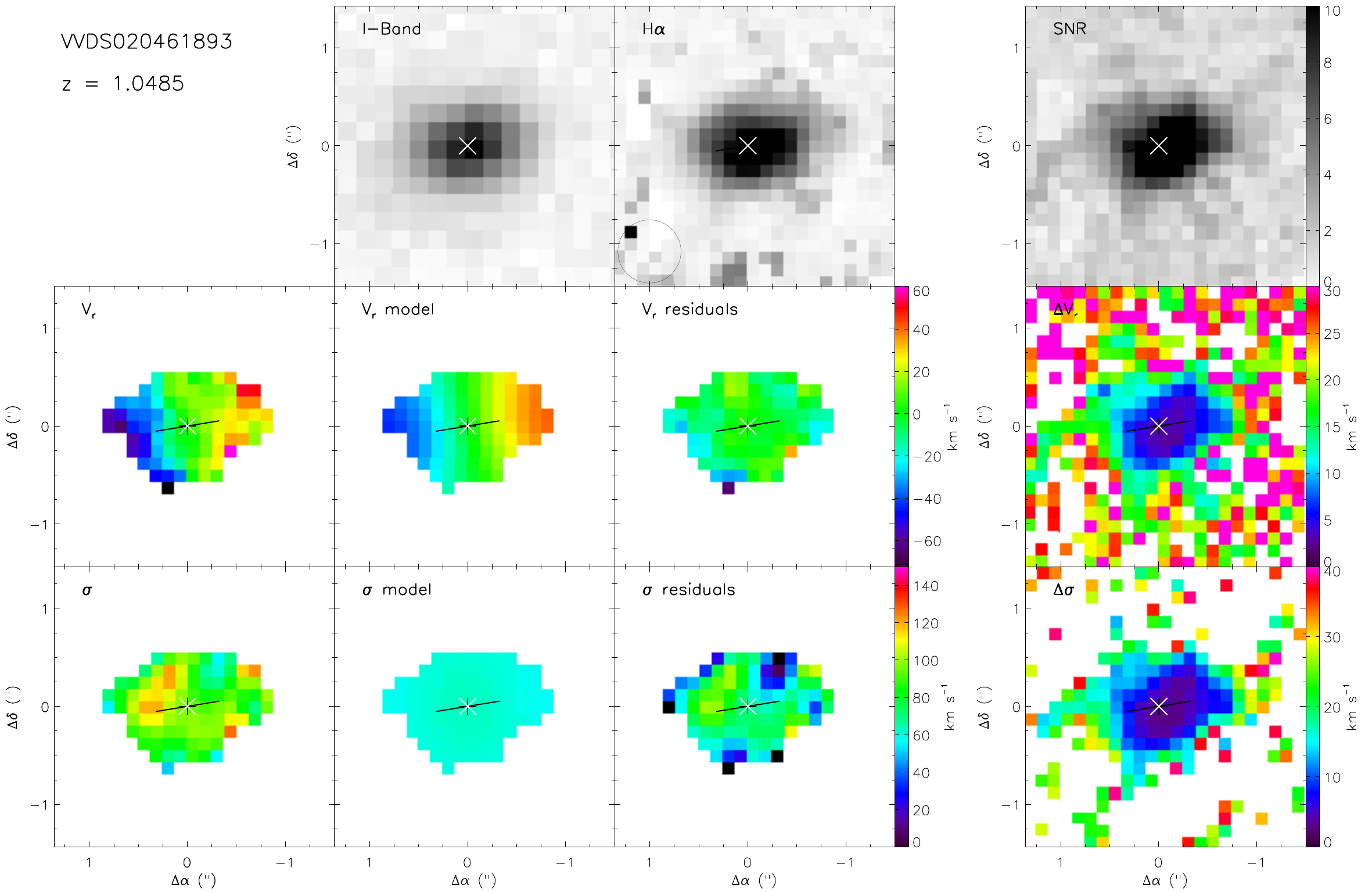}
\caption{Maps for VVDS020461893. Same caption as Figure \ref{maps_caption}.}
\end{center}
\end{figure}
\begin{figure}
\begin{center}
\includegraphics[width=16cm]{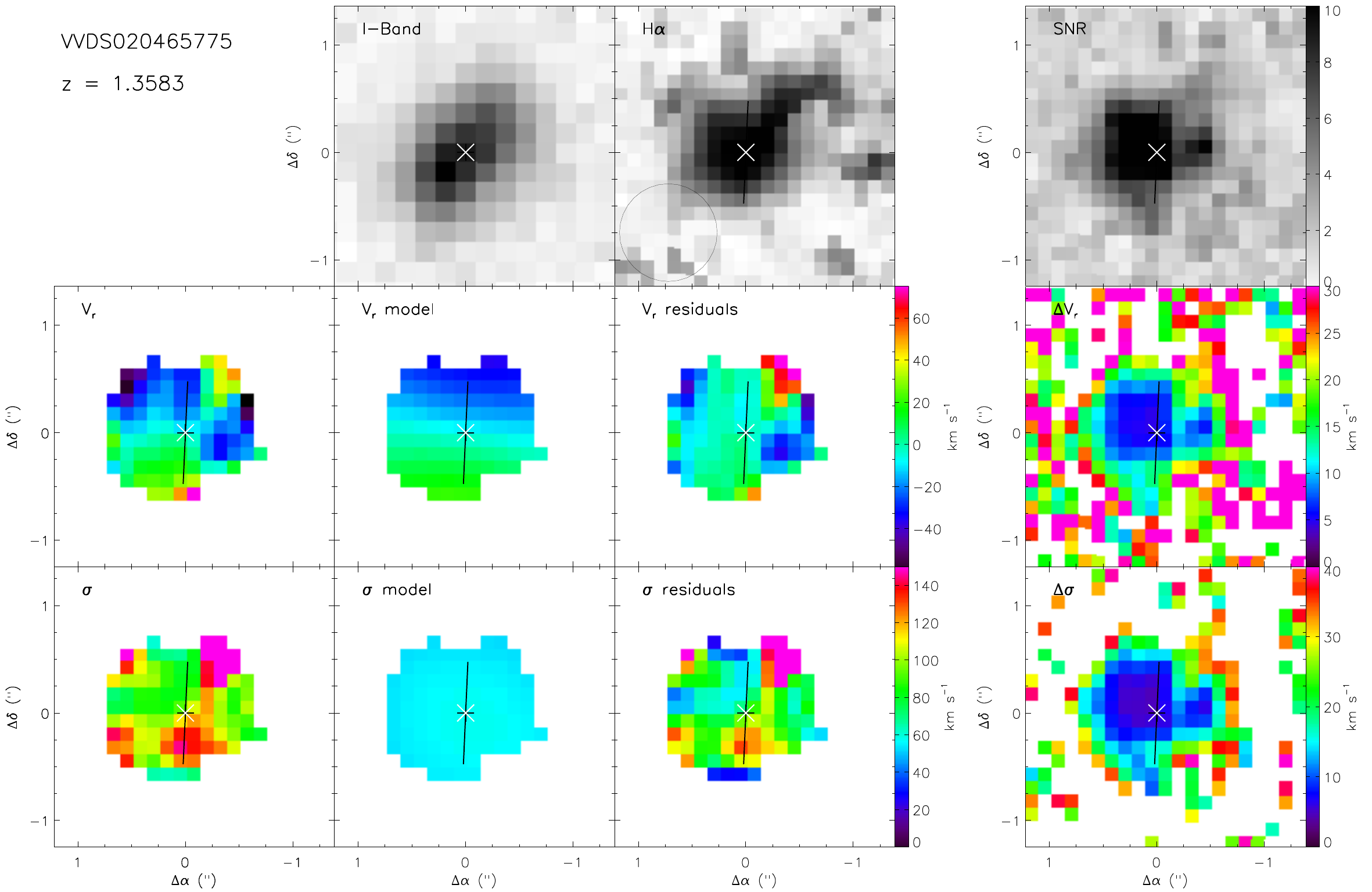}
\caption{Maps for VVDS020465775. Same caption as Figure \ref{maps_caption}.}
\end{center}
\end{figure}

\clearpage

\begin{figure}
\begin{center}
\includegraphics[width=16cm]{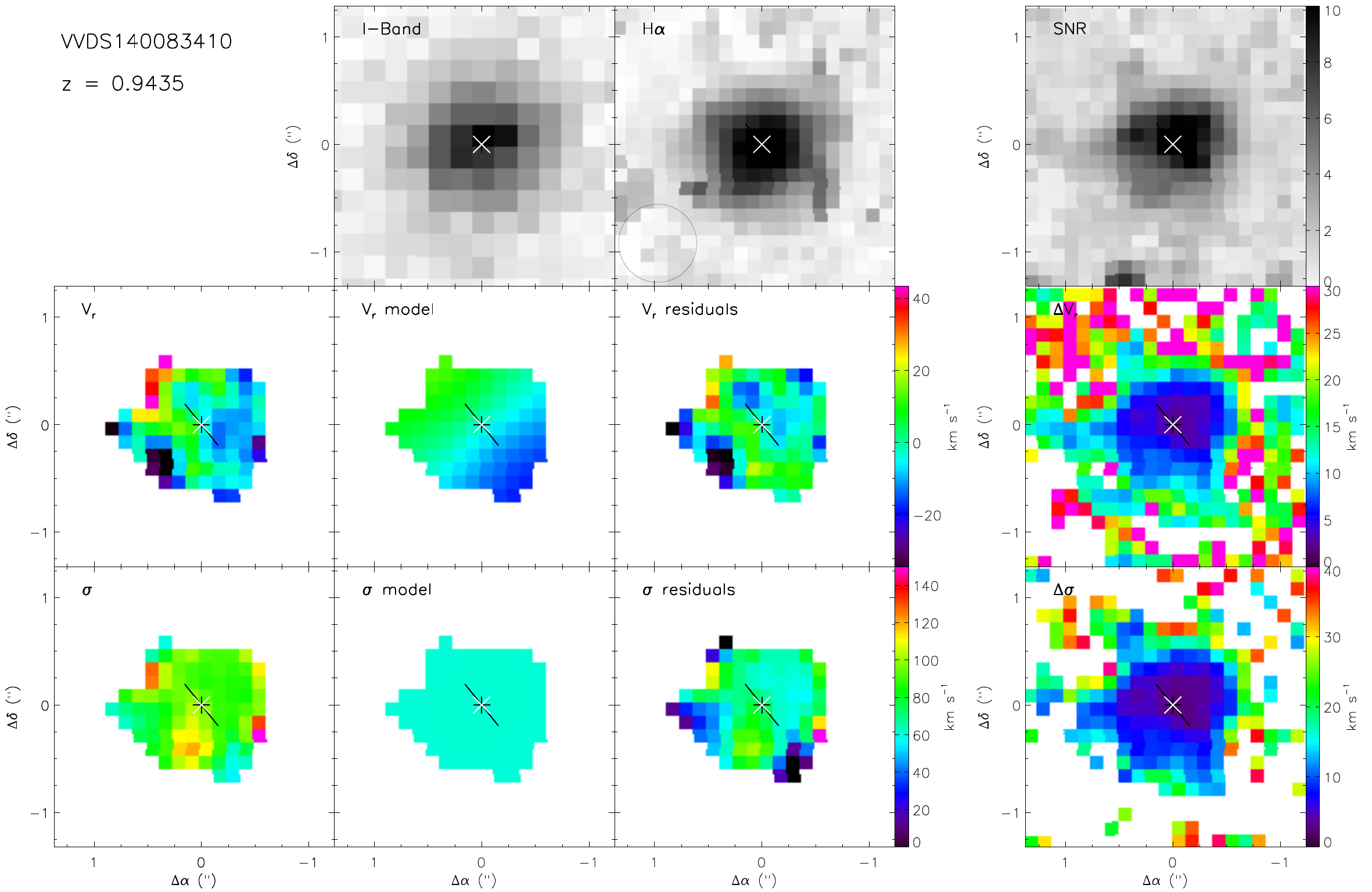}
\caption{Maps for VVDS140083410. Same caption as Figure \ref{maps_caption}.}
\end{center}
\end{figure}
\begin{figure}
\begin{center}
\includegraphics[width=16cm]{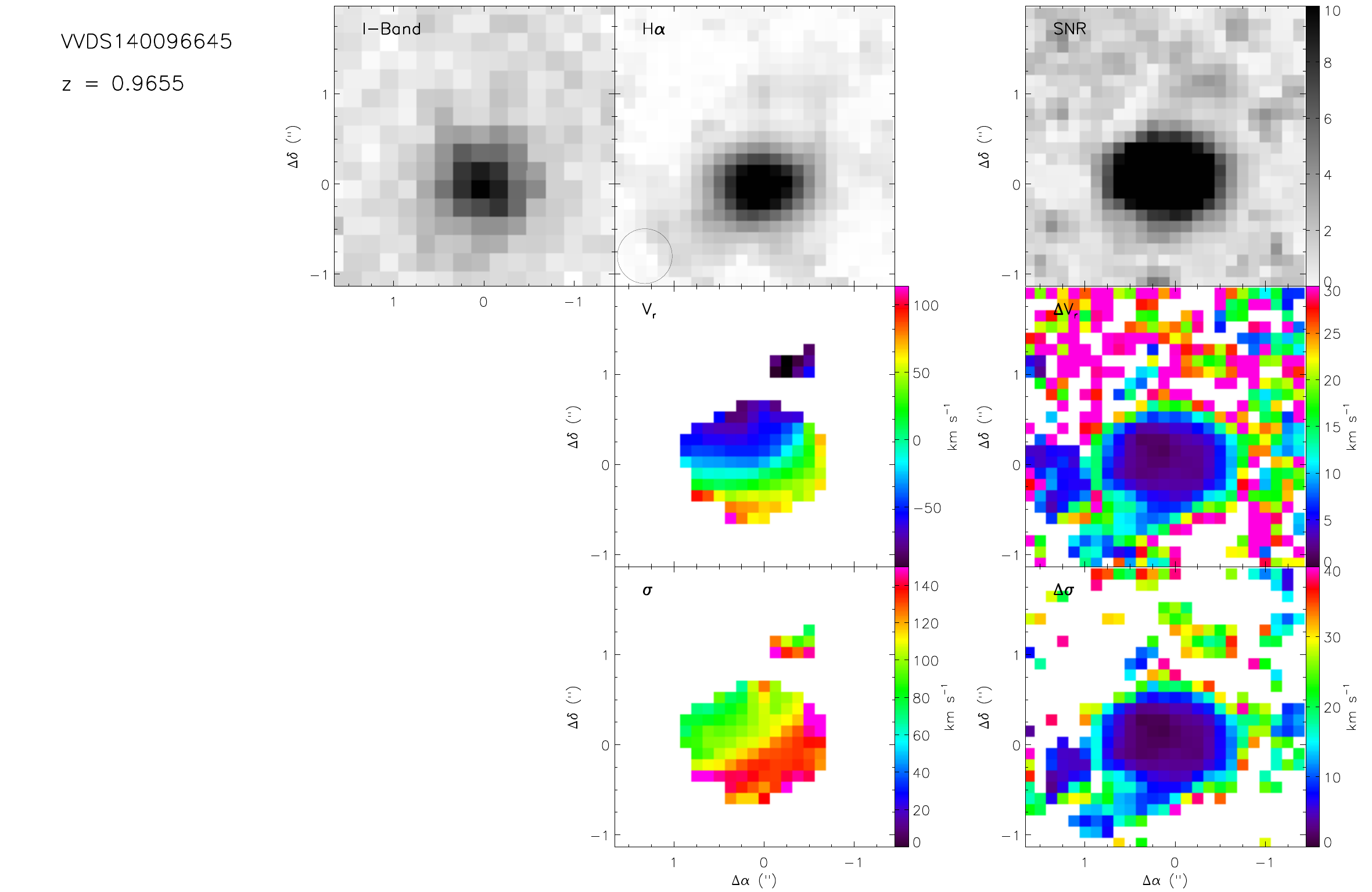}
\caption{Maps for VVDS140096645 system. Same caption as Figure \ref{maps_caption2}.}
\end{center}
\end{figure}

\begin{figure}
\begin{center}
\includegraphics[width=16cm]{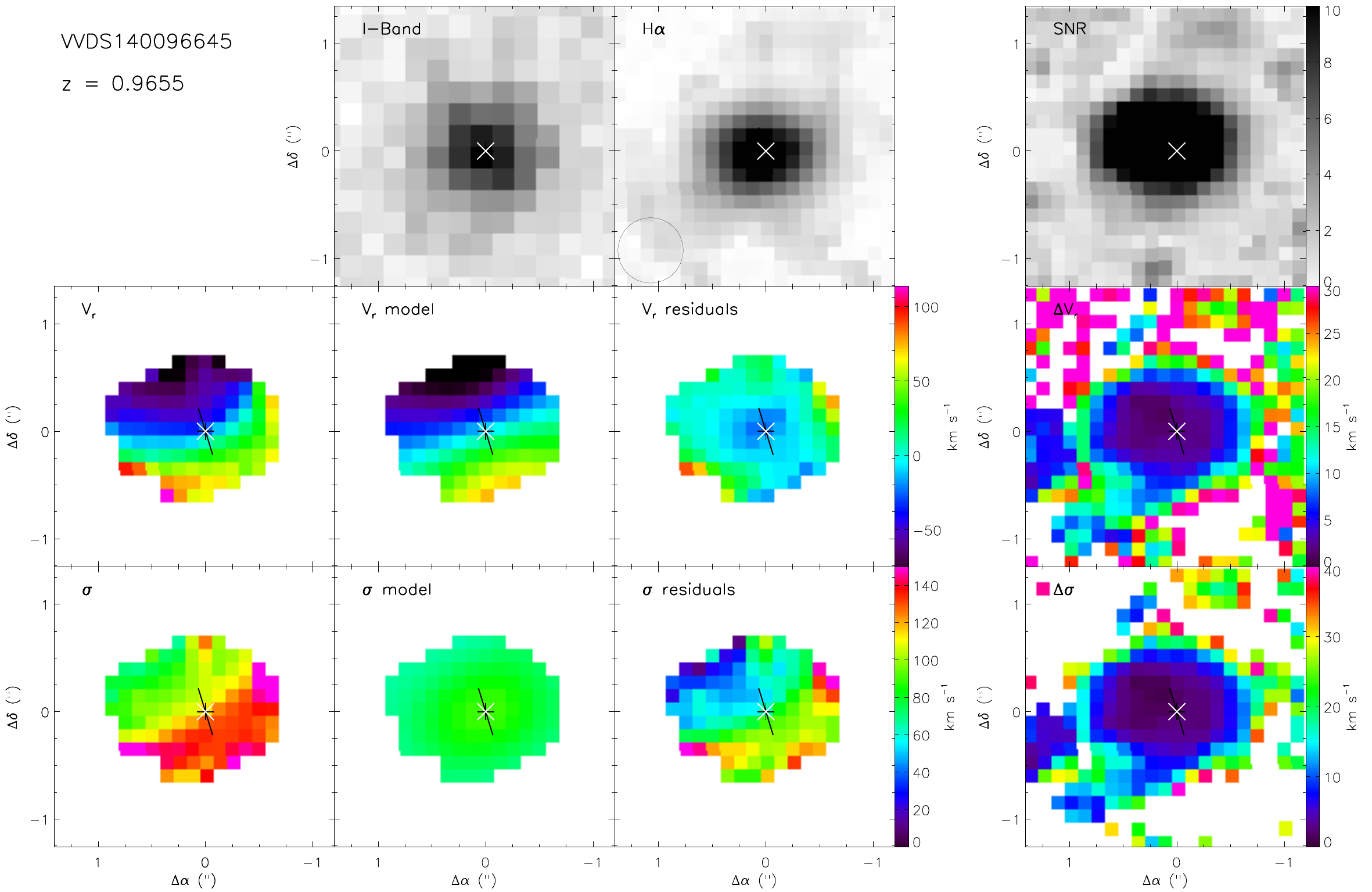}
\caption{Maps for VVDS140096645. Same caption as Figure \ref{maps_caption}.}
\end{center}
\end{figure}
\begin{figure}
\begin{center}
\includegraphics[width=16cm]{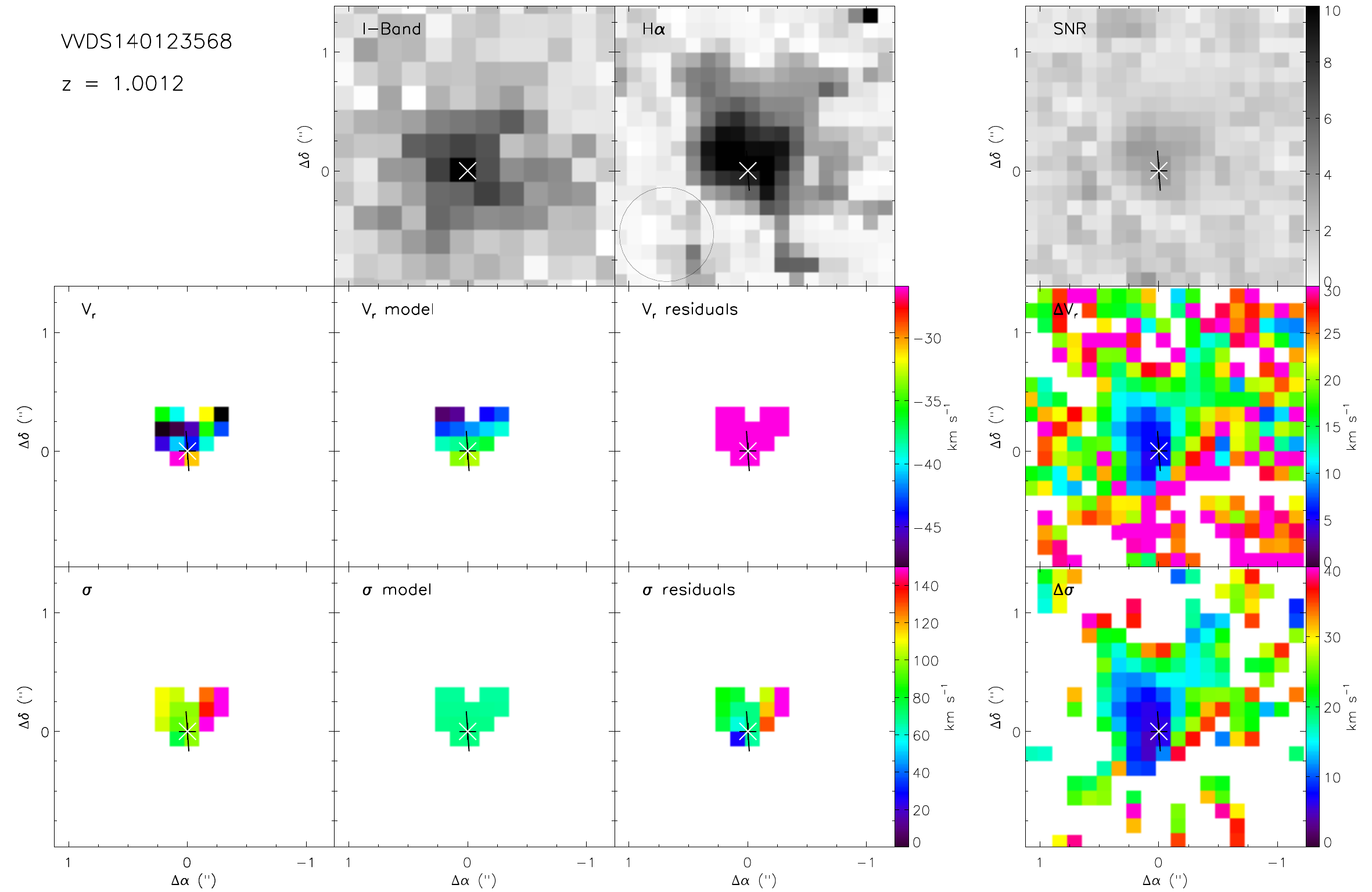}
\caption{Maps for VVDS140123568. Same caption as Figure \ref{maps_caption}.}
\end{center}
\end{figure}

\begin{figure}
\begin{center}
\includegraphics[width=16cm]{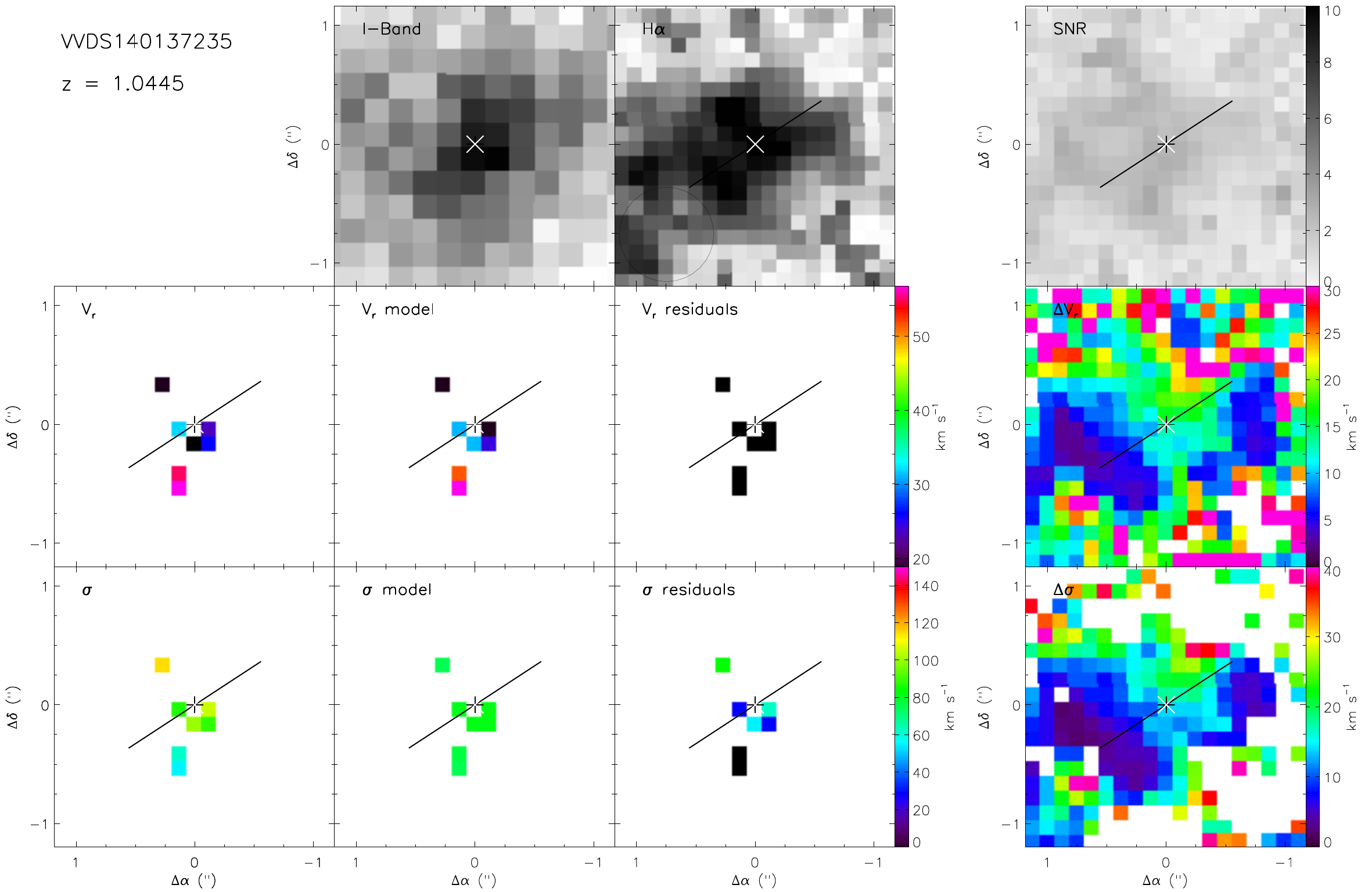}
\caption{Maps for VVDS140137235. Same caption as Figure \ref{maps_caption}.}
\end{center}
\end{figure}
\begin{figure}
\begin{center}
\includegraphics[width=16cm]{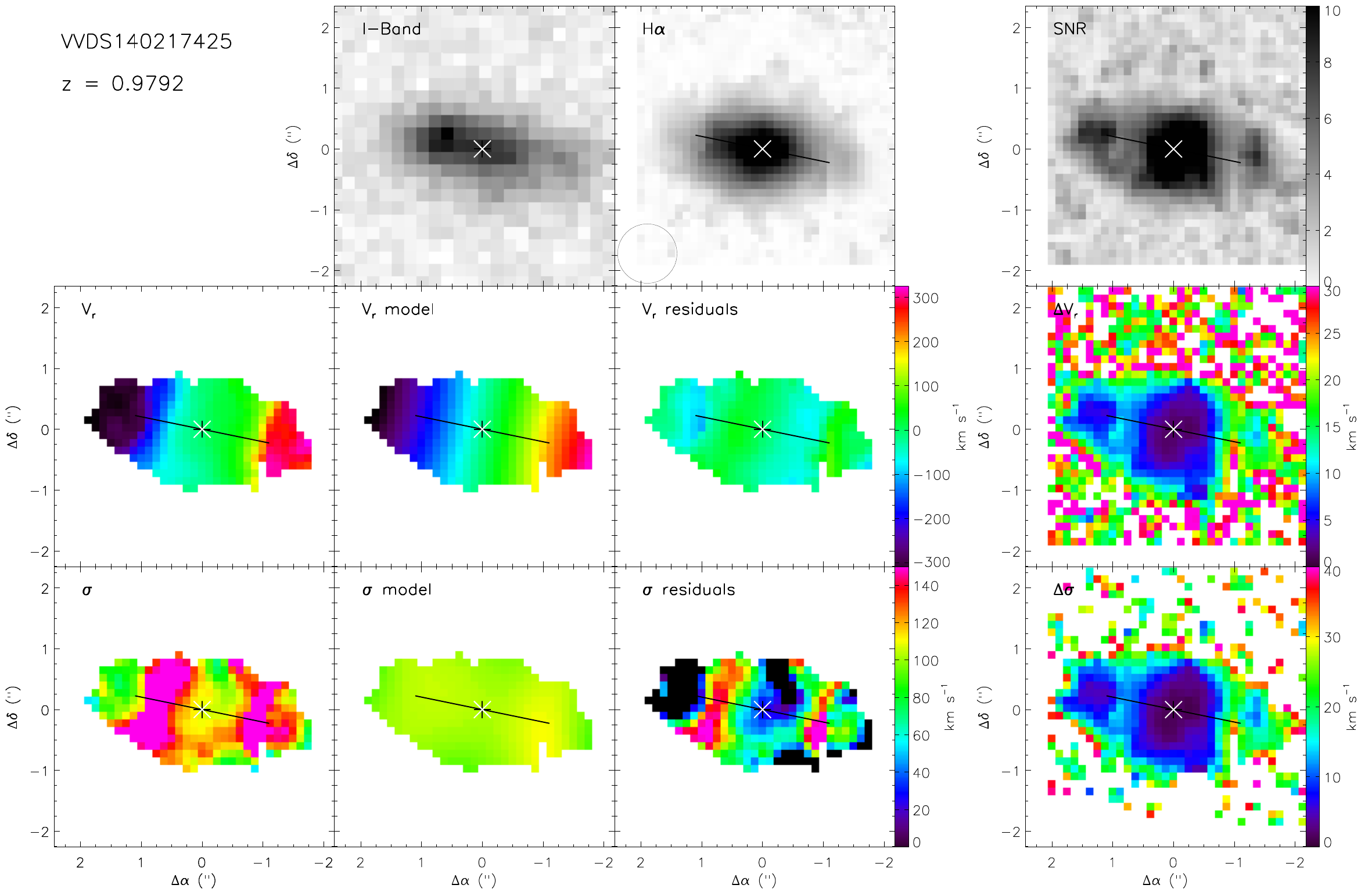}
\caption{Maps for VVDS140217425. Same caption as Figure \ref{maps_caption}.}
\end{center}
\end{figure}

\begin{figure}
\begin{center}
\includegraphics[width=16cm]{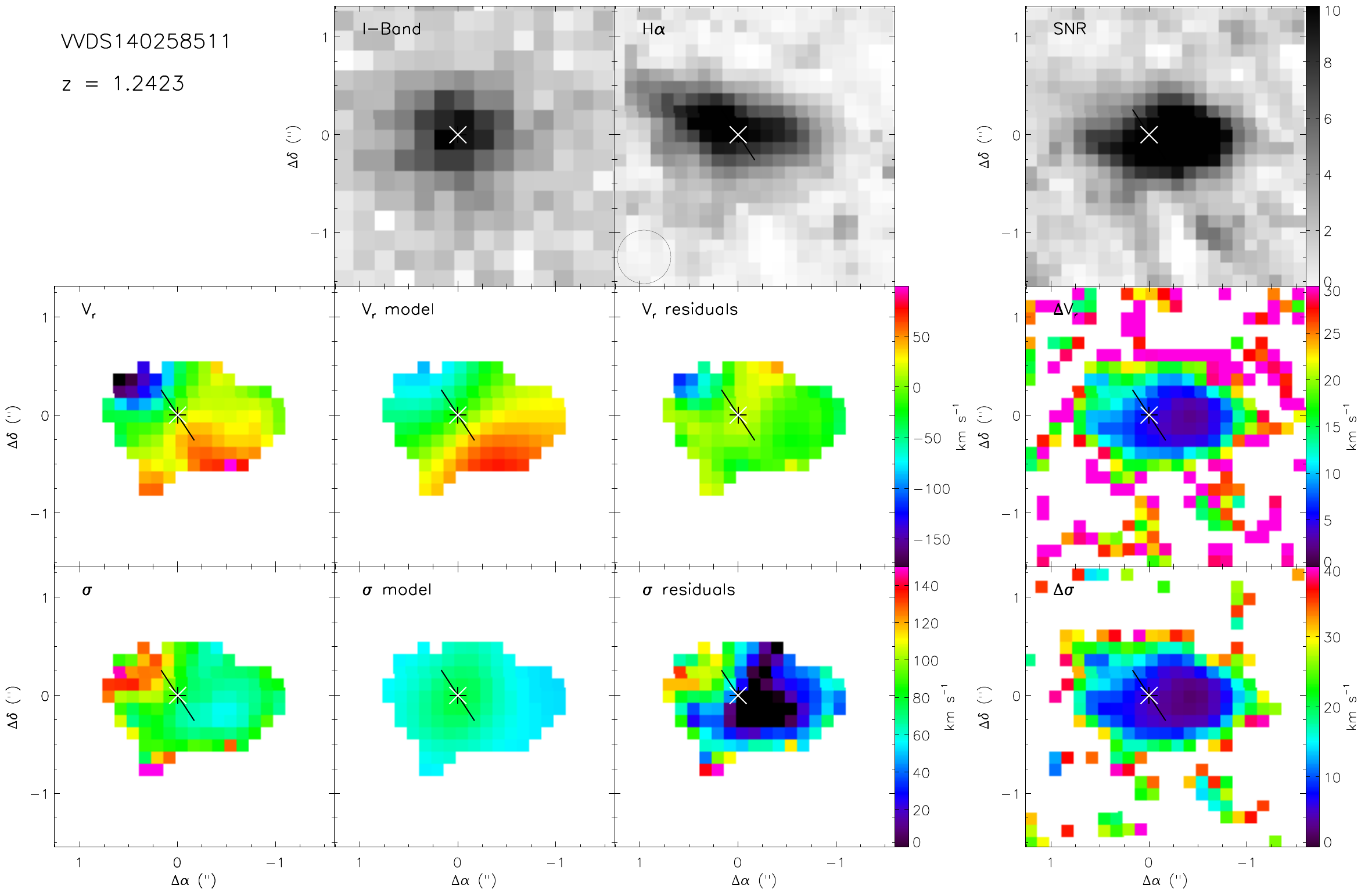}
\caption{Maps for VVDS140258511. Same caption as Figure \ref{maps_caption}.}
\end{center}
\end{figure}
\begin{figure}
\begin{center}
\includegraphics[width=16cm]{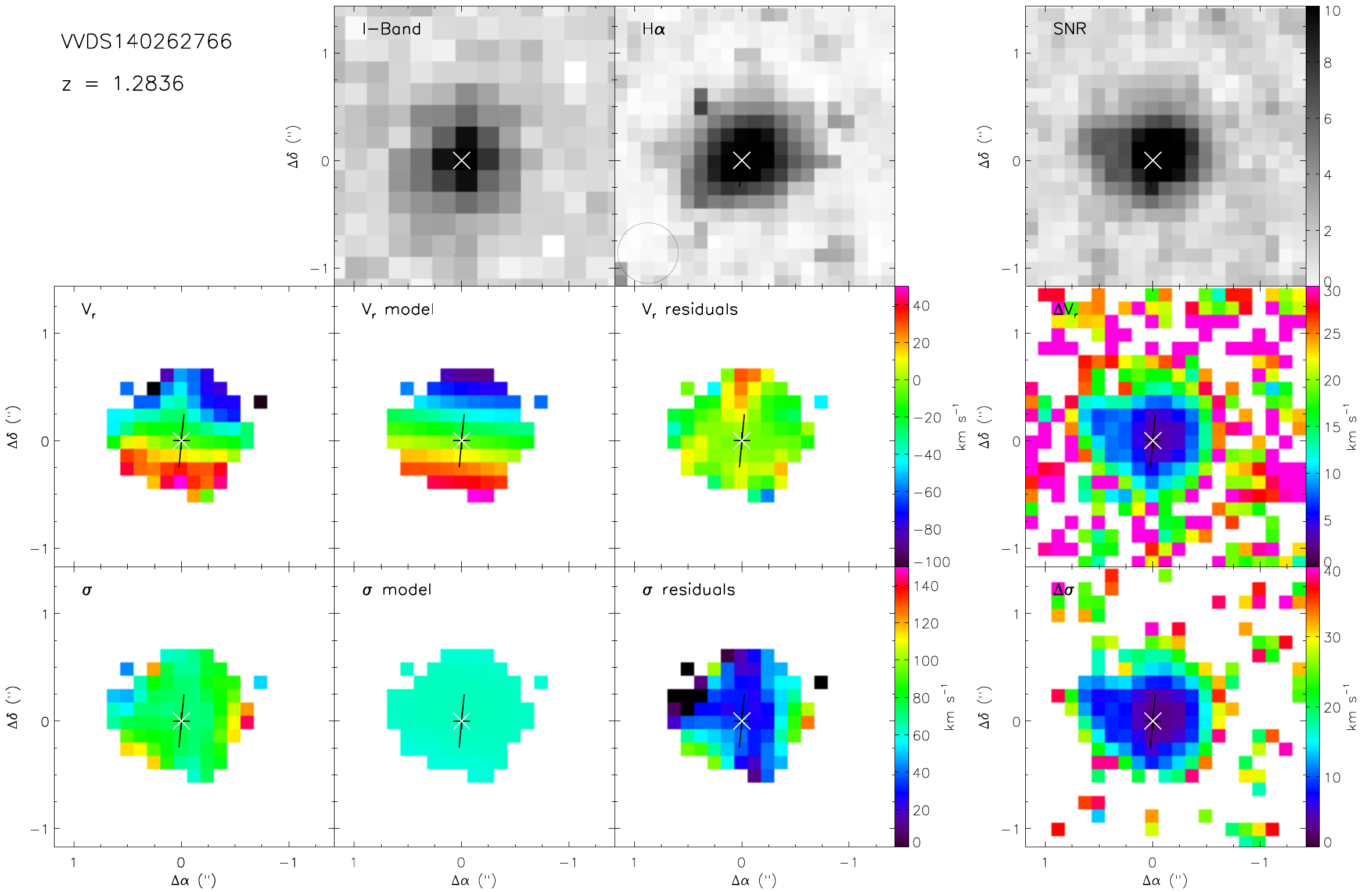}
\caption{Maps for VVDS140262766. Same caption as Figure \ref{maps_caption}.}
\end{center}
\end{figure}

\begin{figure}
\begin{center}
\includegraphics[width=16cm]{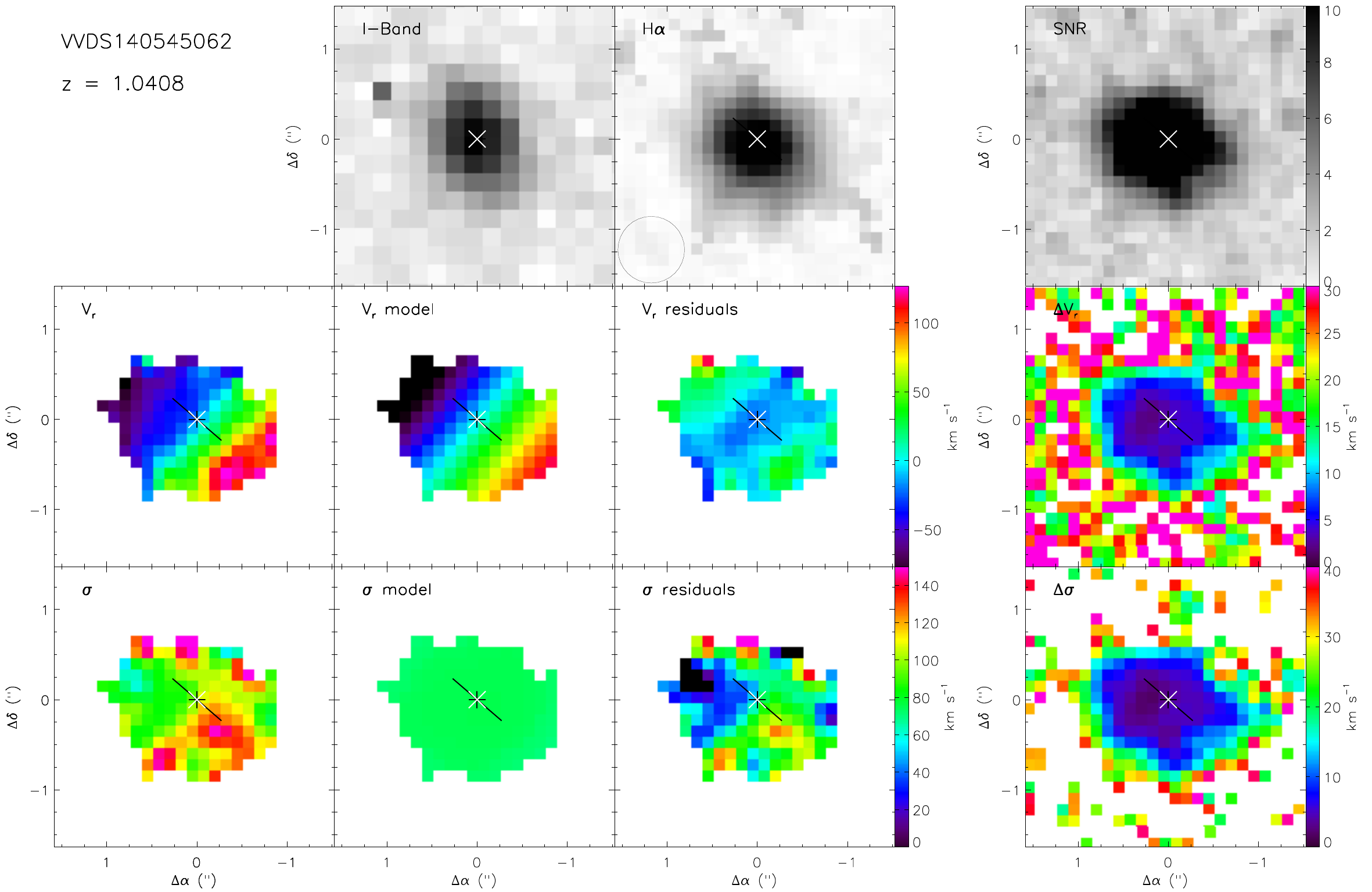}
\caption{Maps for VVDS140545062. Same caption as Figure \ref{maps_caption}.}
\end{center}
\end{figure}
\begin{figure}
\begin{center}
\includegraphics[width=16cm]{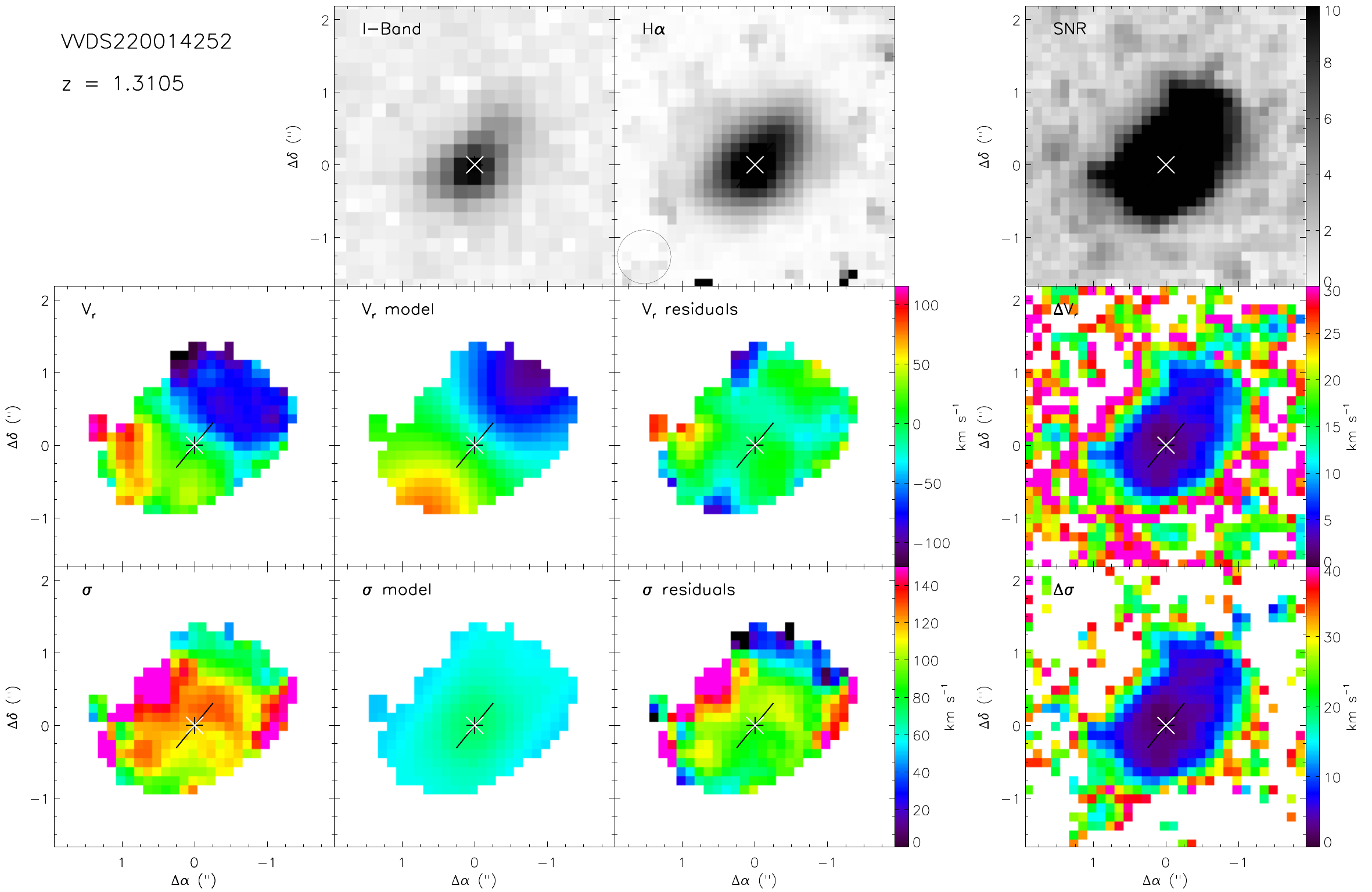}
\caption{Maps for VVDS220014252. Same caption as Figure \ref{maps_caption}.}
\end{center}
\end{figure}

\clearpage

\begin{figure}
\begin{center}
\includegraphics[width=16cm]{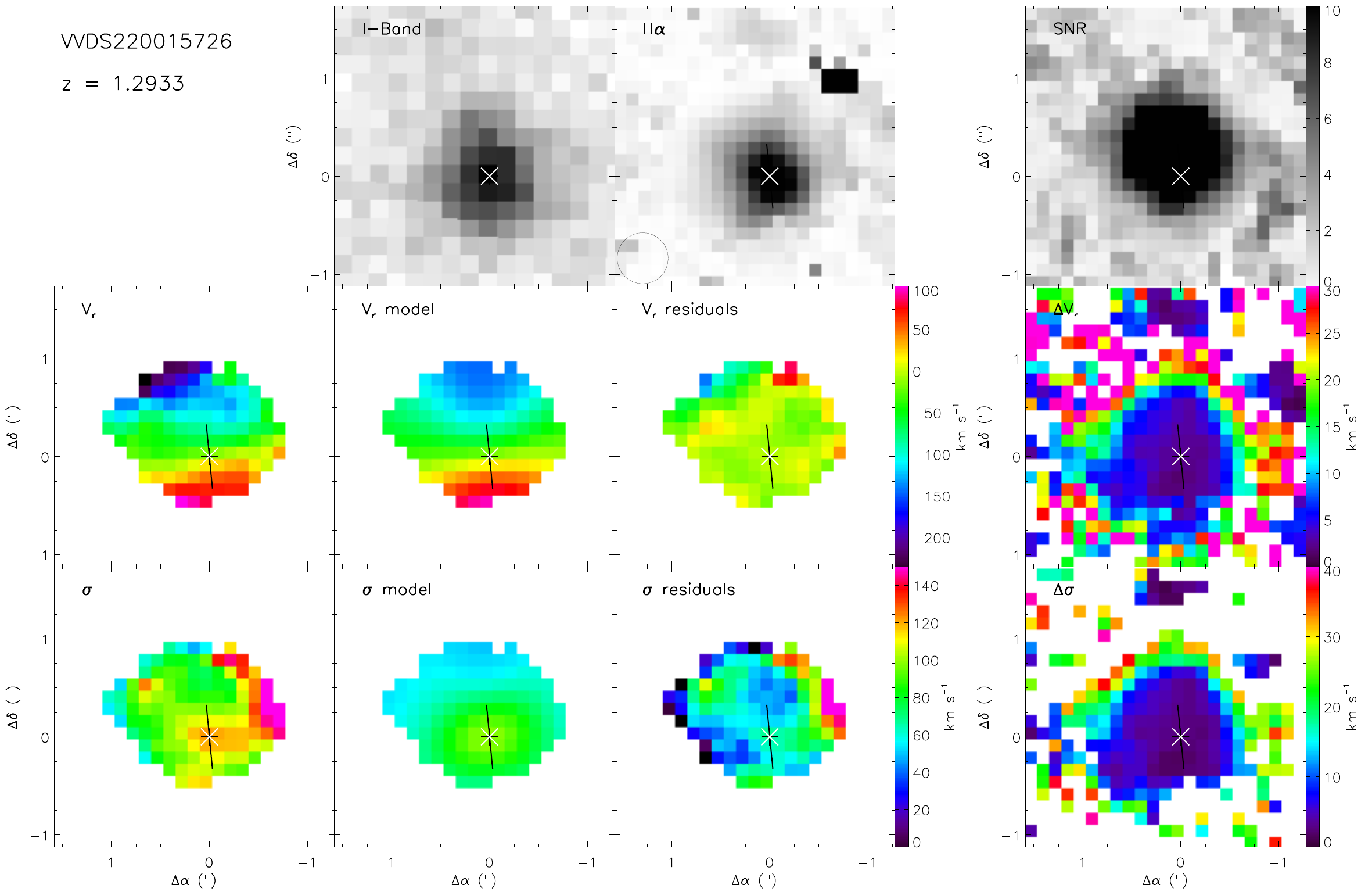}
\caption{Maps for VVDS220015726. Same caption as Figure \ref{maps_caption}.}
\end{center}
\end{figure}
\begin{figure}
\begin{center}
\includegraphics[width=16cm]{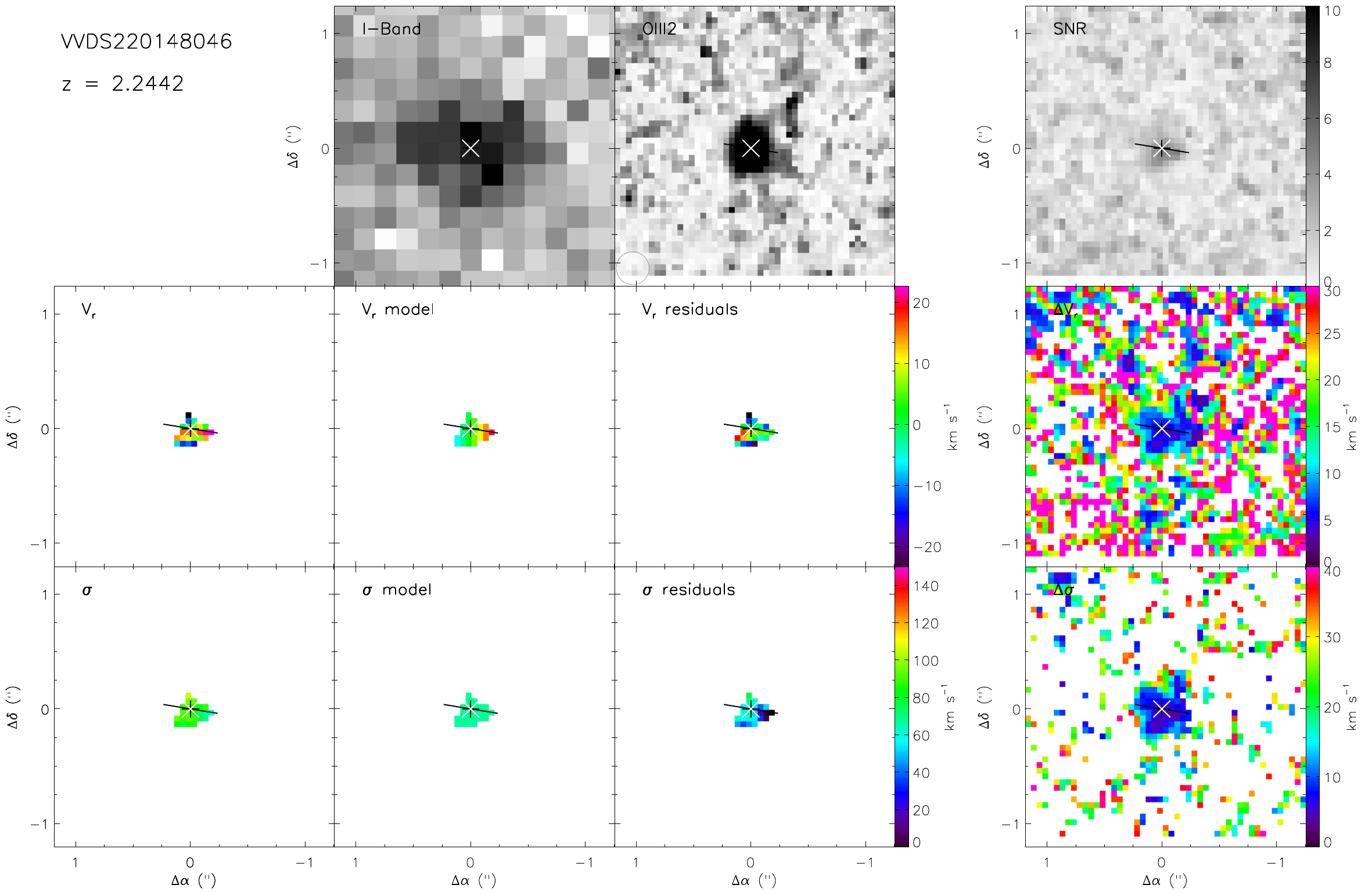}
\caption{Maps for VVDS220148046. Same caption as Figure \ref{maps_caption}.}
\end{center}
\end{figure}

\begin{figure}
\begin{center}
\includegraphics[width=16cm]{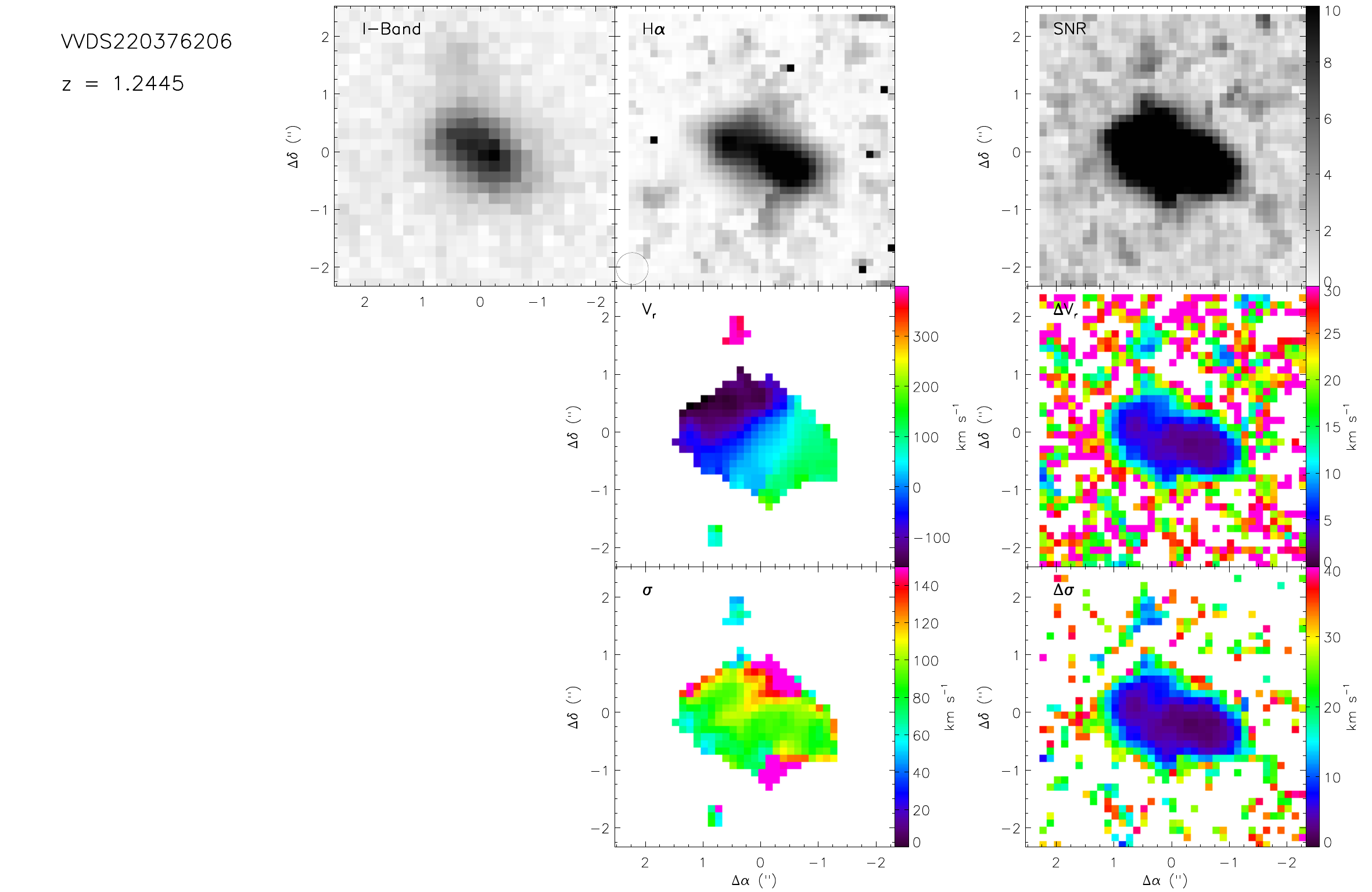}
\caption{Maps for VVDS220376206 system. Same caption as Figure \ref{maps_caption2}.}
\end{center}
\end{figure}
\begin{figure}
\begin{center}
\includegraphics[width=16cm]{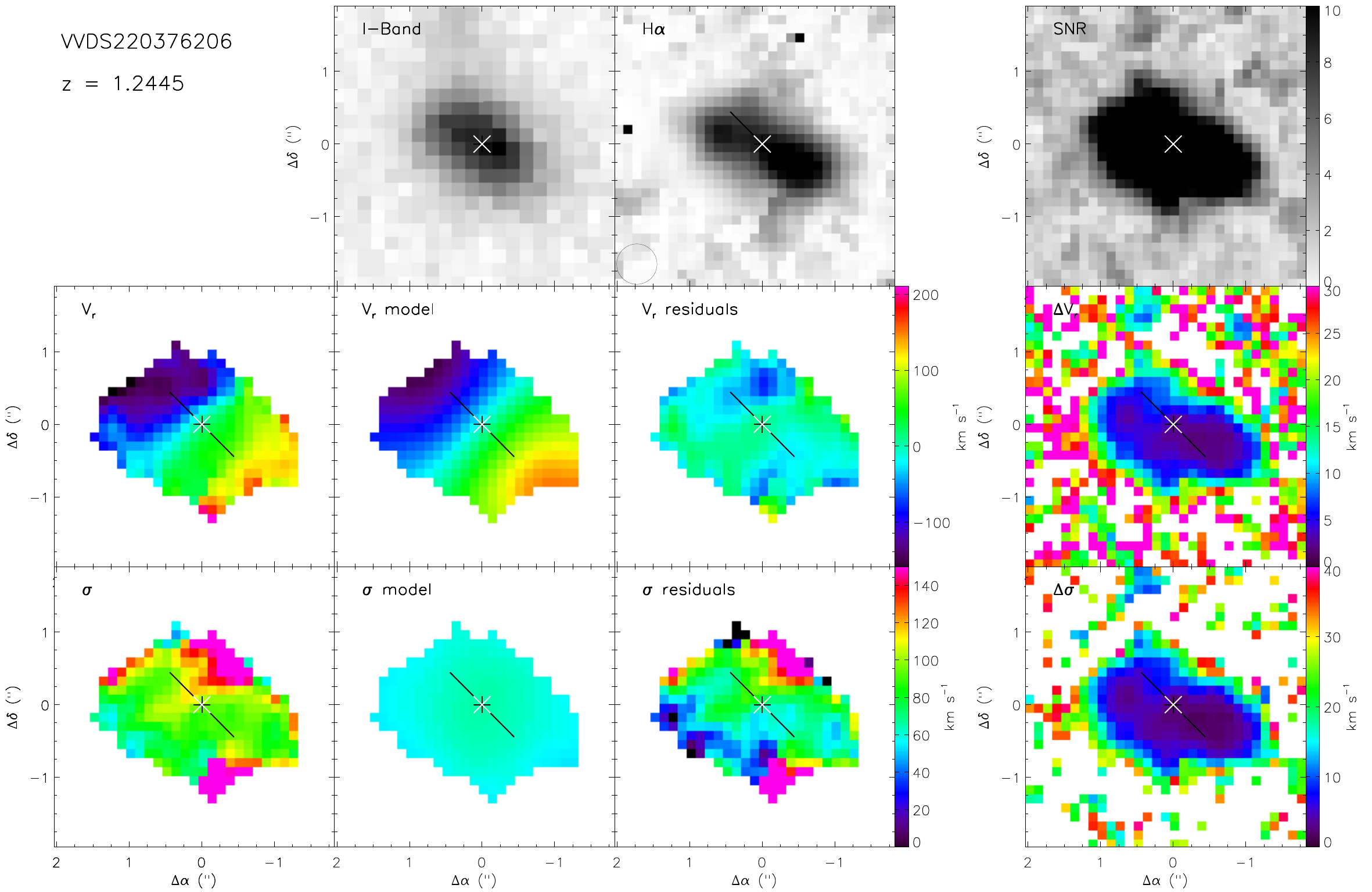}
\caption{Maps for VVDS220376206. Same caption as Figure \ref{maps_caption}.}
\end{center}
\end{figure}

\begin{figure}
\begin{center}
\includegraphics[width=16cm]{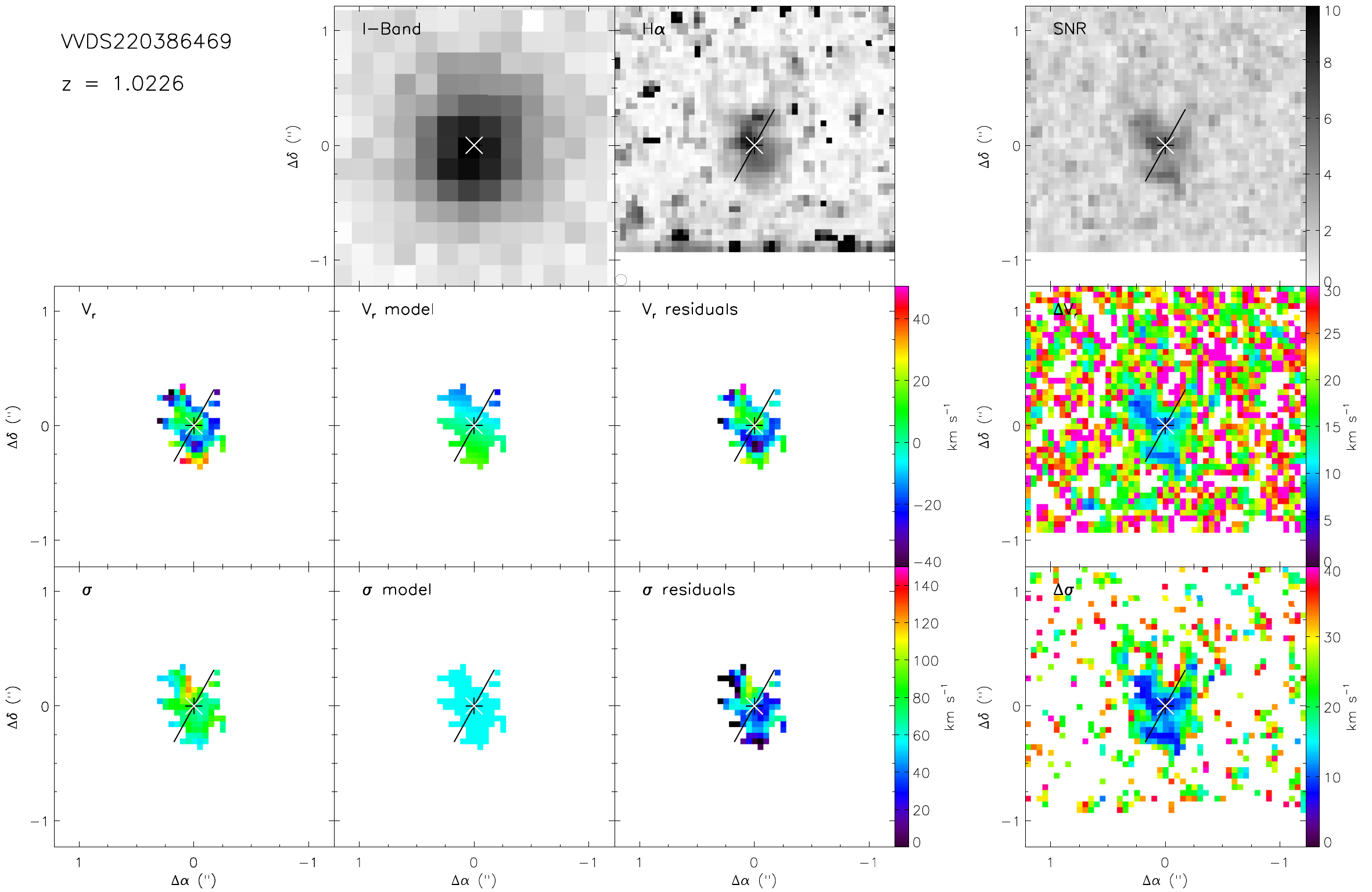}
\caption{Maps for VVDS220386469. Same caption as Figure \ref{maps_caption}.}
\end{center}
\end{figure}
\begin{figure}
\begin{center}
\includegraphics[width=16cm]{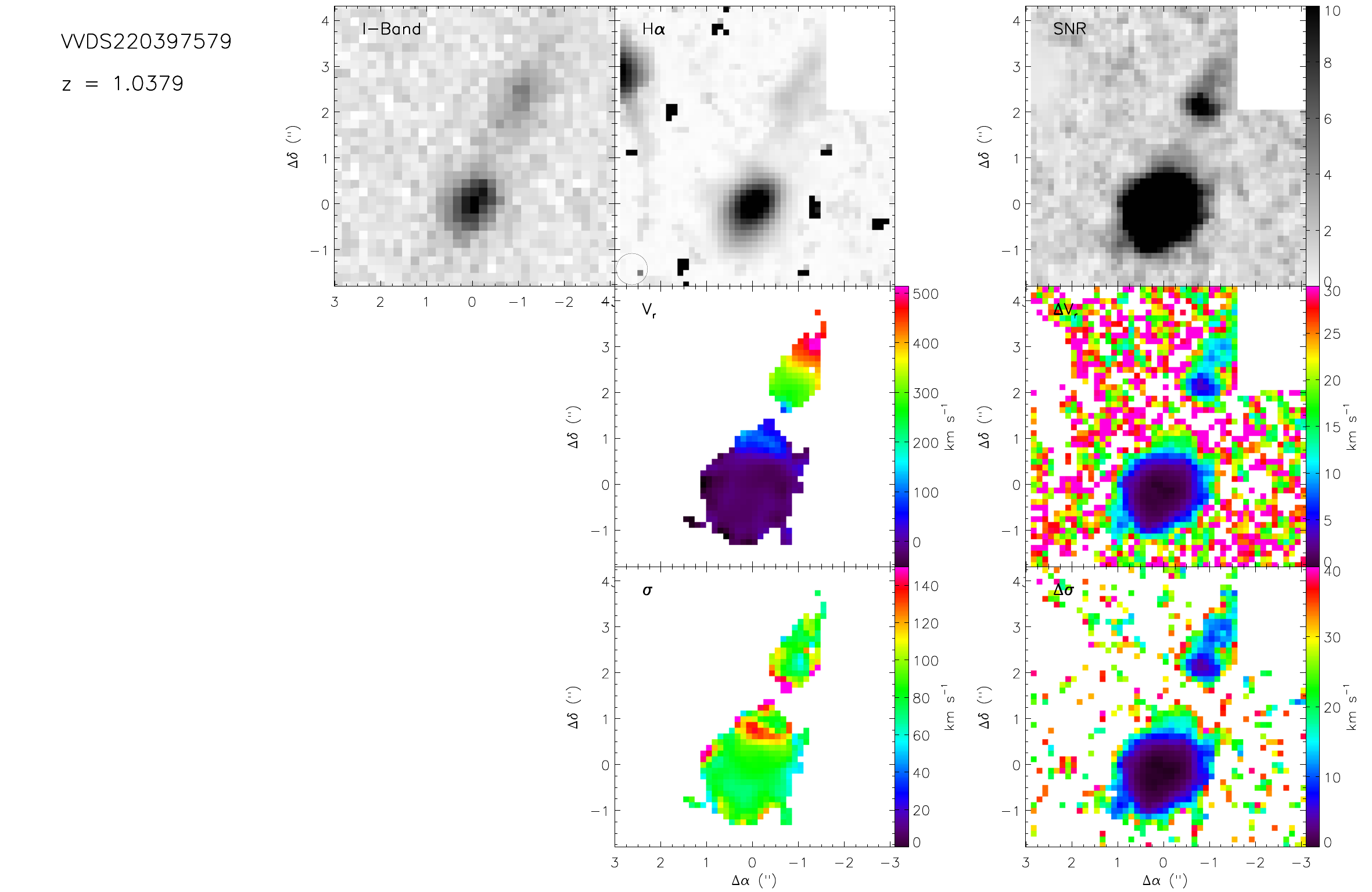}
\caption{Maps for VVDS220397579 system. Same caption as Figure \ref{maps_caption2}.}
\end{center}
\end{figure}

\begin{figure}
\begin{center}
\includegraphics[width=16cm]{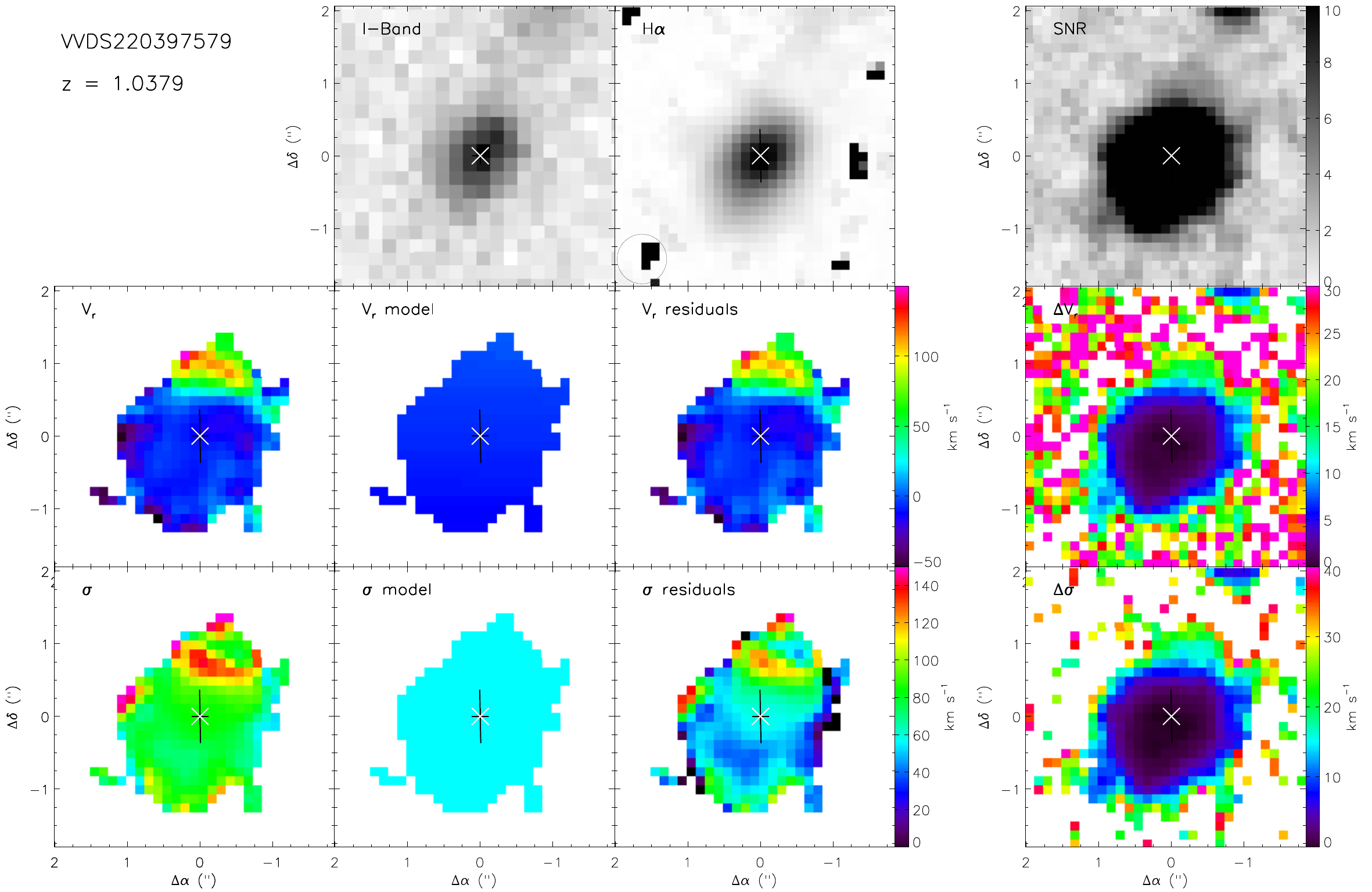}
\caption{Maps for VVDS220397579. Same caption as Figure \ref{maps_caption}.}
\end{center}
\end{figure}
\begin{figure}
\begin{center}
\includegraphics[width=16cm]{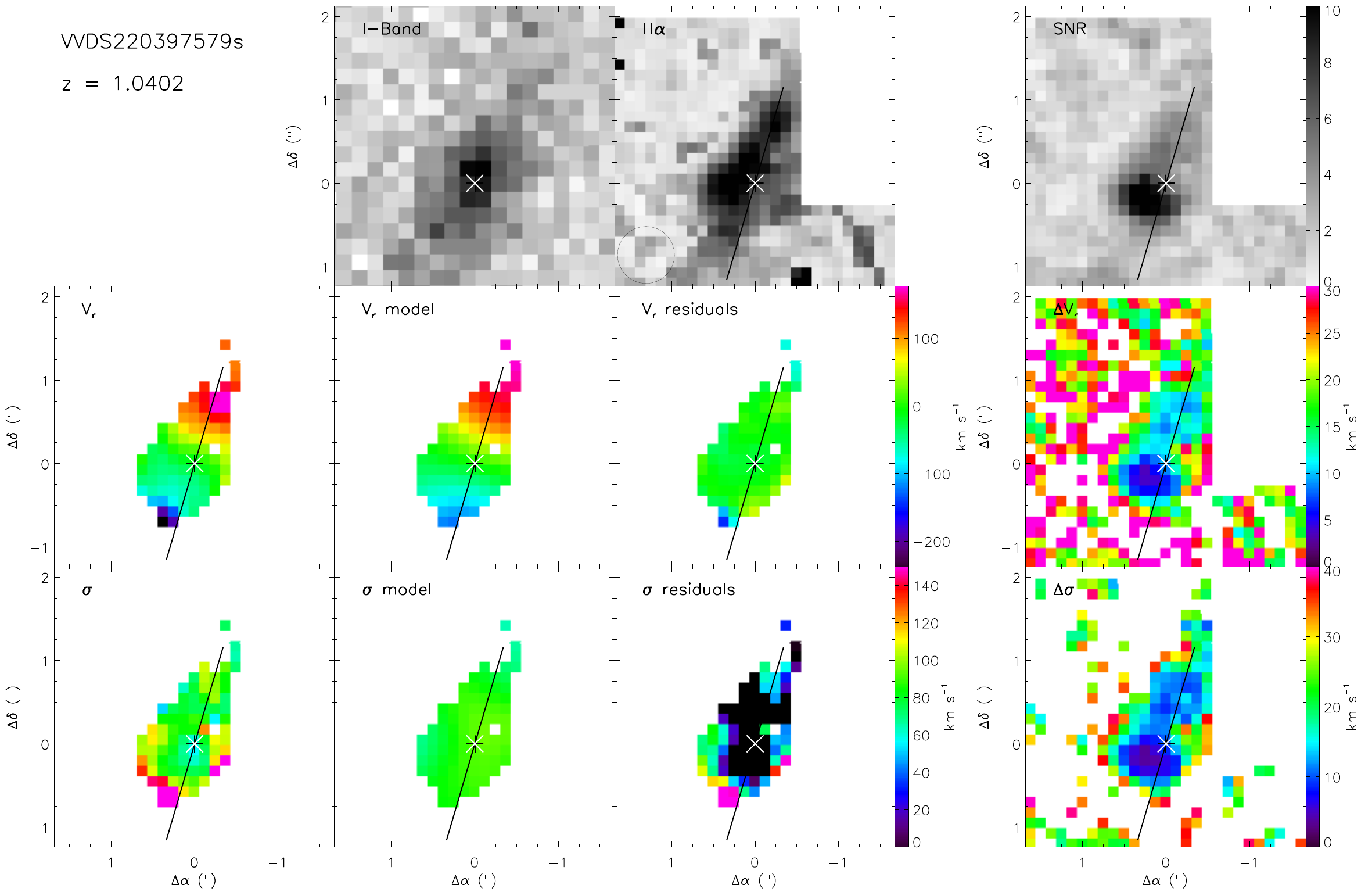}
\caption{Maps for VVDS220397579 companion. Same caption as Figure \ref{maps_caption}.}
\end{center}
\end{figure}

\begin{figure}
\begin{center}
\includegraphics[width=16cm]{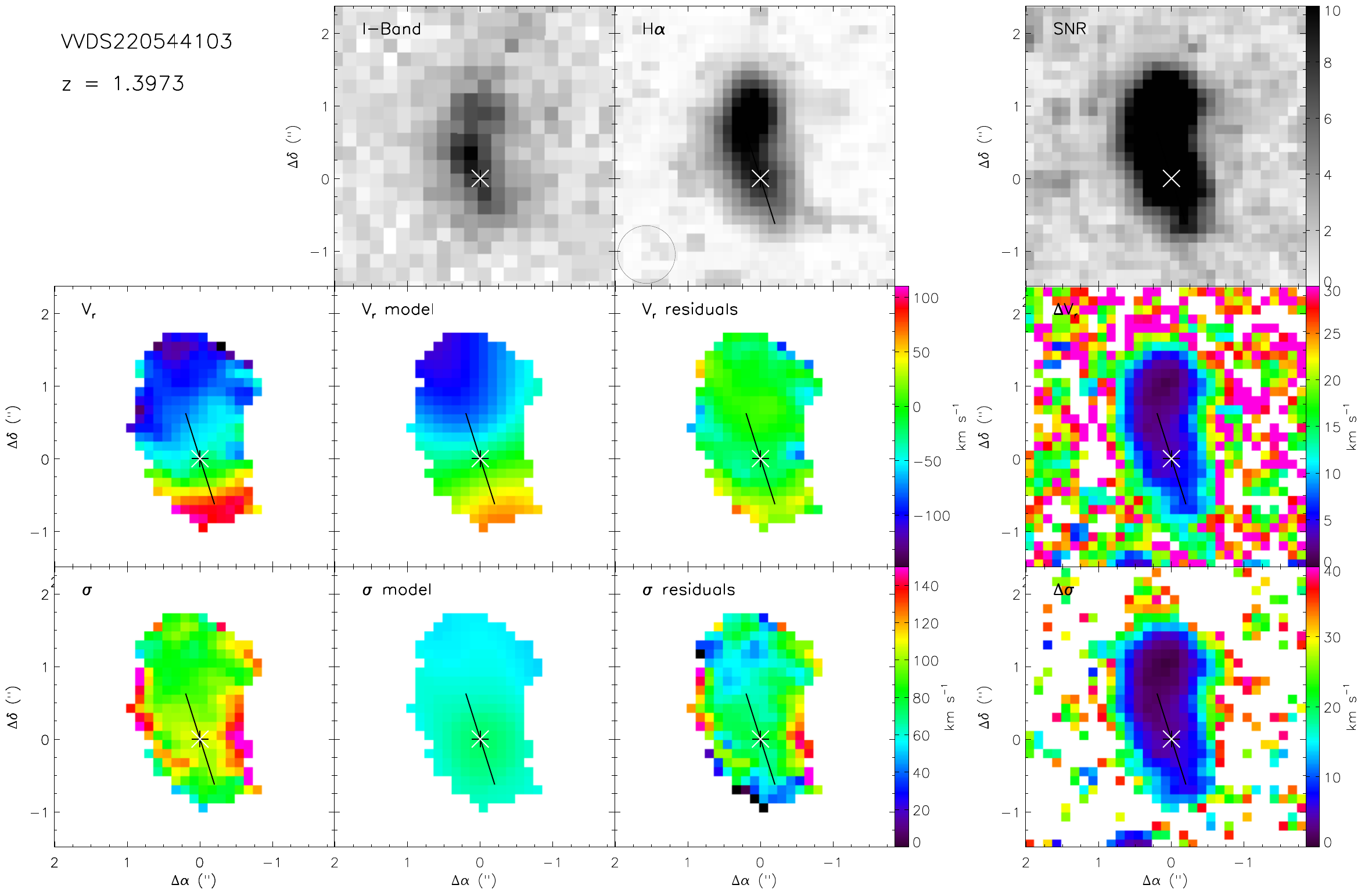}
\caption{Maps for VVDS220544103. Same caption as Figure \ref{maps_caption}.}
\end{center}
\end{figure}
\begin{figure}
\begin{center}
\includegraphics[width=16cm]{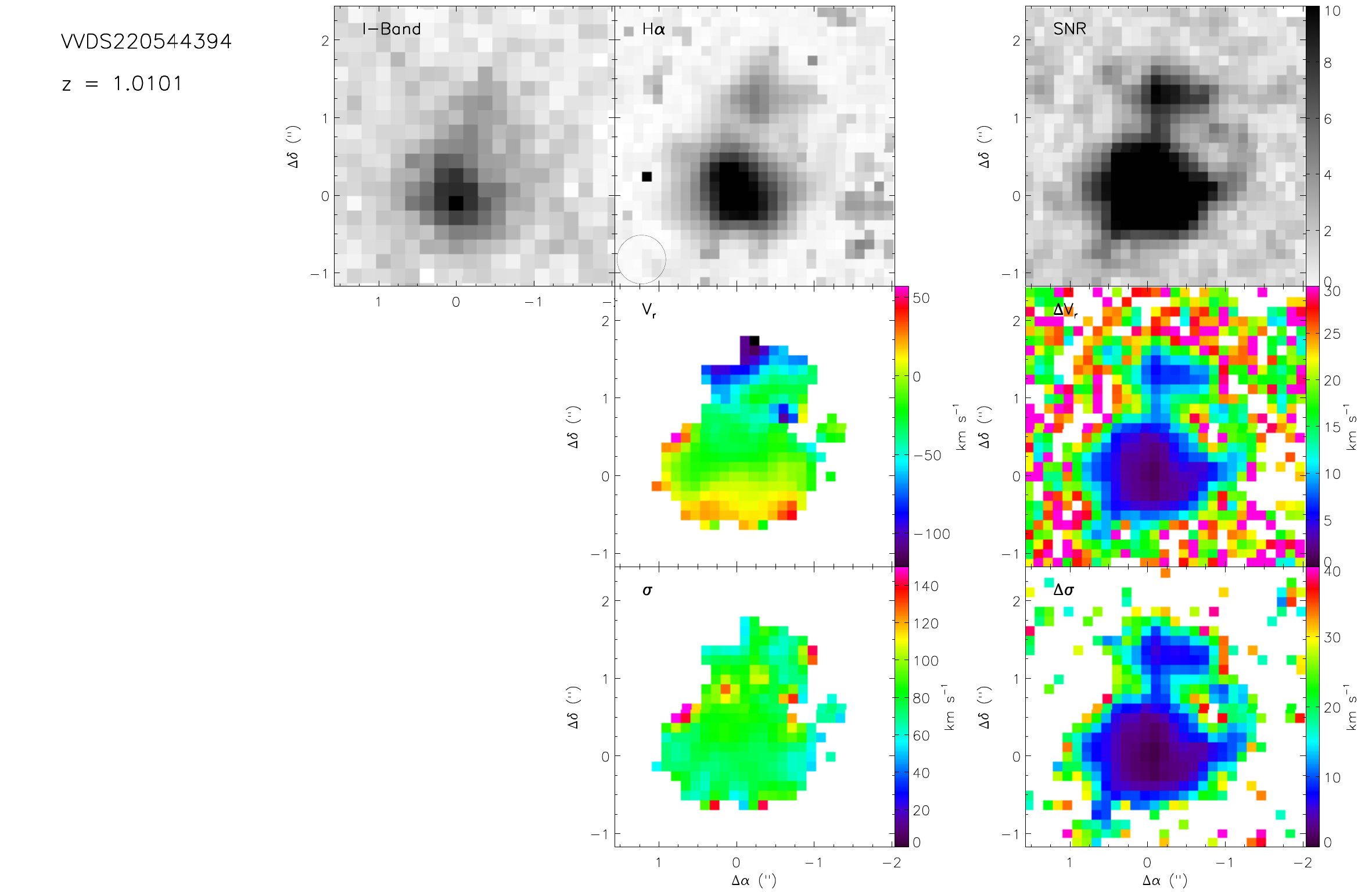}
\caption{Maps for VVDS220544394 system. Same caption as Figure \ref{maps_caption2}.}
\end{center}
\end{figure}

\clearpage

\begin{figure}
\begin{center}
\includegraphics[width=16cm]{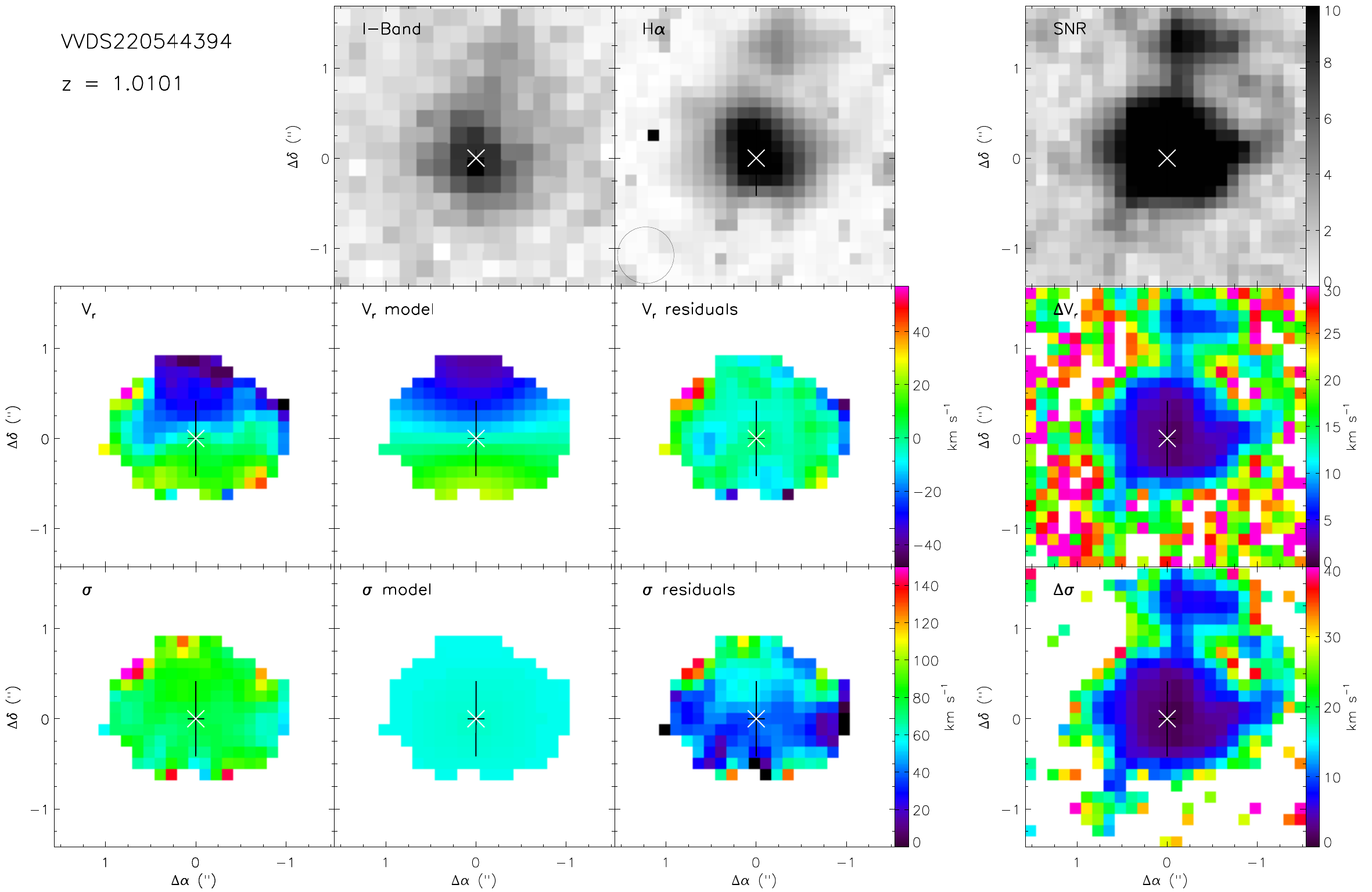}
\caption{Maps for VVDS220544394. Same caption as Figure \ref{maps_caption}.}
\end{center}
\end{figure}
% \begin{figure}
% \begin{center}
% \includegraphics[width=16cm]{figures/kinematics/companions/maps_220544394_paper}
% \caption{Maps for VVDS. Same caption as Figure \ref{maps_caption}.}
% \end{center}
% \end{figure}
\begin{figure}
\begin{center}
\includegraphics[width=16cm]{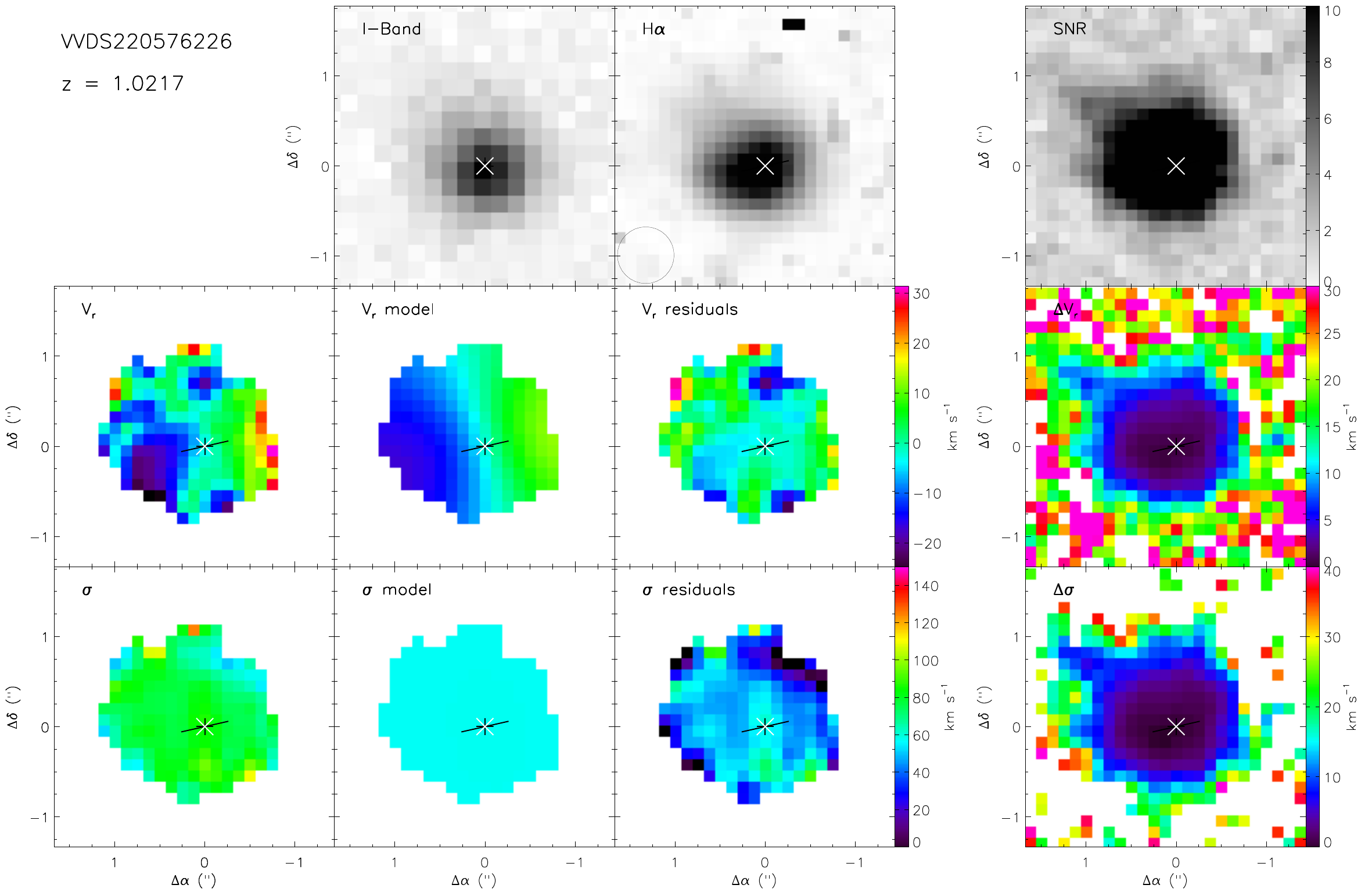}
\caption{Maps for VVDS220576226. Same caption as Figure \ref{maps_caption}.}
\end{center}
\end{figure}

\begin{figure}
\begin{center}
\includegraphics[width=16cm]{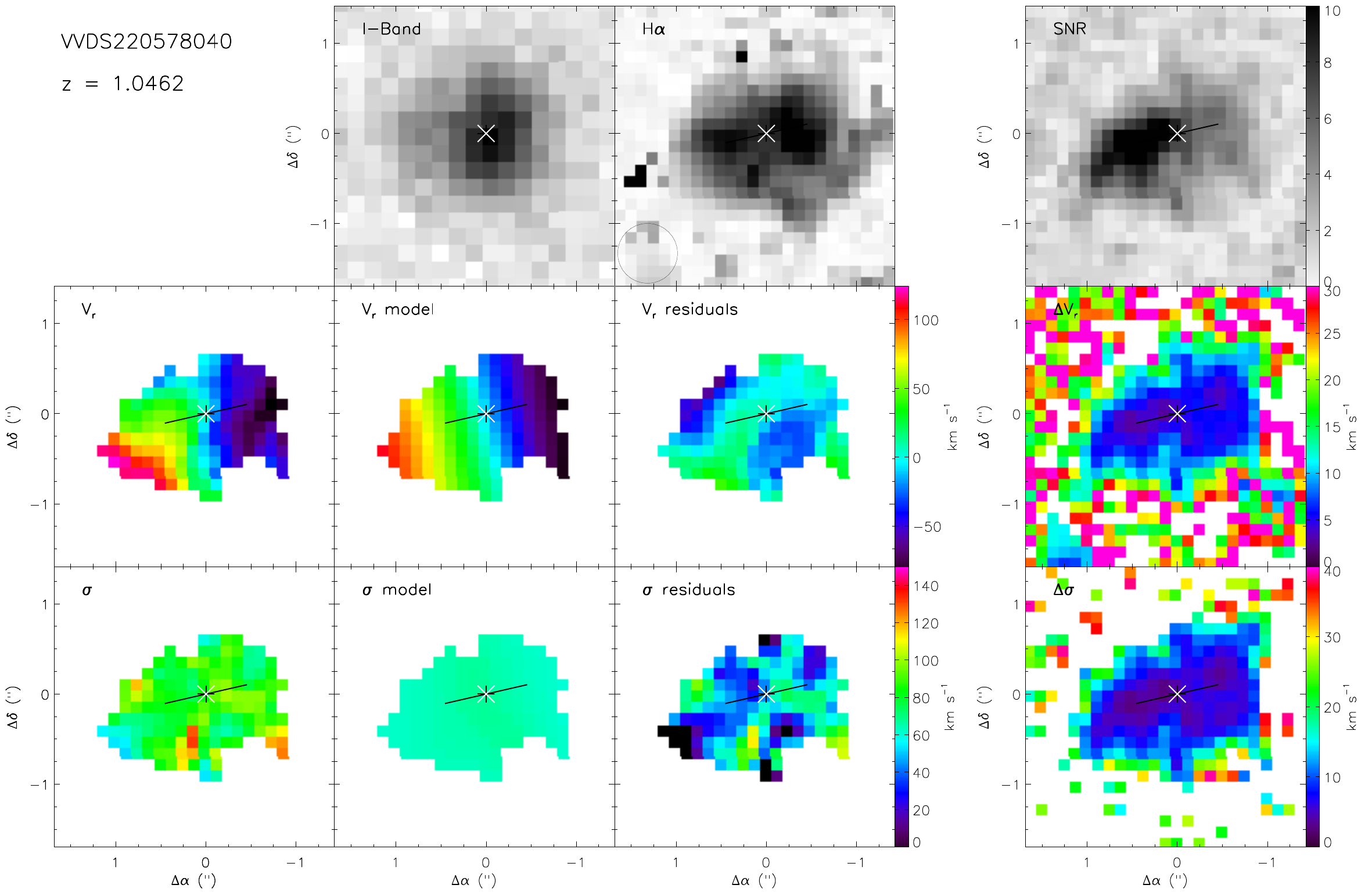}
\caption{Maps for VVDS220578040. Same caption as Figure \ref{maps_caption}.}
\end{center}
\end{figure}
\begin{figure}
\begin{center}
\includegraphics[width=16cm]{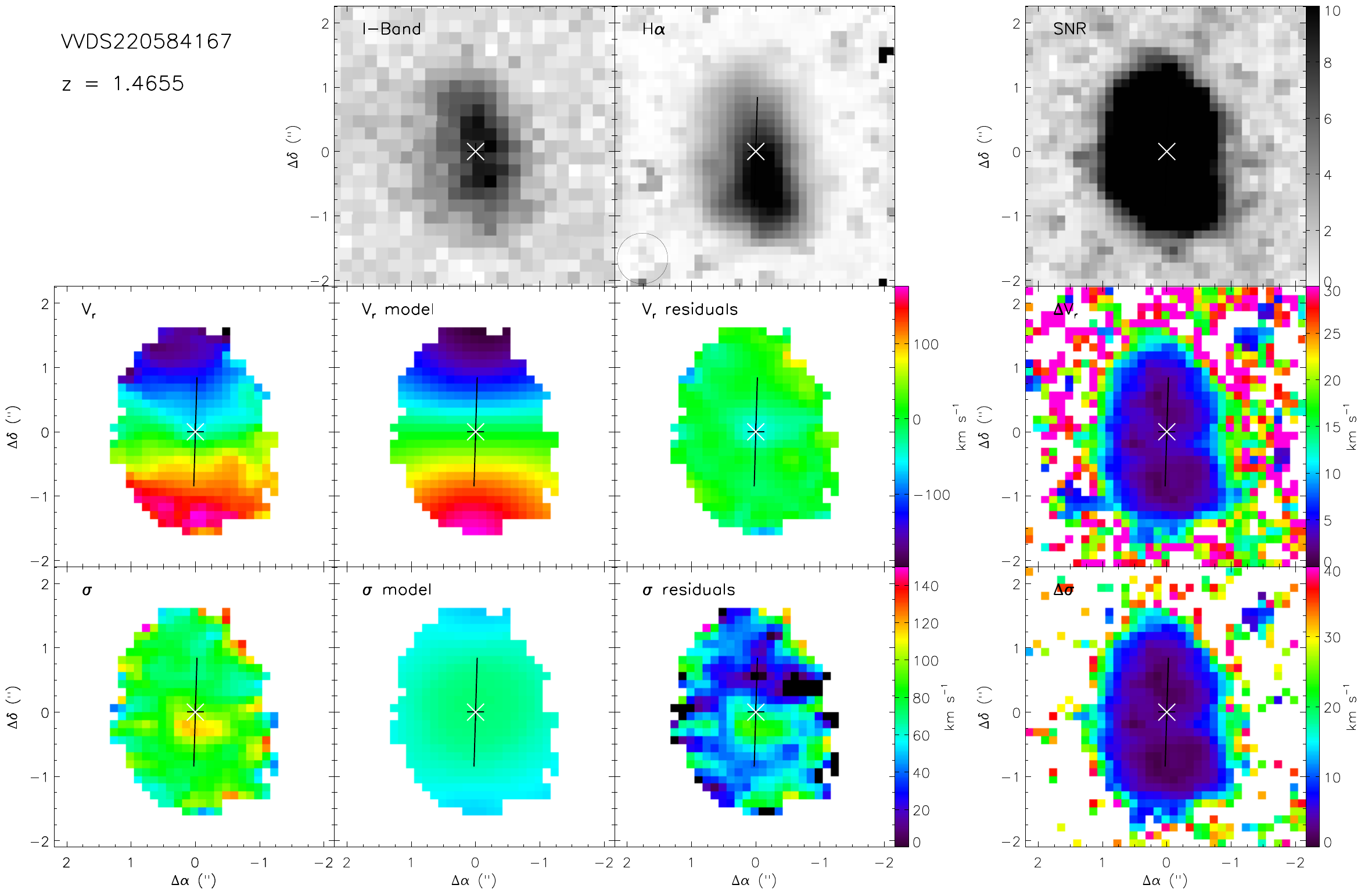}
\caption{Maps for VVDS220584167. Same caption as Figure \ref{maps_caption}.}
\end{center}
\end{figure}

\begin{figure}
\begin{center}
\includegraphics[width=16cm]{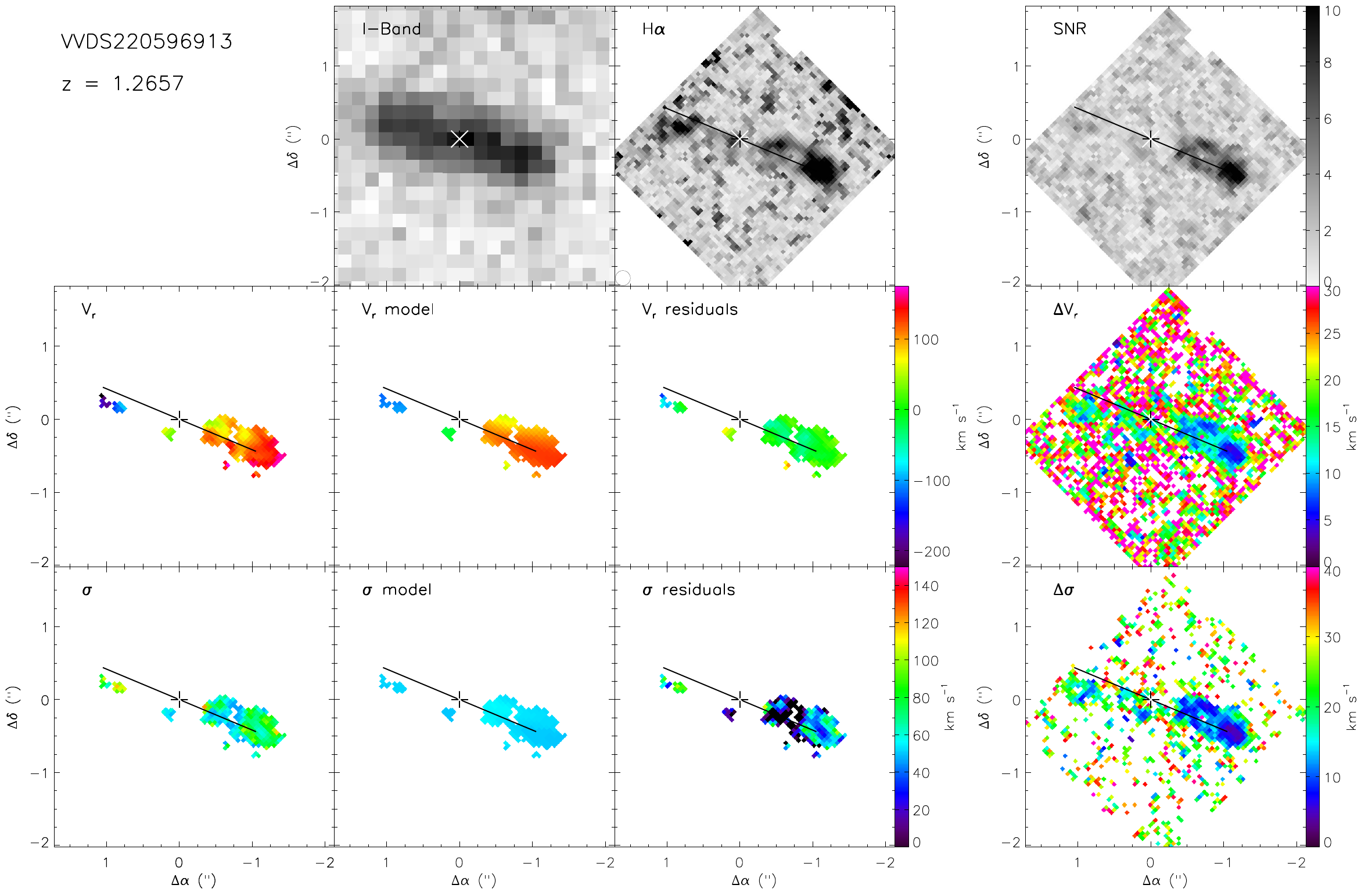}
\caption{Maps for VVDS220596913. Same caption as Figure \ref{maps_caption}.}
\end{center}
\end{figure}
\begin{figure}
\begin{center}
\includegraphics[width=16cm]{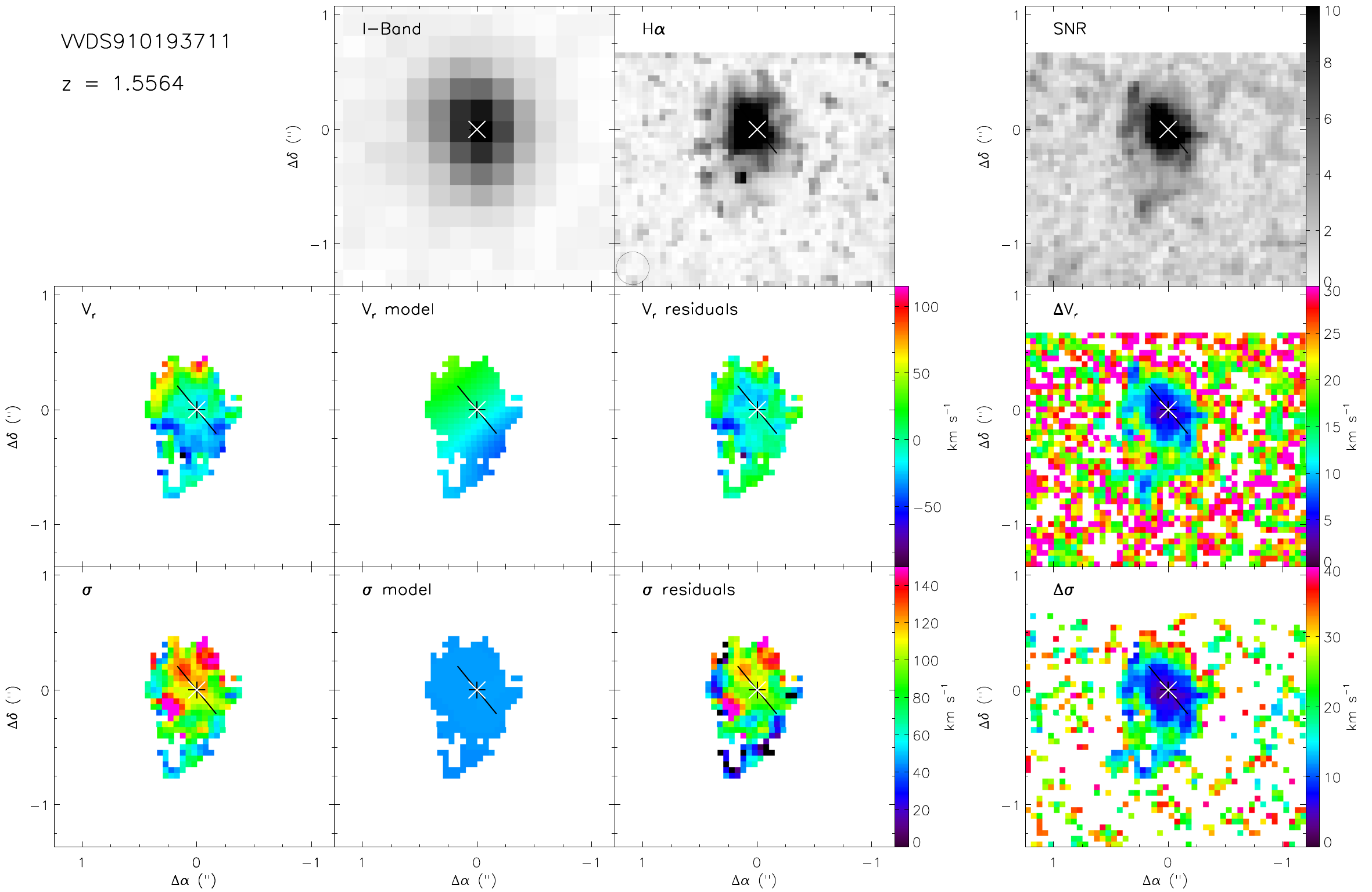}
\caption{Maps for VVDS910193711. Same caption as Figure \ref{maps_caption}.}
\end{center}
\end{figure}

\begin{figure}
\begin{center}
\includegraphics[width=16cm]{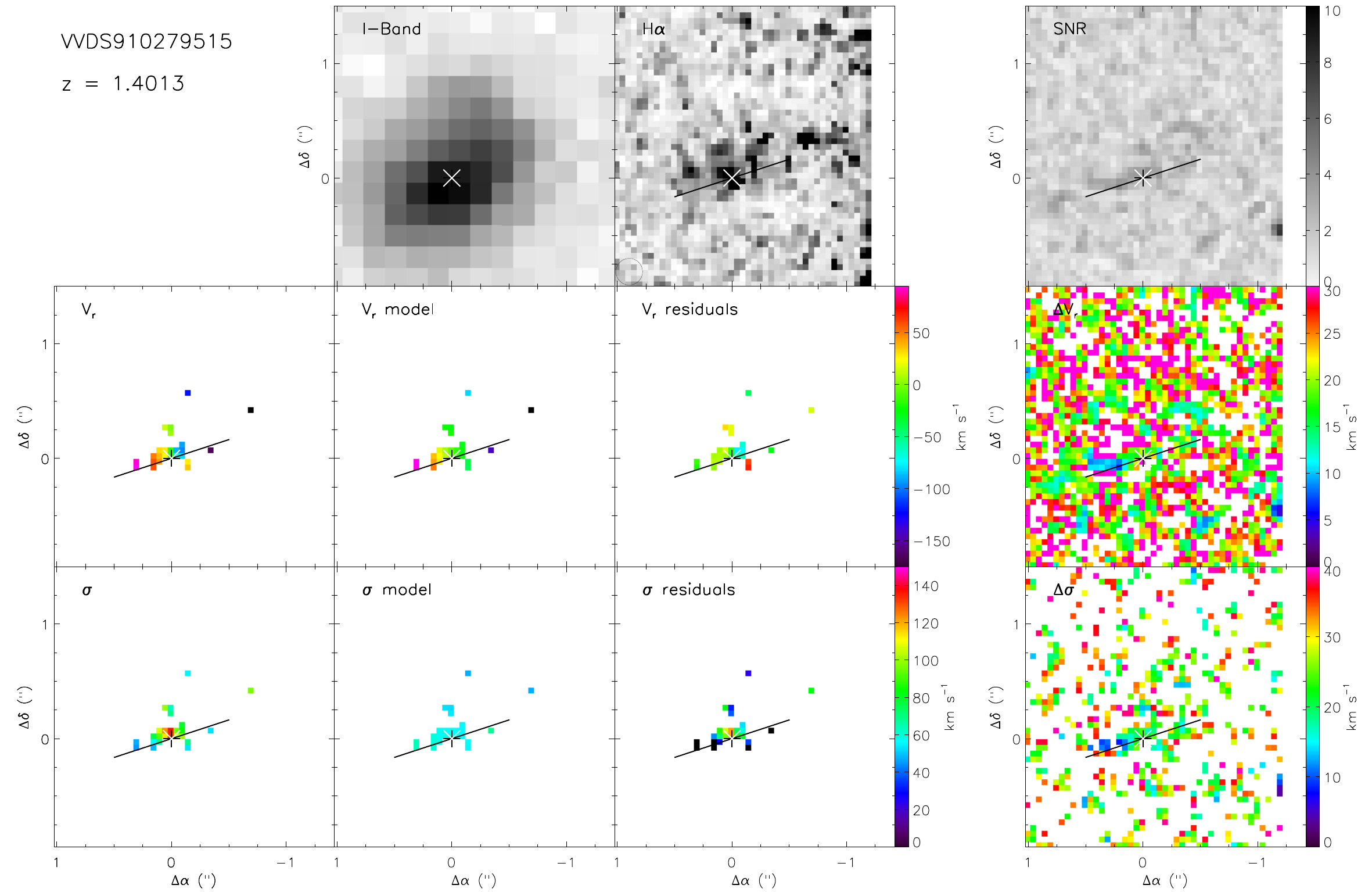}
\caption{Maps for VVDS910279515. Same caption as Figure \ref{maps_caption}.}
\end{center}
\end{figure}

\end{appendix}

\end{document}